\documentclass[11pt,a4paper]{article}
\pdfoutput=1
\usepackage{jheppub}

\usepackage{amsmath}
\usepackage{amssymb}
\usepackage{graphicx}
\usepackage{enumitem}
\usepackage{mathtools}
\usepackage{pgfplots}
\usepackage{pifont}
\usepackage[tight]{subfigure}
\usepackage[normalem]{ulem}
\usepackage{textcomp}
\usepackage{arcs}
\usepackage{dsfont}

\usepackage{tikz}

\usetikzlibrary{calc,fadings,decorations.pathreplacing,shapes,shapes.multipart,arrows,shapes.misc,intersections,positioning}

\newcommand{\norm}[1]{\left\lVert#1\right\rVert}
\newcommand{\twist}{\mathcal{T}}
\newcommand{\be}{\begin{equation}}
\newcommand{\ee}{\end{equation}}
\newcommand\eea{\end{eqnarray}}
\newcommand\bea{\begin{eqnarray}}

\newcommand{\tr}{\text{tr}}

\usepackage{bm}

\usepackage[font={footnotesize}]{caption}
\usepackage{hyperref}
\hypersetup{pdftex,colorlinks=true,allcolors=blue}
\usepackage{hypcap}

\graphicspath{{./}{./images/}}

\title{Barrier from chaos: operator entanglement dynamics of the reduced density matrix}
 
\author[a]{Huajia Wang, Tianci Zhou}
\affiliation[a]{Kavli Institute for Theoretical Physics, University of California, Santa Barbara, CA 93106, USA}

\emailAdd{huajia@kitp.ucsb.edu}
\emailAdd{tzhou@kitp.ucsb.edu}
 
\abstract{It is believed that thermalization drives the reduced density matrix of a subsystem to approach a short-range entangled operator. If the initial state is also short-range entangled, it is possible that the reduced density matrix remains low-entangled throughout thermalization; or there could exist a barrier with high operator entanglement between the initial and thermalized reduced density matrix. In this paper, we study such dynamics in three classes of models: the rational CFTs, the random unitary circuit, and the holographic CFTs, representing systems of increasing quantum chaoticity. We show that in all three classes of models, the operator entanglement (or variant of) exhibits three phases, a linear growth phase, a plateau phase, and a decay phase.  The plateau phase characterized by volume-law operator entanglement corresponds to the barrier in operator entanglement. While it is present in all three models, its persistence and exit show interesting distinctions among them. The rational CFTs have the shortest plateau phase, followed by the slowest decay phase;  the holographic CFTs mark the opposite end, i.e. having the longest plateau phase followed by a discontinuous drop; and the random unitary circuit shows the intermediate behavior. We discuss the mechanisms underlying these behaviors in operator entanglement barriers, whose persistence might serve as another measure for quantum chaoticity.}


\begin{document}
\maketitle 

\section{Introduction}

Quantum entanglement has played a central role in elucidating many recent progresses of theoretical physics, ranging from being an order parameter for topological orders \cite{topo_order_1, topo_order_2} to an explicit probe for  bulk geometries in AdS/CFT correspondence \cite{RT,HRT}. Away from the static settings, the dynamics of quantum entanglement is also a crucial and universal ingredient in understanding systems out of equilibrium \cite{nandkishore_many_2015,calabrese_entanglement_2009}. It encodes the information about thermalization and entropy generation in the process of regaining equilibrium, and probes some surprising aspects of the systems such as pre-thermalization\cite{kollar_generalized_2011,berges_prethermalization_2004,abanin_rigorous_2017,gring_relaxation_2012}, many body quantum chaos\cite{hosur_chaos_2016,srednicki_chaos_1994}, quantum scars \cite{turner_weak_2018,ho_periodic_2018,khemani_signatures_2018}, and in the context of AdS/CFT even the black hole interior behind the horizon \cite{hartman:2013}.

While the entanglement properties are usually defined and studied with respect to a state $|\psi\rangle$, one can also study the entanglement properties of an operator $U: \mathcal{H}\to\mathcal{H}$ that maps within a Hilbert space $\mathcal{H}$. Operationally we can pick a set of orthonormal basis in $\mathcal{H}$ and its dual $\overline{H}$ : $\{|i \rangle \}$ and $\{ \langle j | \}$, and write $U = \sum_{ij} U_{ij} | i \rangle \langle  j |$. The operator state can then be obtained by \footnote{Strictly speaking, one needs to specify a mapping $f: \bar{H} \rightarrow H$ in order to define the operator state. Different choices of $f$ could lead to an ambiguity in the form of a unitary transformation $V$ acting on the second copy of the operator state defined. This could affect the operator entanglement. Here we are picking a particular $f$ by requiring $( \langle i |, f( \langle j | ) ) = \delta_{ij} $. To remove such ambiguity, we could restrict the basis $\{|i \rangle \}$ and $\{ \langle i | \} $  to have the form of tensor product states on A and B, in which the ambiguous unitary transformation factorize $V=V_A\otimes V_B$ and does not affect entanglement.}
\begin{equation}
| U \rangle  = \frac{1}{\sqrt{\tr( U U^{\dagger} )} }\sum_{ij}  U_{ij} |i \rangle  | j \rangle 
\end{equation}

A convenient example to familiarize the set up is to imagine the operator being the thermal density matrix $\rho_{\beta}$, and the corresponding operator state of $\rho_\beta^{\frac{1}{2}}$ being the thermal field double (TFD) state. The \emph{operator entanglement} then refers to the entanglement of the operator state $|U\rangle$, defined with respect to a subsystem $A\subset \mathcal{H}\otimes \mathcal{H}$ \cite{prosen_operator_2007,bandyopadhyay_entangling_2005,prosen_chaos_2007,pizorn_operator_2009}. 

\begin{figure}[h]
  \centering
  \subfigure[]{
    \includegraphics[width=0.3\columnwidth]{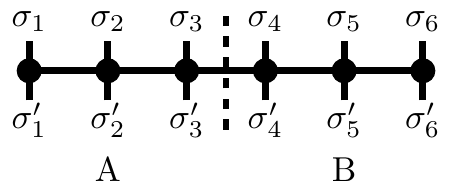}
    \label{fig:mpo}
  }\hspace{20pt}
  \subfigure[]{
    \includegraphics[width=0.55\columnwidth]{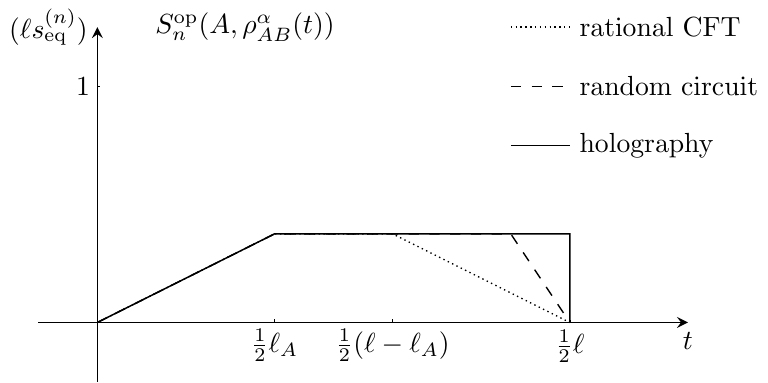}
    \label{fig:growth-plateau-drop}
  }
  \caption{Operator entanglement of the reduced density matrix. (a) Viewing the reduced density matrix as a matrix product operator. The operator entanglement is roughly logarithm of the bond dimension. The physics indices $\sigma$ and $\sigma'$ are the indices in the operator Hilbert space. (b) The time dependence operator entanglement or reduced density matrix in 3 representative systems. Subsystem A is contiguous and has size $\ell_A$. Total system has length $\ell$. The Entanglement quantities are displayed in units of the corresponding state equilibrium entropy density $s^{(n)}_{\rm eq}$ in each system, see section \ref{sec:discuss} for more detailed explanations.}
\end{figure}

In this paper, we study a particular class of dynamical operator entanglement: those associated with the reduced density matrices obtained from quenched states\footnote{There are related works recently on the operator entanglement \cite{nie_signature_2018} and negativity \cite{kudler-flam_quantum_2019} of the unitary evolution matrix, which are different from the operator entanglement of {\it reduced density matrix} studied in this work.} \footnote{Shortly before the post of this work, we became aware of a new preprint \cite{kusuki_dynamics_2019}, which contained the result of the operator entanglement in a local operator quench.}. Let us lay out the set up step by step. Given a quenched state $|\psi(t)\rangle$ that starts with only short-range entanglement at $t=0$, and is defined on the total system $ABC$, we first obtain the reduced density matrix on the subsystem $AB$ by tracing out $\mathcal{H}_C$: 
\begin{equation} 
\rho_{AB}(t) = \text{Tr}_{\mathcal{H}_C} |\psi(t)\rangle \langle \psi(t)|
\end{equation}
For the interest of this paper, we take $C$ to be semi-infinite that surrounds the finite subsystem $AB$ like a heat-bath. The wave-functional of the corresponding (unnormalized) operator state $|\rho_{AB}(t)\rangle$ on the doubled subsystem Hilbert space $\mathcal{H}_{AB}\otimes \mathcal{H}_{AB}$ (see Fig.~\ref{fig:mpo}) is then given by the density matrix element $(\rho_{AB}(t))_{ij}  = \langle i | \rho_{AB}(t) | j \rangle $ as \cite{choi_completely_1975,jamiolkowski_linear_1972}:
\begin{equation}
  | \rho_{AB}(t) \rangle = \sum_{ij} ({\rho_{AB}(t)})_{ij} | i \rangle  | j \rangle
\end{equation}
where $\{ |i \rangle \}$ is an orthonormal basis in $\mathcal{H}_{AB}$. 

To normalize the state, we need the norm of the $|\rho_{AB}(t)\rangle$, which is the purity of the state
\begin{equation}
\norm{\rho_{AB}(t)}^2 = { \langle \rho_{AB}(t)  | \rho_{AB} (t) \rangle }  = \tr( \rho^2_{AB}(t) ) 
\end{equation}
It is in general less than $1$ and $t$-dependent when region AB is entangled with C. 

We are interested in computing the operator R\'enyi entropy of the normalized operator state $|\rho_{AB}(t)\rangle/ \norm{\rho_{AB}(t)}$ on the subsystem $A\otimes A$:
\begin{eqnarray}
\label{eq:S_op_two_term_1}
  &&S_{n}^{ \rm op}\left(A, \rho_{AB}(t)\right)  =  \frac{1}{1 -n} \log \tr_{\mathcal{H}_A\otimes \mathcal{H}_A}\left[\tr_{\mathcal{H}_B\otimes \mathcal{H}_B}( |\rho_{AB}(t) \rangle \langle \rho_{AB}(t) | / \norm{\rho_{AB}(t)}^2 )\right]^n \nonumber\\
  &=&   \frac{1}{1 - n} \log \tr_{\mathcal{H}_A\otimes \mathcal{H}_A}\left[\tr_{\mathcal{H}_B\otimes \mathcal{H}_B}( |\rho_{AB}(t) \rangle \langle \rho_{AB} (t)| )\right]^n - \frac{n}{1-n} S_2 \left(AB,\psi(t) \right)
\end{eqnarray}
where we use $S^{\text{op}}_n\left(A, \rho_{AB}\right)$ to denote the operator R\'enyi entropy of $\rho_{AB}$, and $S_n\left(AB,\psi\right)$ as the state entanglement of the quenched state $|\psi\rangle$. As $|\psi(t)\rangle$ evolves in time $t$, so does $\rho_{AB}(t)$ and consequently $S_n^{\rm op}\left(A, \rho_{AB}(t)\right)$. Finding out the $t$-dependent dynamics of $S^{\text{op}}_n\left(A, \rho_{AB}(t)\right)$ is the main task of this paper.

An important motivation to study the dynamics of such operator entanglement comes from the following question that is both theoretically interesting and of potential practical significance: is it possible to simulate the density matrix efficiently in the matrix product operator (MPO) framework {\it throughout} the thermalization process? Since large operator entanglement indicates large bond dimension in a MPO (Fig.~\ref{fig:mpo}), the question can be quantified as weather or not to have low operator entanglement all the way to thermalization.

Various proposals  \cite{pizorn_operator_2009,haegeman_time-dependent_2011,haegeman_unifying_2016,leviatan_quantum_2017} have been made for directly simulating the reduced density matrix to study the evolution to the thermal states. The operator entanglement, crudely speaking, measures the logarithm the bond dimension to represent the reduced density matrix by a matrix product operator (Fig.~\ref{fig:mpo}). In the process of quenched thermalization, the initial state is short-range entangled, and as a result the operator state is also short-range entangled with a low operator entanglement. On the other hand, if the system thermalizes, the final reduced density matrix will be close to an identity operator at high temperature. Thus while the state entanglement increases and saturates to volume-law, the operator entanglement is again very small. Assuming this feature of low operator entanglement persists in between, numerical algorithms that use variational approaches \cite{haegeman_time-dependent_2011,haegeman_unifying_2016,leviatan_quantum_2017} at the low bond dimension manifold should be enough to reproduce all the correlation functions faithfully.

The question is whether the feature of low operator entanglement persists or not. In general, the time-dependence of operator entanglement may result in intermediate ``bumps". If the values at the bumps are comparable to system size, then we should think of them as barriers in operator entanglement, i.e. barriers for efficient simulation of the density matrix.  

In this paper we address this question in three classes of models, the two dimensional rational CFTs, the random unitary circuit, and the holographic CFTs. They represent systems with increasing quantum chaoticity \cite{maldacena_bound_2015}. All of the three models give a ``growth-plateau-drop'' pattern as shown in Fig.~\ref{fig:growth-plateau-drop}. The plateau after the initial growth is of volume-law value, and thus is a barrier that prevents efficient simulation using the MPO representation. The results are briefly summarized in the following outline of the paper. 

In section \ref{sec:CFT}, we study the problem in 2-dimensional CFTs. The computation was performed in Ref.~\cite{dubail_entanglement_2017} under the implicit assumption that the theory is rational. We redo the computation and allow the operator to be the powers of the reduced density matrix, i.e. $|\rho^\alpha_{AB}\rangle, \alpha \in \mathbb{N}$. We find that the story is similar to what is raised up in Ref.~\cite{asplund_entanglement_2015}: the rational CFT assumption entails additional OPE regimes among wist operators, and a quasi-particle interpretation which gives the result in Fig.~\ref{fig:growth-plateau-drop}. For irrational CFTs, the corresponding OPE regime disappears and explicit calculations become difficult. 

In section \ref{sec:random}, we perform the computation in a random unitary circuit using tools developed in Refs.~\cite{zhou_emergent_2018,jonay_coarse-grained_2018,nahum_quantum_2017}. The random unitary circuit is a recently proposed quantum circuit model that is believed to capture the universal features of chaotic dynamics with local interactions. The circuit averaged entanglement in this model is mapped to a statistical mechanical problem of domain walls. At different stages of the evolution, three different types of the domain wall configurations dominate, which gives rise to the three continuous segments in Fig.~\ref{fig:growth-plateau-drop}. 

The holographic CFTs occupy the maximally chaotic corner of irrational CFTs, in which explicit calculations become plausible again via AdS/CFT. In section \ref{sec:holo}, we study the dynamics of entanglement entropy in holographic CFTs for the operator state $|\rho^{1/2}_{AB}\rangle$. This coincides with the reflected entropy considered in Ref.~\cite{tom:2019}, where it was found to closely relate to the entanglement wedge cross section \cite{EoP} in the static case. Using the picture derived in \cite{tom:2019}, the dynamics of operator entanglement in this case is driven by the interplay between two HRT surfaces homologous to $AB$ and $A$ respectively in the bulk of the original state. The three different dominant configurations of the HRT surfaces then give rise to the three phases for the operator entanglement. In particular, the holographic plot in Fig.~\ref{fig:growth-plateau-drop} is obtained from explicit calculations for $\text{AdS}_3/\text{CFT}_2$ in appendix \ref{app:AdS3},  which extends the results of section \ref{sec:CFT} into regimes inaccessible by OPE analysis.

While in all three models there exists a plateau barrier phase characterized by volume-law operator entanglement, thus denying the possibility of efficiently simulating these density matrices throughout thermalization, the length/persistence of these barriers seem to be correlated with how chaotic the systems are, suggesting it as another measure for quantum chaoticity. We elaborate this point further in the discussion section \ref{sec:discuss}.

\section{Two Dimensional CFTs}
\label{sec:CFT}
\begin{figure}[h]
\centering
\includegraphics[width=0.8\columnwidth]{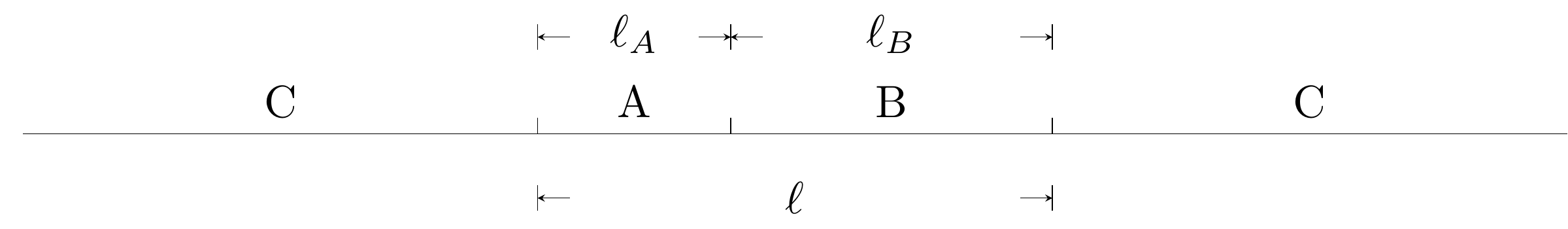}
\caption{The space partition in the 1+1d CFT. We take subsystem A, B to be contiguous with size $\ell_A$ and $\ell_B$. In general we take A to be smaller: $\ell_A \le \ell_B$. The total system size is taken to be much larger than $\ell = \ell_A + \ell_B$, so that thermalization will take place at late time.}
\label{fig_cft_setup}
\end{figure}

In this section, we investigate the operator R\'enyi entropy in two dimensional conformal field theories defined on an infinite line. The partition into subsystems ABC is illustrated in Fig.~\ref{fig_cft_setup} .

\subsection{Replica structure and twist operators}\label{subsec:renyi_CFT}
When computing state R\'enyi entropy, one can work in an orbifold theory by introducing pairs of twist operators that is equivalent to the replica structure \cite{calabrese_entanglement_2004,calabrese_quantum_2016}. In two dimensional CFTs the computation becomes especially tractable since the twist operators become quasi-local with known conformal dimensions. This has resulted in many progresses for computing state entanglements in 2d CFTs both in the ground state and quench scenarios\cite{calabrese_entanglement_2009,calabrese_quantum_2016}. 

The standard procedure to compute the state entropy is to represent the state by a Euclidean path-integral with an open boundary. The reduced density $\rho_A$ is then obtained by sewing along the compliment $A^c$. Quantities such as $\tr( \rho_A^n )$ then connect neighboring copies of $\rho_A$ by gluing the upper and lower rims of the open slits along $A$, leading to a partition function on a branched manifold of the form a staircase geometry. This is equivalent to the insertion of twist fields at the entanglement cut. 

The operator R\'enyi entropy works in an analogous picture albeit with more complicated permutation structures. We consider the more general state $ | \rho_{AB}^\alpha \rangle  / \norm{\rho_{AB}^\alpha }, \alpha \in \mathbb{N}$, i.e. operator state associated with the $\alpha$-th power of the reduced density matrix.


\begin{figure}[h]
\centering
\subfigure[]{
\includegraphics[width=0.95\columnwidth]{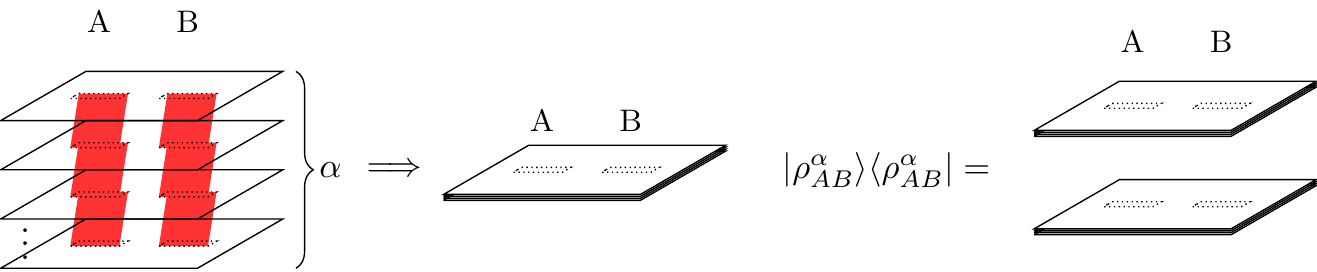}
\label{fig:stair_1}
}
\subfigure[]{
\includegraphics[width=0.35\columnwidth]{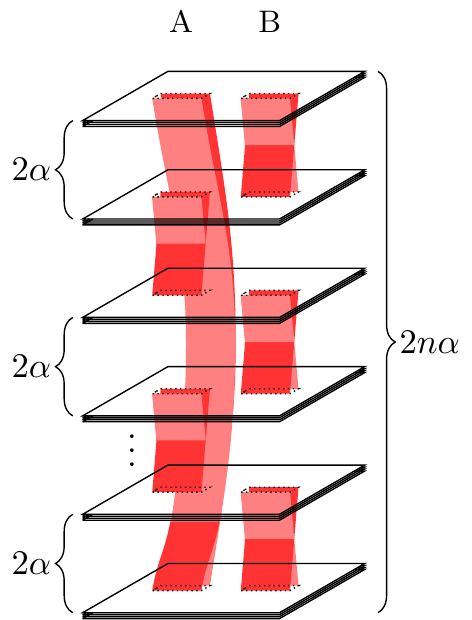}
\label{fig:stair_2}
}
\hspace{20pt}
\subfigure[]{
\includegraphics[width=0.35\columnwidth]{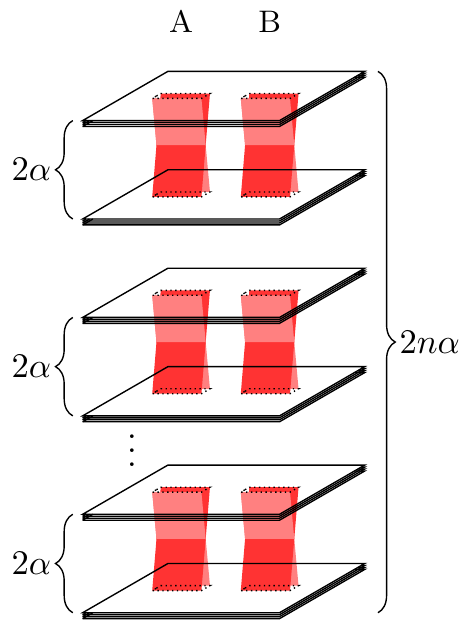}
\label{fig:stair_3}
}\\
\subfigure[]{
\includegraphics[width=0.35\columnwidth]{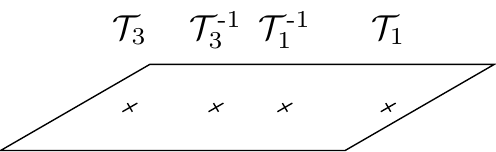}
\label{fig:stair_2_twist}
}
\hspace{20pt}
\subfigure[]{
\includegraphics[width=0.35\columnwidth]{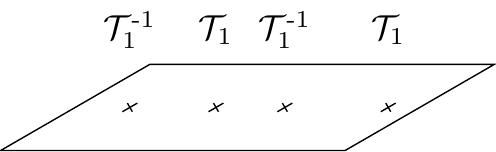}
\label{fig:stair_3_twist}
}
\caption{The Euclidean path integrals for evaluating the operator R\'enyi entropy,the two intervals represent $A$ and $B$ respectively. (a): The operator state $| \rho^{\alpha} (t) \rangle$ (left) and the operator density matrix $| \rho_{AB}^{\alpha} \rangle \langle \rho_{AB}^{\alpha} | $ (right).  (b): The non-trivial part of the operator R\'enyi entropy, where the way to connects layers is different in A and B. (c): The normalization of the operator density matrix, which is mapped to the state R\'enyi entropy. (d)/(e): The corresponding twist field insertion for (b)/(c).} 
\label{fig:stair}
\end{figure}

The reduced density matrix $\rho_{AB}$ is represented by a path-integral with a slit along $A\cup B$. The power of the reduced density matrix $\rho_{AB}^\alpha$ is given by gluing the upper and lower rims of the slits in a staggered manner out of $\alpha$ copies of $\rho_{AB}$, thus leaving two rims, one from the first copy and one from the $\alpha$-th copy. Placing its complex conjugate produces the unnormalized {\it operator} density matrix, see Fig.~\ref{fig:stair_1}.

To compute the corresponding $n$-th operator R{\'enyi} entropy, we generalize Eq.~\eqref{eq:S_op_two_term_1}:
\begin{eqnarray}
\label{eq:S_op_two_term_2} 
S^{\rm op}_n(A, \rho^\alpha_{AB} (t)) &=&  \frac{1}{1 - n} \log \tr_{\mathcal{H}_A\otimes \mathcal{H}_A}\tr^n_{\mathcal{H}_B\otimes \mathcal{H}_B}( |\rho^{\alpha}_{AB}(t) \rangle \langle \rho^{\alpha}_{AB} (t)| )\nonumber\\
 &-& \frac{n(2\alpha -1)}{n-1} S_{2\alpha} \left(AB,\psi(t) \right)
\end{eqnarray}

It proceeds in a few steps. The first term in Eq.~\eqref{eq:S_op_two_term_2} involves partial tracing in Fig.~\ref{fig:stair_2} over $B$, i.e. gluing along B within each operator density matrix. The resulting path-integral is then replicated $n$ times, whose slits along $A$ are then glued staggeringly, see Fig.~\ref{fig:stair_2}. The second term in Eq.~\eqref{eq:S_op_two_term_2} comes from the norm of the operator state,  see Fig.~\ref{fig:stair_3}, where the trace is performed in $A\cup B$. 

According to the descriptions above, one can work out the equivalent twist fields to be inserted (more details can be referred to Ref.~\cite{tom:2019}):
\begin{equation}
\label{eq:twist_field}
\begin{aligned}
&\twist^{-1}_3 = ( \alpha+1, \cdots, 3\alpha ) ( 3\alpha + 1, \cdots, 5\alpha ) \cdots \left( (2n-1) \alpha +1, \cdots,  2n\alpha, 1, \cdots, \alpha \right) \\
&\twist_2 = \left(\alpha+1, 3\alpha+1, \cdots, (2n-1)\alpha +1 \right)^{-1} \left(1, 2\alpha+1, \cdots, (2n-2)\alpha +1 \right)  \\
&\twist_1 = ( 123 \dots 2\alpha ) ( 2\alpha + 1 \cdots 4\alpha ) \cdots \left( (2n-2)\alpha+1 \cdots, 2n\alpha \right) \\
\end{aligned}
\end{equation}
Fig.~\ref{fig:stair_2_twist} and Fig.~\ref{fig:stair_3_twist} demonstrate the twist field insertions that reproduce Fig.~\ref{fig:stair_2} and Fig.~\ref{fig:stair_3} respectively. Our set up corresponds to the limit A and B being adjacent to each other, thus the two twist operators in the middle are fused via OPE to give $\twist^{-1}_3 \twist^{-1}_1 \to \twist_2,\;\twist_1 \twist^{-1}_1\to \mathds{1}$. Define the scaling dimension of size $n$ cyclic permutation to be $\mathfrak{h}_n$, then the dimensions of the twist fields are:
\begin{equation}
h_1 = h_3 = n \mathfrak{h}_{2\alpha} \quad h_2 = 2 \mathfrak{h}_n 
\end{equation}

\subsection{Boundary state set up}
We set up the quench dynamics by considering the following initial state: 
\be 
|\psi\rangle = e^{-H\frac{\beta}{4}} |B\rangle
\end{equation}
where $|B\rangle$ is a conformal boundary state. The imaginary evolution makes the state normalizable with short-range entanglement at length of order $\beta$. We follow Cardy and Calabrese \cite{calabrese_quantum_2016} to set up the calculation for operator entanglement. This has been done for the operator state $| \rho_{AB} \rangle / \norm{\rho_{AB}} $ in \cite{dubail_entanglement_2017}. We repeat the analysis of \cite{dubail_entanglement_2017} for general $\alpha\in\mathbb{N}$, and compute $S^{\rm op}_n(A, \rho^\alpha_{AB} (t))$.

We shall emphasize that the results are only valid for rational CFTs. A similar problem also occurs in the two interval entanglement \cite{asplund_entanglement_2015}, where the behaviors of the four point function distinguish the integrable and chaotic models. We will clarify this as we proceed.

\begin{figure}[h]
\centering

\subfigure[]{
\includegraphics[width=0.4\columnwidth]{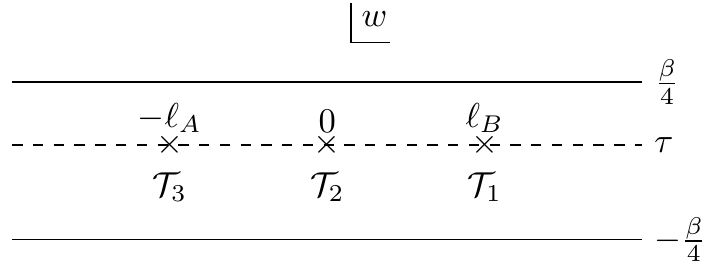}
\label{fig:3_twist} 
}
\subfigure[]{
\includegraphics[width=0.4\columnwidth]{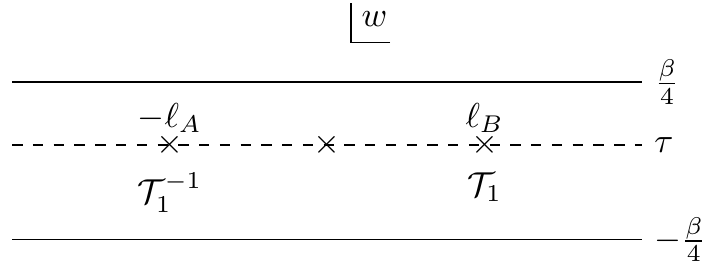}
\label{fig:3_twist_de} 
}

\caption{Locations of the twist field in an infinite stripe. (a) The correlation function of the numerator in Eq.~\eqref{eq:corr_quotient} (b) The correlation function of the denominator in Eq.~\eqref{eq:corr_quotient}. $\langle \psi(\tau)|$ and $| \psi( \tau)\rangle $ are defined by the upper and lower portion of the Euclidean path-integrals.}
\label{fig:twist}
\end{figure}

For CFTs, it is easier to first wick rotate to Euclidean time $t\to -i\tau$:
\be 
|\psi(\tau)\rangle = e^{-H\tau-H\frac{\beta}{4}}|B\rangle,\;\;\langle \psi(\tau)| = \langle B | e^{H\tau-H\frac{\beta}{4}}
\end{equation}
and analytically continue back later. Both $|\psi(\tau)\rangle$ and $\langle\psi(\tau)|$ can be constructed from Euclidean path integrals with boundaries. The procedure for computing the operator R\'enyi entropy discussed in subsection \ref{subsec:renyi_CFT} simply amounts to inserting the appropriate twist operators when gluing the two path integrals for $|\psi(\tau)\rangle$ and $\langle\psi(\tau)|$. The setup is shown in Fig.~\ref{fig:twist}: we have a strip geometry separated by the thermal length and three twists fields inserted for the unnormalized state $| \rho_{AB} \rangle$ and two twist fields for the normalization $\norm{\rho_{AB}^\alpha}$ (Fig.~\ref{fig:twist}). 

The operator entanglement we are interested is the ratio of correlation functions
\begin{equation}
\label{eq:corr_quotient}
e^{- (n-1 ) S_n } = \frac{\langle  \twist_1 ( w_1 , \bar{w}_1)   \twist_2 ( w_2, \bar{w}_2 )  \twist_3 ( w_3, \bar{w}_3 ) \rangle_{\text{strip}}}{\langle  \twist_1 ( w_1, \bar{w}_1 )   \twist_1 ( w_3, \bar{w}_3 )\rangle_{\text{strip}}}
\end{equation}
where
\begin{equation}
w_1 = i\tau +\ell_B,\;\; w_2=i\tau, \; \; w_3=i\tau -\ell_A
\end{equation}
We proceed by conformally mapping to the upper half plane using
\begin{equation}
z = i \exp{\left[ -  \frac{2\pi }{\beta} w \right]} 
\end{equation}
The conformal boundary condition of $|B\rangle$ then means that the correlation functions satisfy the same ward-identities as holomorphic correlation functions on the full complex plane, but with doubled number of points coming from the images: 
\begin{equation}
\label{eq:corr_quotient_uhp}
\begin{aligned}
  e^{ - (n -1 ) S_n } &= \left(\frac{\beta}{2\pi} \right)^{2 h_2 }  \frac{\langle  \twist_1 ( z_1, \bar{z}_1  )   \twist_2 ( z_2, \bar{z}_2 )  \twist_3 ( z_3, \bar{z}_3 ) \rangle_{\rm UHP}}{\langle   \twist_1 ( z_1, \bar{z}_1 )   \twist_1 ( z_3, \bar{z}_3 ) \rangle_{\rm UHP} } \\
 &\sim \left(\frac{\beta}{2\pi}  \right)^{2 h_2 }  \frac{  \langle  \twist_1 ( z_1 )  \twist_2 ( z_2 )  \twist_3 ( z_3 ) \twist_1 ( z_4)  \twist_2 ( z_5 )  \twist_3 ( z_6 ) \rangle_{\rm \mathbb{C}} }{\langle   \twist_1 ( z_1 ) \twist_1(z_4 )   \twist_1 ( z_3 ) \twist_1 ( z_6 )  \rangle_{\rm \mathbb{C}} }
\end{aligned}
\end{equation}
where $z_4=\bar{z}_1, z_5=\bar{z}_1, z_6=\bar{z}_3$ are the images $z_1, z_2 $ and $ z_3$. When continued to real time, their coordinates become:
\begin{equation}
\begin{aligned}
  z_1 &=&i \exp\left[\frac{2\pi }{\beta} ( \ell_B -t ) \right]    \quad z_2 &=& i \exp\left[-\frac{2\pi }{\beta}t \right]  \quad z_3 &=& i \exp\left[\frac{2\pi }{\beta} ( - \ell_A -t )\right ] \\
  z_4 &=& - i \exp\left[\frac{2\pi }{\beta} ( \ell_B+ t) \right] \quad z_5 &=& -i \exp\left[\frac{2\pi }{\beta} t \right]  \quad z_6 &=& -i \exp\left[\frac{2\pi }{\beta} ( - \ell_A+ t) \right]   \\
\end{aligned}
\end{equation}
\subsection{Evaluation via OPE analysis}
In taking the high temperature limit $\beta \rightarrow 0 $, the holomorphic conformal invariants of the correlation functions are pushed towards the boundary of moduli space for generic $t$. As a result the evaluation of correlation functions are controlled by particular OPE channels and the corresponding singularities, provided that they exist. In this subsection we proceed by assuming that all singularities arising from ``chiral" OPEs exist, a point we shall come back to comment in the next subsection.

As $t$ evolves, the conformal invariants bounce from one boundary of moduli space to another, and the dominant OPE channels switch accordingly. As we summarize in the following, there are 4 phases of Eq.~\eqref{eq:corr_quotient_uhp} with different OPE structure (see Fig.~\ref{fig:fusion}), they correspond to the segments of the "growth-plateau-drop" pattern in Fig.~\ref{fig:growth-plateau-drop}, also see review in Ref.~\cite{asplund_entanglement_2015}

\subsubsection{$0<t<\ell_A/2$}
During this regime, the OPE structure is as follows:
\be \label{eq:OPE_1}
\left(\left(\left((z_3  \leftarrow z_6) \leftarrow z_2\right) \leftarrow z_5\right) \leftarrow z_1\right) \leftarrow z_4 
\end{equation}
The fusion proceeds in the following steps: 
\bea
\twist_1(z_3) \twist_1(z_6)\rightarrow (z_6)^{-2 h_1}\mathds{1},\;\twist_2(z_2) \twist_2(z_5)\rightarrow (z_5)^{-2 h_2} \mathds{1},\;\twist_3(z_1) \twist_3(z_4)\rightarrow (z_4)^{-2 h_1}\mathds{1}\nonumber
\eea
for the numerator and 
\begin{equation}
 \twist_1(z_3) \twist_1(z_6)\rightarrow (z_6)^{-2 h_1}\mathds{1},\;\twist_1(z_1) \twist_1(z_4)\rightarrow (z_4)^{-2 h_1}\mathds{1}\nonumber
\end{equation}
for the denominator. Thus we can evaluate the numerator and denominator respectively as:
\bea
\text{6-point function} &\approx & (z_4) ^{ -2h_1 } (z_5) ^{ -2h_2 } (z_6) ^{ -2h_1 } \nonumber\\
\text{4-point function}&\approx & (z_4) ^{-2h_1 } (z_6) ^{-2h_3}
\eea
under which Eq.~\ref{eq:corr_quotient_uhp} evaluates to
\begin{equation}
e^{-(n-1)S_n} \approx \left( \frac{\beta}{2\pi}  \right)^{ 2h_2 }(z_5)^{-2 h_2 }=\left( \frac{\beta}{2\pi}  \right)^{ 2h_2 }\exp{\left[-2h_2\left(\frac{2\pi}{\beta} t\right) \right]}
\end{equation}
Thus
\bea
S^{\rm op}_n(A, \rho^{\alpha}_{AB} (t))  &\approx & \frac{ 4 \mathfrak{h}_n }{n - 1} \left( \ln \left(\frac{2\pi}{\beta}\right) + \frac{2\pi}{\beta}  t \right ) 
\eea
It gives the linear growth with the rate same as that of the state entanglement. 

\subsubsection{$\ell_A/2<t<\ell_B/2$}
In this regime, the OPE structure switches to:
\be \label{eq:OPE_2}
\left(\left(\left((z_3  \leftarrow z_2) \leftarrow z_6\right) \leftarrow z_5\right) \leftarrow z_1\right) \leftarrow z_4 
\end{equation}
Now the fusion proceeds in the following steps: 
\bea
 &&\twist_3(z_3 ) \twist_2(z_2 ) \rightarrow (z_2) ^{-h_2}\twist_1 (z_3 ),\;\;\twist_1(z_3)\twist_3(z_6)\rightarrow (z_6)^{h_2-2h_1}\twist_2(z_3)\nonumber\\
 &&\twist_2(z_3)\twist_2(z_5)\rightarrow (z_5)^{-2h_2}\mathds{1},\;\;\twist_1(z_1)\twist_1(z_4)\rightarrow (z_4)^{-2h_1}\mathds{1}\nonumber
\eea
for the numerator. The denominator stays the same. The numerator changes to:
\bea
\text{6-point function} \approx ( z_4 )^{ -2h_1 }(z_2) ^{-h_2 } (z_6) ^{h_2- 2h_1 }(z_5)^{-2h_2}
\eea
Eq.~\ref{eq:corr_quotient_uhp} now evaluates to
\begin{equation}
e^{-(n-1)S_n} \approx \left( \frac{\beta}{2\pi} \right)^{ 2h_2 } \left(\frac{z_6}{z_2 z_5^2}\right) ^{h_2 } = \left( \frac{\beta}{2\pi}  \right)^{ 2h_2 }  \exp{\left(-\frac{2\pi}{\beta}\ell_A h_2\right)}
\end{equation}
This gives the operator entanglement:
\bea
S^{\rm op}_n(A, \rho^{\alpha}_{AB} (t)) \approx \frac{ 4 \mathfrak{h}_n }{n - 1} \left( \ln \left(\frac{2\pi}{\beta}\right)+ \frac{2\pi}{\beta}  \frac{\ell_A}{2} \right ) 
\eea
It contains a time-independent volume-law term, and corresponds to the plateau barrier. 
\subsubsection{$\ell_B/2<t<\ell/2$}
In this regime, the OPE structure switches to:
\be \label{eq:OPE_3}
\left(\left(\left((z_3  \leftarrow z_2) \leftarrow z_6\right) \leftarrow z_1\right) \leftarrow z_5\right) \leftarrow z_4 
\end{equation}
Now the fusion proceeds in the following steps: 
\bea
 &&\twist_3(z_3 ) \twist_2(z_2 ) \rightarrow (z_2) ^{-h_2} \twist_1 (z_3 ),\;\;\twist_1(z_3)\twist_3(z_6)\rightarrow (z_6)^{h_2-2h_1}\twist_2(z_3)\nonumber\\
 &&\twist_2(z_3)\twist_1(z_1)\rightarrow (z_1)^{-h_2}\twist_3(z_3),\;\;\twist_3(z_3)\twist_2(z_5)\rightarrow (z_5)^{-h_2}\twist_1(z_3),\;\;\twist_1(z_3)\twist_1(z_4)\rightarrow (z_4)^{-2h_1}\mathds{1}\nonumber
\eea
for the numerator, while the denominator remains unchanged. So the numerator further changes to:
\begin{equation}
\text{6-point function} \approx \left( \frac{z_6}{ z_1 z_2 z_5}\right)^{ h_2 } \left( \frac{1}{z_6 z_4}\right)^{ 2h_1} 
\end{equation}
Eq.~\ref{eq:corr_quotient_uhp} now evaluates to
\begin{equation}
e^{-(n-1)S_n} \approx \left( \frac{\beta}{2\pi} \right)^{ 2h_2 }  \left( \frac{z_6}{ z_1 z_2 z_5}\right)^{ h_2 } = \left( \frac{\beta}{2\pi} \right)^{ 2h_2 } \exp{\left[-\frac{2\pi}{\beta}(\ell-2t) h_2\right]}
\end{equation}
This gives the operator entanglement:
\bea
S^{\rm op}_n(A, \rho^{\alpha}_{AB} (t))  \approx \frac{ 4 \mathfrak{h}_n }{n - 1} \left[ \ln \left(\frac{2\pi}{\beta}\right) + \frac{2\pi}{\beta}  \left(\frac{\ell}{2}-t\right) \right ]
\eea
which corresponds to the drop phase. 
\subsubsection{$t>\ell/2$}
In this last regime, the OPE structure settles down to:
\be \label{eq:OPE_4}
\left(\left(\left((z_3  \leftarrow z_2) \leftarrow z_1\right) \leftarrow z_6\right) \leftarrow z_5\right) \leftarrow z_4 
\end{equation}
The fusion process decompose into two clusters: 
\bea
 &&\twist_3(z_3 ) \twist_2(z_2 ) \rightarrow (z_2) ^{-h_2} \twist_1 (z_3 ),\;\;\twist_1(z_3)\twist_1(z_1)\rightarrow (z_1)^{-2h_1}\mathds{1}\nonumber\\
 &&\twist_3(z_6 ) \twist_2(z_5 ) \rightarrow (z_5) ^{-h_2} \twist_1 (z_6 ),\;\;\twist_1(z_6)\twist_1(z_4)\rightarrow (z_4)^{-2h_1}\mathds{1}\nonumber
\eea
In this phase, the denominator also undergoes a phase transition, whose fusion process now takes the form:
\bea
 &&\twist_1(z_3 ) \twist_1(z_1 ) \rightarrow (z_1)^{-2h_1} \mathds{1},\;\;\twist_1(z_6 ) \twist_1(z_4 ) \rightarrow (z_4)^{-2h_1} \mathds{1}\nonumber
\eea
The numerator and denominator is then evaluated to be:
\bea
\text{6-point function} &\approx & \left(z_2 z_5\right)^{-h_2}\left( z_1 z_4\right)^{-2h_1}\nonumber\\
\text{4-point function}&\approx & \left(z_1 z_4\right)^{-2h_1} 
\eea
Eq.~\ref{eq:corr_quotient_uhp} now evaluates to
\begin{equation}
e^{-(n-1)S_n} \approx \left( \frac{\beta}{2\pi} \right)^{ 2h_2 } (z_2 z_5)^{-h_2}=\left( \frac{\beta}{2\pi}  \right)^{ 2h_2 }
\end{equation}
This gives the short-range operator entanglement:
\bea
S^{\rm op}_n(A, \rho^{\alpha}_{AB} (t))  &\approx & \frac{ 4 \mathfrak{h}_n }{n - 1} \left( \ln \left(\frac{2\pi}{\beta}\right) \right ) 
\eea

\begin{figure}[h]
\centering

\subfigure[]{
\includegraphics[width=0.18\columnwidth]{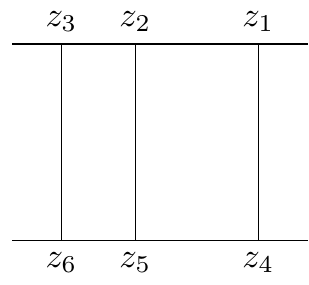}
\label{fig:fusion_1}
}
\hspace{10pt}
\subfigure[]{
\includegraphics[width=0.18\columnwidth]{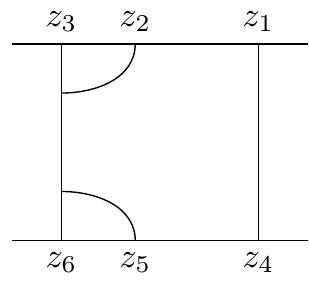}
\label{fig:fusion_2}
}
\hspace{10pt}
\subfigure[]{
\includegraphics[width=0.18\columnwidth]{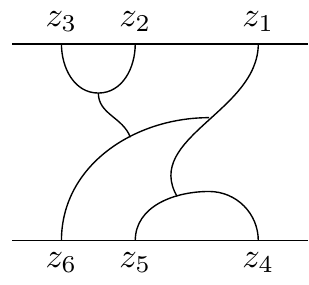}
\label{fig:fusion_3}
}
\hspace{10pt}
\subfigure[]{
\includegraphics[width=0.18\columnwidth]{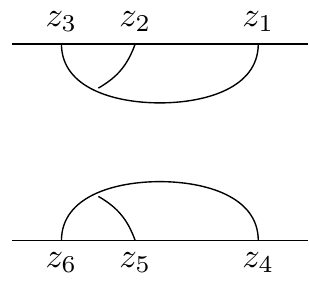}
\label{fig:fusion_4}
}

\caption{The ``fusion channels'' of the 6-point chiral correlation functions. (a) $t \le \ell_A / 2 $, (b) $\ell_A / 2 < t < \ell_B / 2$, (c) $\ell_B / 2 < t < \ell / 2$, (d) $\ell / 2< t $.  }
\label{fig:fusion}
\end{figure}
\begin{figure}[h]
\centering
\includegraphics[width=0.55\columnwidth]{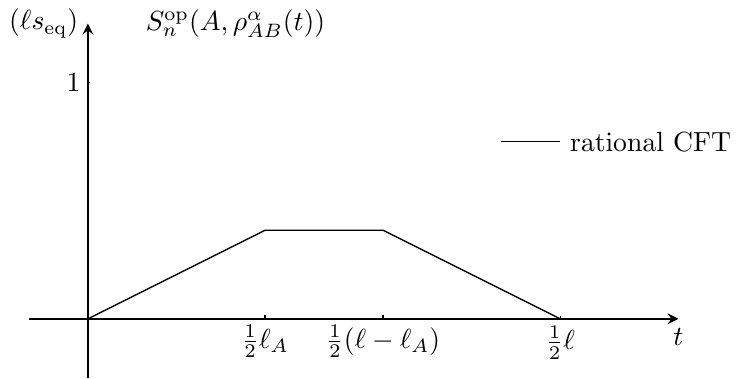}
\caption{Results of rational conformal field theory.}
\label{fig:cft_res}
\end{figure}

The results for all four regimes are thus summarized as below (see Fig.~\ref{fig:cft_res})
\begin{equation}
  \label{eq:cft_qp_res}
  S_n^{\rm op}( A, \rho_{AB}^{\alpha} (t) )
  = \frac{ 4 \mathfrak{h}_n }{n - 1} \left\lbrace 
    \begin{aligned}
      & \ln\left(\frac{2\pi}{\beta}\right)+ \frac{2\pi}{\beta}  t & \quad  t < \frac{\ell_A}{2}  \\
      &  \ln\left(\frac{2\pi}{\beta}\right) + \frac{2\pi}{\beta}  \frac{\ell_A }{2}  & \quad \frac{\ell_A }{2} < t < \frac{\ell_B }{2} \\
      & \ln \left(\frac{2\pi}{\beta}\right) + \frac{2\pi}{\beta}  \left(\frac{\ell}{2} - t\right) & \quad \frac{\ell_B }{2} < t < \frac{\ell}{2} \\
      & \ln \left(\frac{2\pi}{\beta}\right)  & \quad \frac{\ell}{2} < t 
    \end{aligned} \right.
\end{equation}

\subsection{OPE singularities and quasi-particle picture}\label{subsec:OPE_CFT}
The 4 segments of the results are fixed by the 4 ``holomorphic" fusion channels of the 6-point correlation function Eq.~\eqref{eq:corr_quotient_uhp}, see Fig.~\ref{fig:fusion}, and the corresponding singular behaviors therein. However, as was pointed out in \cite{asplund_entanglement_2015}, in generic CFTs the class of singular behaviors is only a subset of those produced by chiral OPEs as in the last subsection. The holomorphic correlation function with images inserted only captures the conformal transformation property of the boundary state; the OPE occurs in the UHP and is controlled by both the holomorphic and anti-holomorphic coordinates $(z,\bar{z})$. In particular, there are only two types of OPEs in the UHP:
\bea\label{eq:UHP_OPE_1}
\mathcal{O}_1\left(z_1,\bar{z}_1\right)\mathcal{O}_2\left(z_2, \bar{z}_2\right)&\sim & |z_1-z_2|^{h_3-h_1-h_2} |\bar{z}_1-\bar{z}_2|^{\bar{h}_3-\bar{h}_1-\bar{h}_2}\mathcal{O}_3\left(z_1,\bar{z}_1\right)\\
\mathcal{O}_1 \left(z_1,\bar{z}_1\right)&\sim & |z_1-\bar{z}_1|^{-2h_1}\mathds{1}_b\label{eq:UHP_OPE_2}
\eea
The first line Eq.~\eqref{eq:UHP_OPE_1} corresponds to ordinary OPE $\left(z_1\to z_2,\;\bar{z}_1\to\bar{z}_2\right)$ in the interior of UHP; the second line Eq.~\eqref{eq:UHP_OPE_2} corresponds to the dominant channel when $\mathcal{O}_1$ approaches the boundary $z_1\leftrightarrow\bar{z}_1$, and fuses into a boundary operator (identity in this case). 

One can check that of the 4 channels shown in Fig.~\ref{fig:fusion}, three of them consists solely of legitimate OPEs in the UHP and therefore whose singularities exist in generic CFTs:
\begin{itemize}
\item{$0<t<\ell_A/2$:  $z_i\leftrightarrow \bar{z}_i,\; i = 1,2,3$;}
\item{$\ell_A/2<t<\ell_B/2$:  $\left(z_2\to z_3, \bar{z}_2\to \bar{z}_3\right),\; z_3 \leftrightarrow \bar{z}_3,\; z_1 \leftrightarrow \bar{z}_1$;}
\item{$t>\ell/2$:  $\left(z_2\to z_3, \bar{z}_2\to \bar{z}_3\right),\;\left(z_3\to z_1, \bar{z}_3\to \bar{z}_1\right)$.}
\end{itemize}
On the other hand, the fusion channel in the third phase $\left(\ell_B/2<t<\ell/2\right)$ of Fig.~\ref{fig:fusion} cannot be decomposed into a series of legitimate OPEs in the UHP. Therefore the corresponding singularities giving rise to the linear decline phase in Fig.~\ref{fig:cft_res} do not exist in generic CFTs. It is only in rational CFTs that the existence of these singularities is necessary in order to satisfy the crossing symmetry of correlation function, see \cite{asplund_entanglement_2015}. We thus conclude that the linear decline behavior for the phase $\left(\ell_B/2<t<\ell/2\right)$ is a  consequence of working with rational CFTs. 

\begin{figure}[h]
  \centering

\includegraphics[scale=0.137]{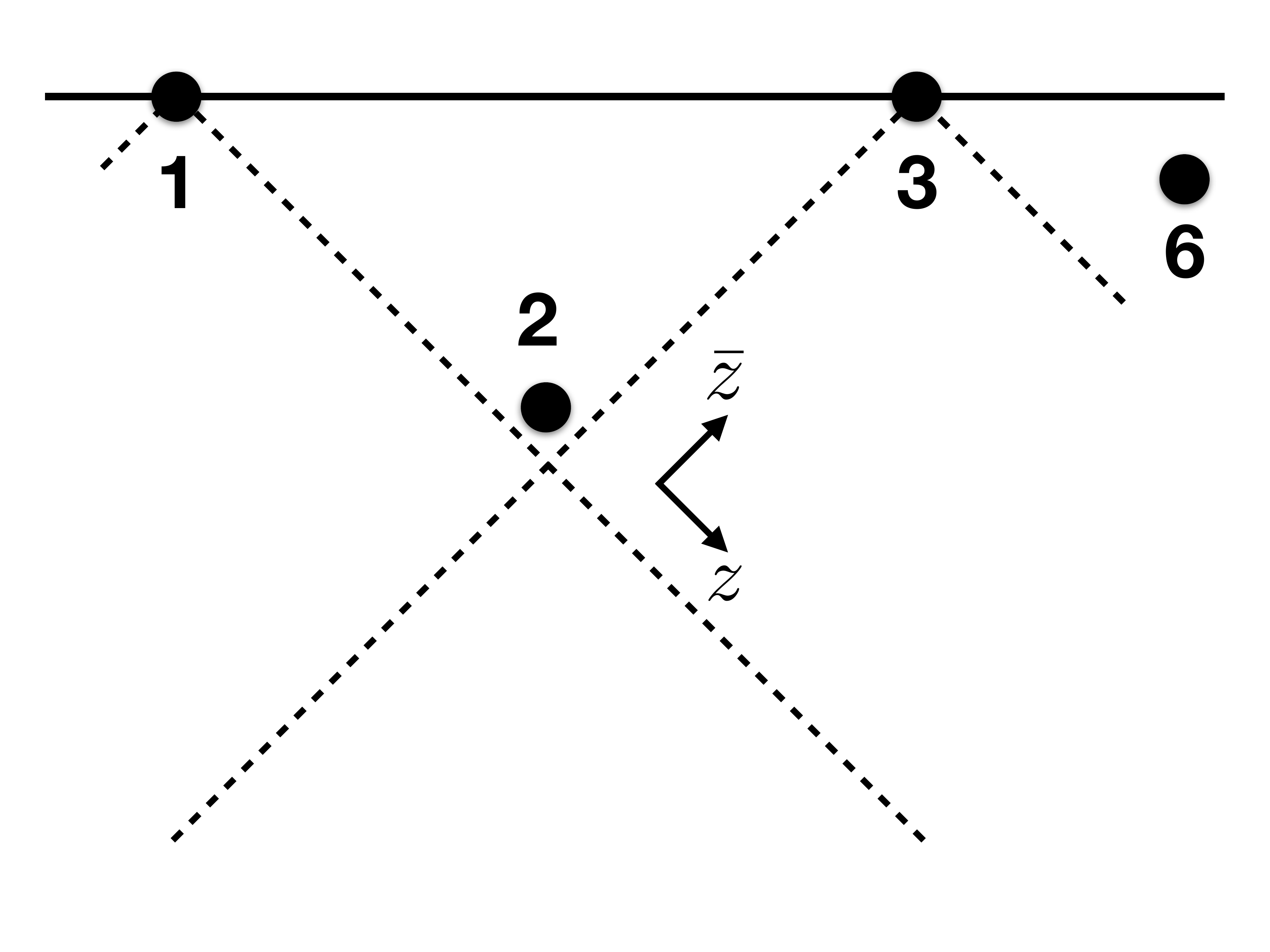}
\includegraphics[scale=0.137]{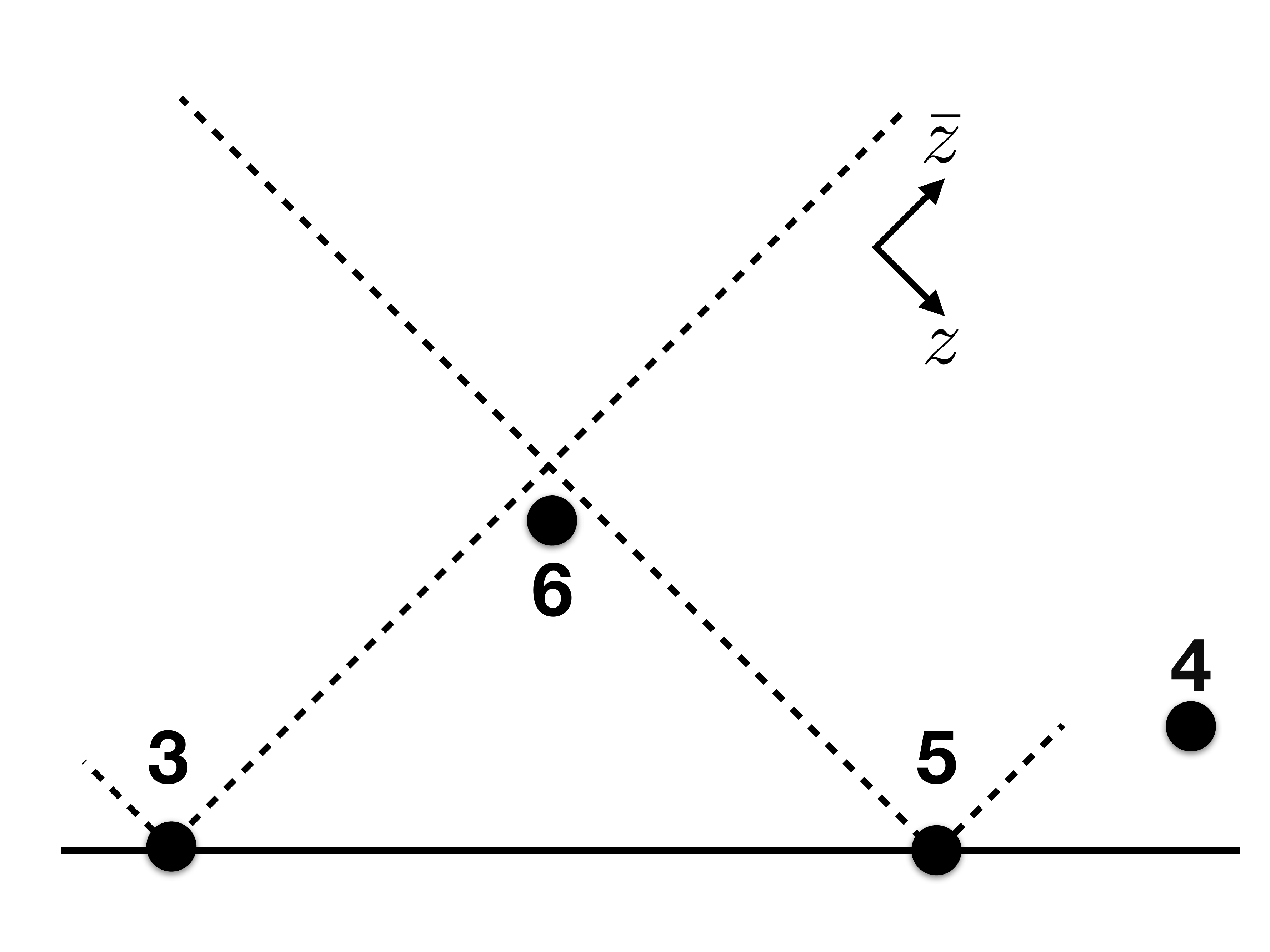}
\includegraphics[scale=0.137]{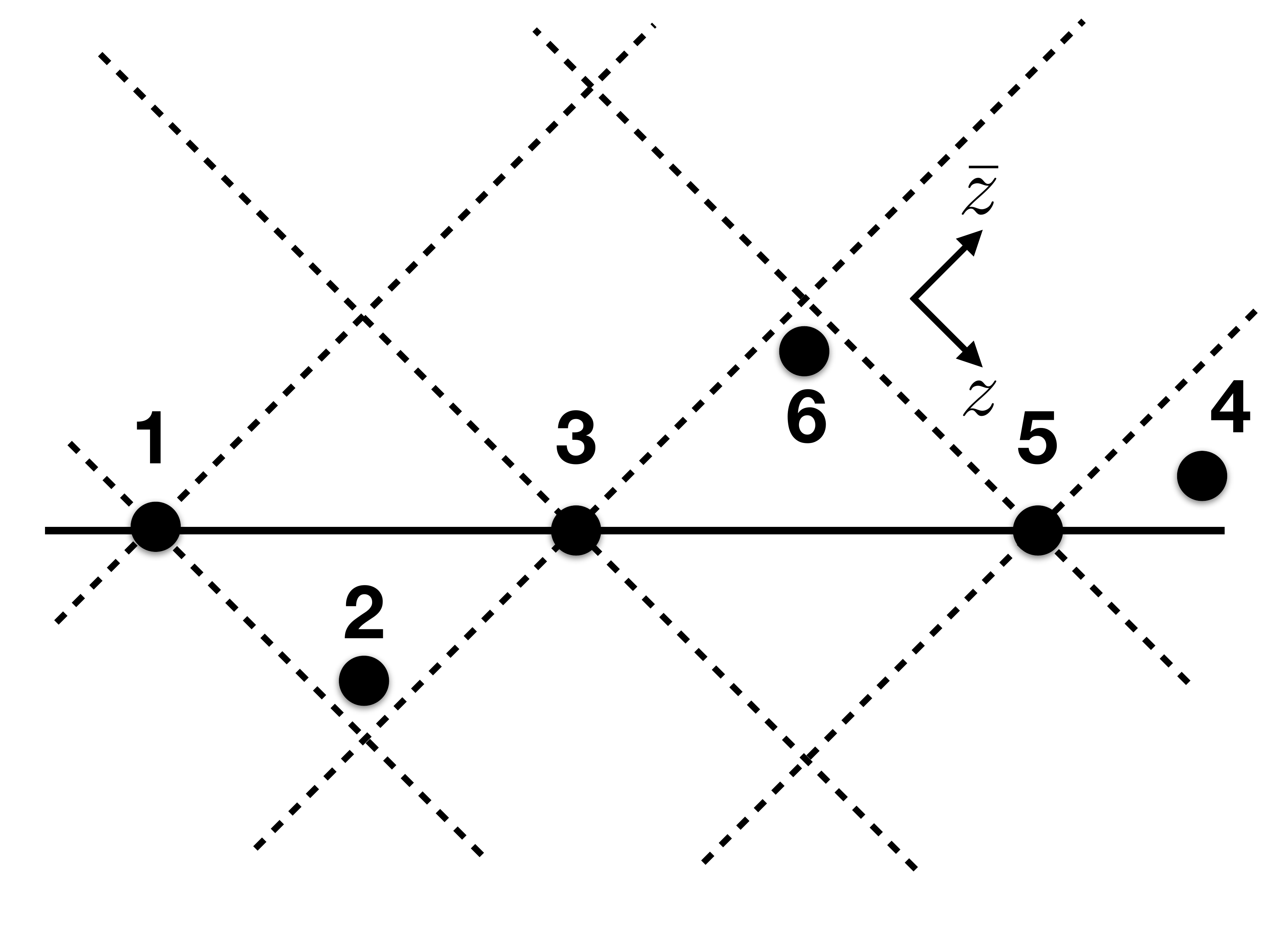}

  \caption{Left: light cone singularity associated with Eq.~\eqref{eq:lightcone_1}. Center: light cone singularity associated with Eq.~\eqref{eq:lightcone_2}. Right: the two light cone singularities are tangled.}
  \label{fig:lightcones}
\end{figure}

The nature of these singularities characterizing rational CFTs becomes more transparent if instead of the boundary state $|B\rangle$, we consider the thermal field double (TFD) state defined in the doubled CFT$_1\times$CFT$_2$: 
\begin{equation}
| \psi \rangle = \sum_n e^{ - \frac{\beta}{2} H  } | n \rangle_1 | n \rangle_2
\end{equation}
where $|n\rangle$ are eigenstates of $H$. The reduced density matrix is thus defined on (AB)$_1\times$(AB)$_2$, and the corresponding operator state is defined on a further double copy of (AB)$_1\times$(AB)$_2$.  The stripe geometry in Fig.~\ref{fig:twist} becomes a cylinder of circumference $\beta$, with another set of twist operators inserted at the images. The operator R\'enyi entropy is then related to the full 6-point correlation function on the complex plane:
\begin{equation}
e^{- (n-1 ) S_n } \propto \frac{\langle  \twist_1 ( z_1 , \bar{z}_1)   \twist_2 ( z_2, \bar{z}_2 )  \twist_3 ( z_3, \bar{z}_3 ) \twist_1 ( z_4 , \bar{z}_4)   \twist_2 ( z_5, \bar{z}_5 )  \twist_3 ( z_6, \bar{z}_6 )\rangle_{\mathbb{C}}}{\langle  \twist_1 ( z_1, \bar{z}_1 )   \twist_1 ( z_3, \bar{z}_3 ) \twist_1 ( z_4,\bar{z}_4 )   \twist_1 ( z_6,\bar{z}_6 )\rangle_{\mathbb{C}} }
\end{equation}
As conjectured and argued in \cite{asplund_entanglement_2015}, the TFD state can be used to reproduce the singularities in the boundary state. In this case, the singularities in $\left(\ell_B/2<t<\ell/2\right)$ takes the form of two ``tangled" light cone singularities. To see what this means, we compute the following two conformal invariants: 
\bea\label{eq:lightcone_1}
\eta_2 &=& \frac{(z_2-z_1)(z_3-z_6)}{(z_2-z_6)(z_3-z_1)}\to 1,\;\; \bar{\eta}_2 = \frac{(\bar{z}_2-\bar{z}_1)(\bar{z}_3-\bar{z}_6)}{(\bar{z}_2-\bar{z}_6)(\bar{z}_3-\bar{z}_1)}\approx e^{-2\pi/\beta\ell_A} \to 0\\
\eta_6 &=& \frac{(z_6-z_3)(z_5-z_4)}{(z_6-z_4)(z_5-z_3)}\approx e^{-2\pi/\beta \ell_A}\to 0,\;\; \bar{\eta}_6 = \frac{(\bar{z}_6-\bar{z}_3)(\bar{z}_5-\bar{z}_4)}{(\bar{z}_6-\bar{z}_4)(\bar{z}_5-\bar{z}_3)}\to 1\label{eq:lightcone_2}
\eea
The first line Eq.~\eqref{eq:lightcone_1} indicates that $\left(z_2,\bar{z}_2\right)$ is approaching the light cone ``tip" of $1$ and $3$, i.e. simultaneously approaching the light cones of both $1$ and $3$; the second line Eq.~\eqref{eq:lightcone_2} indicates that $\left(z_6,\bar{z}_6\right)$ is approaching the light cone tip of $3$ and $5$. Each one corresponds to a light cone singularity discussed in \cite{asplund_entanglement_2015}, see Fig.~\ref{fig:lightcones}. However, we cannot isolate them into two separate ``clusters" as in \cite{asplund_entanglement_2015}; they are tangled together, and this makes it impossible to factorize the 6-point function into a product of lower-point functions. In principle, such light cone limits of correlation functions in chaotic 2d CFTs, including finite $c$ corrections, can be explicitly studied via the structure of fusion matrix \cite{lightcone1, lightcone2, lightcone3}. We leave this for future studies. 

\begin{figure}[h]
  \centering

\includegraphics[width=\columnwidth]{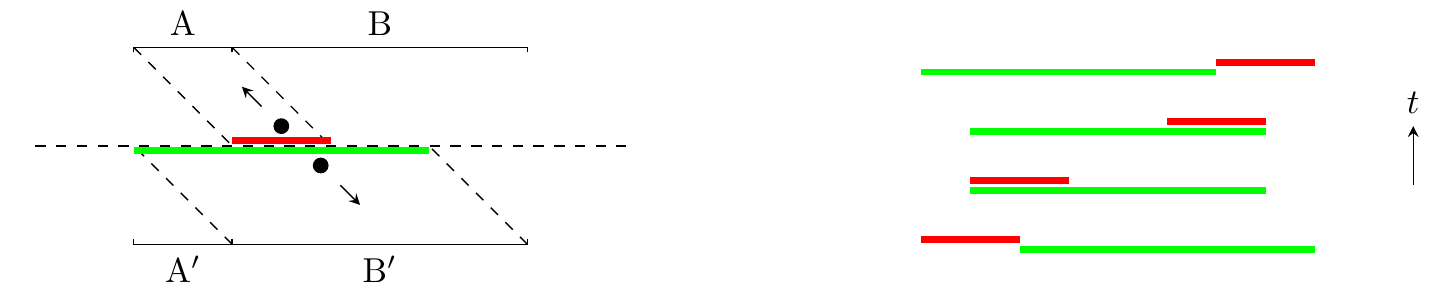}

  \caption{Quasi-particle interpretation of results in Eq.~\eqref{eq:cft_qp_res}. Left: Quasi-particle in thermal field double states. Initially each pair of degrees of freedom in the left and right form an EPR pair. The entanglement is proportional to the overlap of past the light cones. Right: the overlap of the past light cones of $A$ (red) and $B'$(green) in different stages. Figure shows those turning points $t = 0, \frac{\ell_A}{2}, \frac{1}{2}(\ell- \ell_A ), \frac{1}{2} \ell$ in Eq.~\eqref{eq:cft_qp_res} }
  \label{fig:qp}
\end{figure}

The time dependence in Eq.~\eqref{eq:cft_qp_res} describes rational CFTs. As a consistency check, it can be exactly reproduced by the quasi-particle picture, which is believed to model the essential physics regarding entanglement evolution in integrable models. We present this in the thermal field double set up, see Fig.~\ref{fig:qp}. The initial state is represented by the middle line. After the quench, there are coherent pairs of quasi-particles generated and they move to the opposite directions. As time goes on, the overlap regions on the line, which are the area that emitted quasi-particles at time $t$ that now entangled between region $A$ and $B'$, is proportional to the operator entanglement. It gives rise to the 4 segments in Eq.~\eqref{eq:cft_qp_res} and the symmetric trapezoid in Fig.~\ref{fig:cft_res}.

\section{Random Unitary Circuit}
\label{sec:random}

In this section, we introduce the random unitary circuit model and the corresponding domain wall picture for the entanglement. Using this, we compute and interpret the change of the quench operator entanglement as the succession of three dominant domain wall configurations. 

\begin{figure}[h]
\centering
\subfigure[]{
\includegraphics[width=0.25\columnwidth]{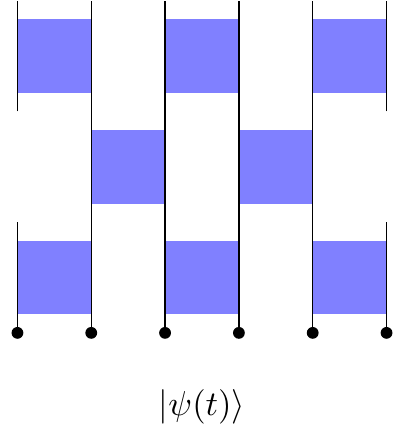}
\label{fig:ruc}

}\hspace{20pt}
\subfigure[]{

\includegraphics[width=0.25\columnwidth]{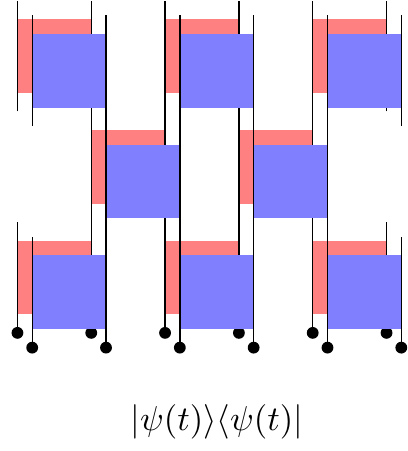}
\label{fig:ruc_rho}
}\hspace{20pt}
\subfigure[]{

\includegraphics[width=0.25\columnwidth]{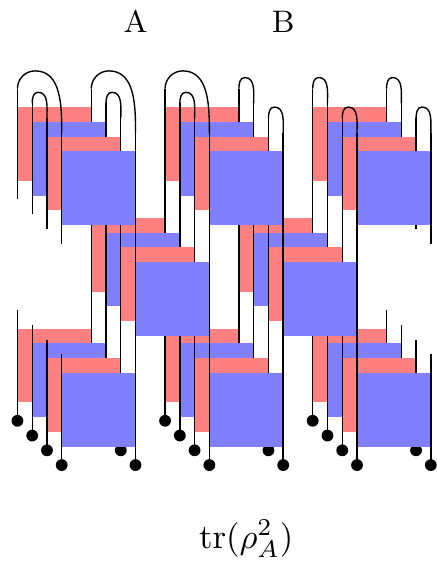}
\label{fig:ruc_purity}
}
\caption{Random unitary circuit: (a) its tensor network structure, where each blue block represents a $q\times q$ Haar random unitary matrix (b) the time dependent density matrix (c) computing $\text{tr}\left(\rho_A^n\right),n=2$ by contracting different copies.}
\label{fig:n_copy}
\end{figure}

Random unitary circuit is a tensor network that models chaotic evolution with local interactions. The structure of the network is shown in Fig.~\ref{fig:ruc}, where the horizontal direction represents the system: a one-dimensional chain with $q$ bits of local degrees of freedom. The dynamics evolves along the vertical direction. Each block is an $q \times q$ {\it independent} Haar random unitary matrix. The network is constructed to involve only local interactions that entangle (almost maximally) nearby sites in one time step. And the locations switch by one lattice spacing in alternating steps as the system evolves.

The circuit retains minimally the features of evolution by a local Hamiltonian -- the local generation and propagation of large entanglement -- yet dispenses particular forms of interactions. The average over quenched randomness is what provides analytic handles on the dynamics of chaos. Recently, there have been fruitful results based on this model and its variants, about the scaling behaviors of out-of-time-order correlator \cite{nahum_operator_2017,von_keyserlingk_operator_2017,khemani_operator_2017}, entanglement growth \cite{jonay_coarse-grained_2018,nahum_dynamics_2017,nahum_quantum_2017,zhou_emergent_2018,chan_solution_2017}, hydrodynamic long tails in $U(1)$ conserved systems \cite{khemani_operator_2017}, etc. Similar {\it non-unitary} random circuit were also proposed to reconstruct the holographic space time as an arguably more explicit tensor network structure. The Ryu-Takayanagi (RT) surface of entanglement, transition of entanglement scalings, etc. were considered \cite{qi_space-time_2018,hayden_holographic_2016,vasseur_entanglement_2018} in this setting.

Among these results, the random unitary circuit enables a ``domain wall'' picture to understand the universal behavior of entanglement. It is motivated by the minimal cut picture of a general tensor network. In this picture, the number of bonds the minimal cut traverses through the tensor network gives an upper bound of the bond dimension to represent the states, and therefore is also an upper bound for the entanglement \cite{swingle2012entanglement}. The domain wall picture is a refinement of the minimal cut by taking unitarity into account \cite{jonay_coarse-grained_2018}. They can be viewed as world-lines of objects carrying permutation data. Unitarity puts causality constraints on their trajectories and possible forms of interactions. More explicitly, computing the circuit averaged entanglement is mapped to a statistical mechanical problem of domain walls, where the minimal free energy gives the entanglement. 

The objects carrying permutation data can be understood from the underlying replica structure defined by $\tr( \rho_A^n )$, see Fig.~\ref{fig:ruc_purity} for $n=2$. They perform permutation operations at the entanglement cut, which is analogous to the twist fields in the field theory definition of $\tr( \rho_A^n )$ using path integrals. In random unitary circuit, the random average yields an effective statistical mechanics model, whose solutions extend the permutations from the entanglement cut into the interior of the tensor network. In this sense, the domain wall gives a geometric picture to the entanglement of the random unitary circuit, in much the same way as the RT surface \cite{ryu_aspects_2006} in holography. In fact, domain walls of permutation in non-unitary random circuit were formulated to be the tensor network analog of the RT surface in attempts \cite{qi_space-time_2018,hayden_holographic_2016} to understand the geometric structure of the space time. In a different language, the formulation has also been studied in the context of machine learning \cite{you_entanglement_2018}. 

\subsection{The effective model}

We now review the effective statistical mechanical model in the random unitary circuit. 

\begin{figure}[h]
\centering
\subfigure[]{
  \includegraphics[width=0.2\columnwidth]{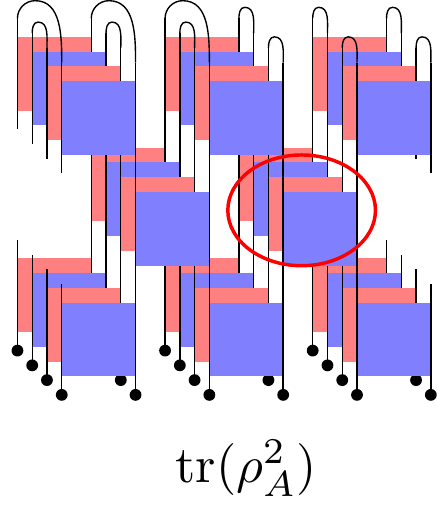}
  \label{fig:circle_block}
}
\subfigure[]{
  \includegraphics[width=0.7\columnwidth]{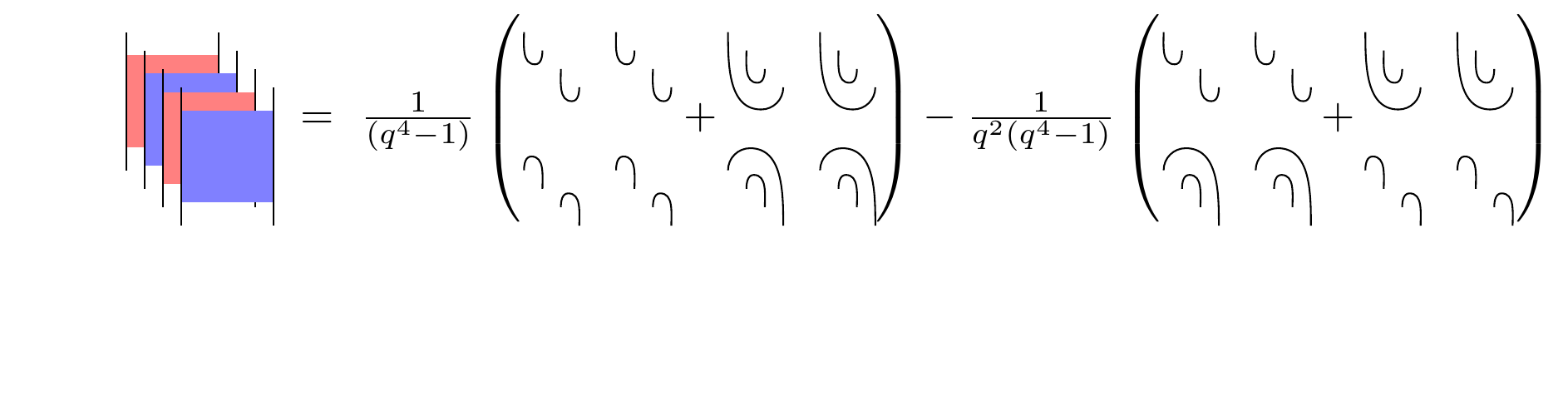}
  \label{subfig:block_aver}
}
\caption{Permutations in the circuit average. The blue and red blocks represents the independent Haar random matrices and their complex conjugate. (a) The circuit structure for $n=2$. Red circle encloses $2n=4$ layers of unitaries, which can be regarded as an $8n$-leg tensor that has $4n$ legs at the bottom as input and $4n$ legs on the top as output. (b) Random average over the $2n$ layers in each block generates a sum over permutation elements with weights.}
\label{fig:block_aver}
\end{figure}

Let us begin by looking at the interior of the tensor network of the quantities $\tr( \rho_{A}^n )$ shown in Fig.~\ref{fig:circle_block}, for illustration we pick $n=2$. The overlapped blue and red squares are $n=2$ copies of the Haar random matrix $U_{q\times q}$ and its complex conjugate $U^*_{q \times q}$. Each block with $U$ (blue in Fig.~\ref{fig:circle_block}) and $U^*$ (red in Fig.~\ref{fig:circle_block}) is replicated $n$ times in the network. Random average gives rise to ``contractions'' between $U$ and $U^*$ among the copies. Each contraction can be represented as a permutation element in the $n$-th symmetric group, see Fig.~\ref{subfig:block_aver}. The permutation elements are what emerge as the dynamical degrees of freedom after random average, and we call them ``spins''. The effective model is represented by the statistical mechanics of these spins\footnote{There are actually negative weights for some of the spin configurations. But if we integrate out half of the spins, all the spin configurations in leading order in $q$ will have positive weight. We can then treat the negative weights as perturbative corrections to the leading order configurations, thus regaining the statistical model interpretation.}. Boundary conditions are imposed at both the top and bottom boundaries of the tensor network. They play different roles: the boundary condition on the top comes from the replica structure for computing the R\'enyi entropy; the boundary condition on the bottom encodes the initial state, see Fig.~\ref{fig:ruc_purity}.

\begin{figure}[h]
\centering
\includegraphics[width=\columnwidth]{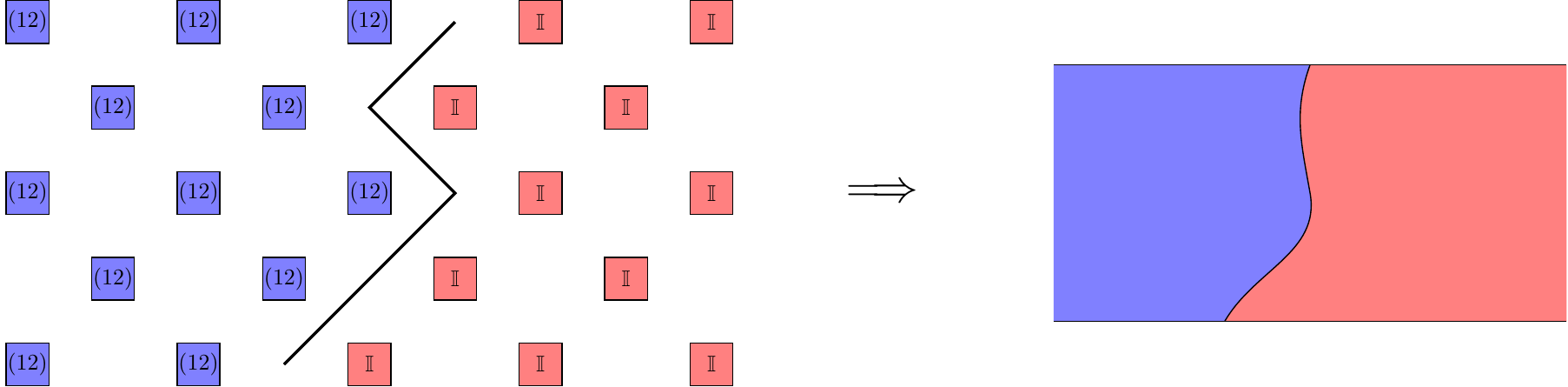}
\caption{Statistical model in terms of domain walls. Left: Domain wall configurations for the 2nd R\'enyi entropy. Right: Coarse grained domain wall configuration. Red and blue represent two domains of permutation extended from the boundary twist imposed by the definition of entanglement.}
\label{fig:spin_to_dw}
\end{figure}

For the purpose of studying entanglement, it is more convenient to describe the configurations in terms of domain walls rather than spins, because the former fits better with the boundary conditions to be imposed and unitarity constraints. Different spin domains are characterized by different channels of contractions between the $n$ copies of $U$ and $U^*$, the domain walls can be understood as the twists of those contractions. For example, in order to evaluate the $n$-th R\'enyi entropy, the contractions operation on the top boundary of Fig.~\ref{fig:ruc_purity} are identities acting on the subsystem A, and permutations $(12\cdots n)$ acting on the subsystem B respectively. The domain wall is a twist between them, inserted at the entanglement cut.  If the spins in the left domain are all $\sigma_1$ and those in the right domain $\sigma_2$, then the domain wall is a permutation $\tau=\sigma^{-1}_1 \sigma_2$. In a general tensor network, these permutations may become other tensors when propagating from the top boundary into the interior. But in random unitary circuits, the domain walls (or permutations) are the only degrees of freedom that survive the random average. For instance the average purity $\overline{\text{tr}\left(\rho_A^2\right)}$ will only contain wandering domain walls as in Fig.~\ref{fig:spin_to_dw}. The R\'enyi entropy thus is given by the freedom energy of the domain walls subject to the boundary conditions at both the top and bottom boundaries. For the product initial state we are considering, we have free boundary condition at the bottom. 

\begin{figure}[h]
\centering

\subfigure[]{
\includegraphics[width=0.4\columnwidth]{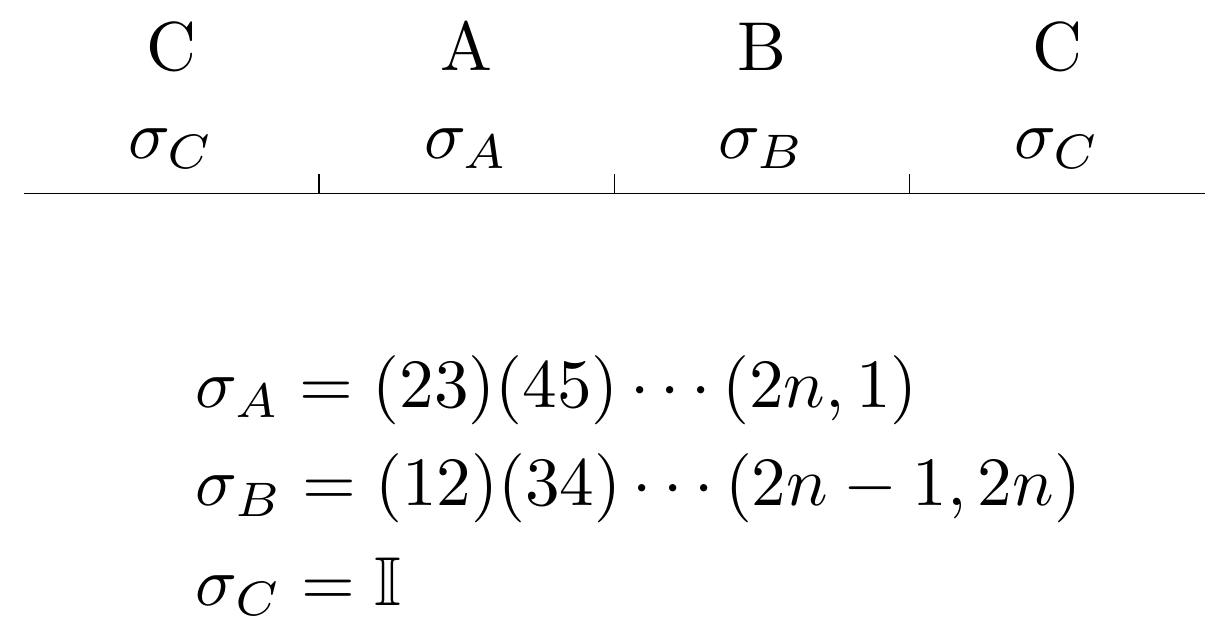}
\label{fig:op_idx_1} 
}
\hspace{20pt}
\subfigure[]{
\includegraphics[width=0.4\columnwidth]{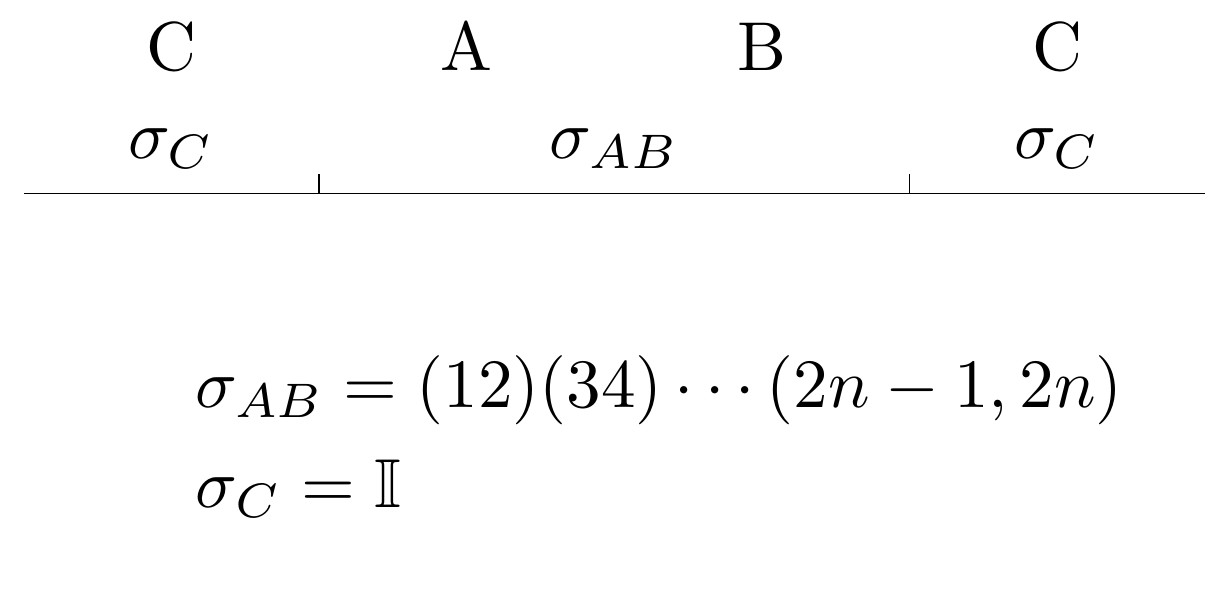}
\label{fig:op_idx_2} 
}\\
\subfigure[]{
\includegraphics[width=0.4\columnwidth]{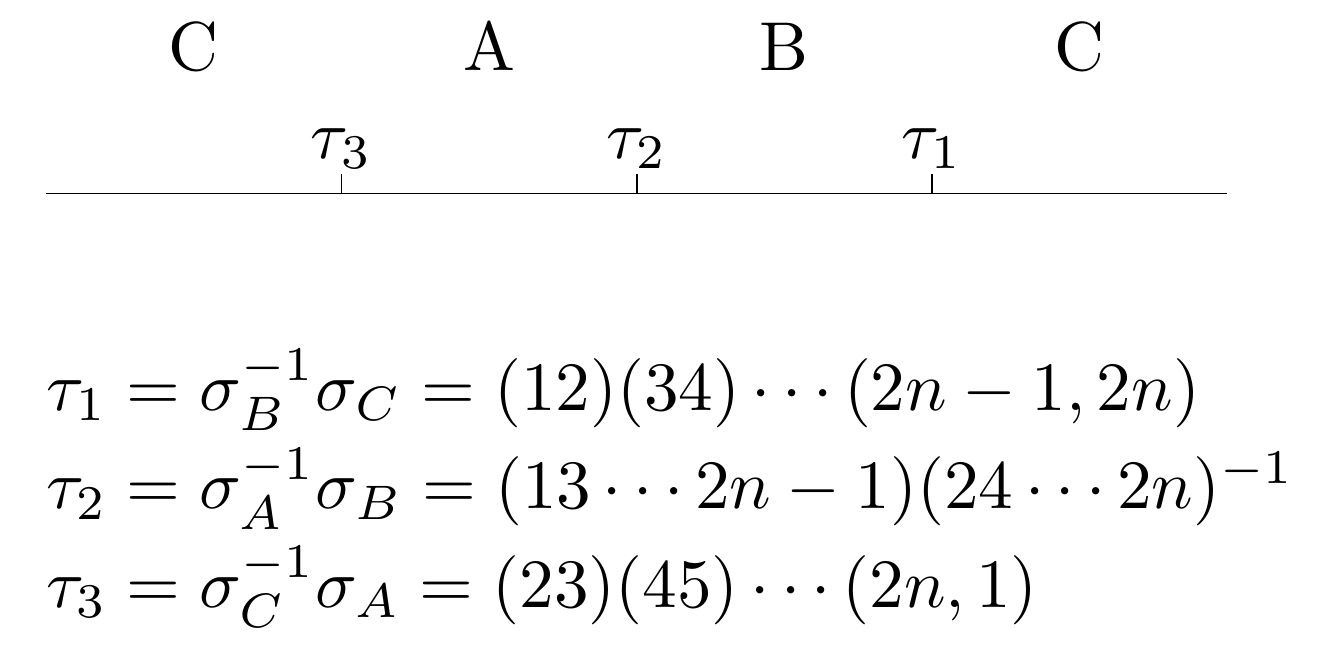}
\label{fig:dw_1} 
}
\hspace{20pt}
\subfigure[]{
\includegraphics[width=0.4\columnwidth]{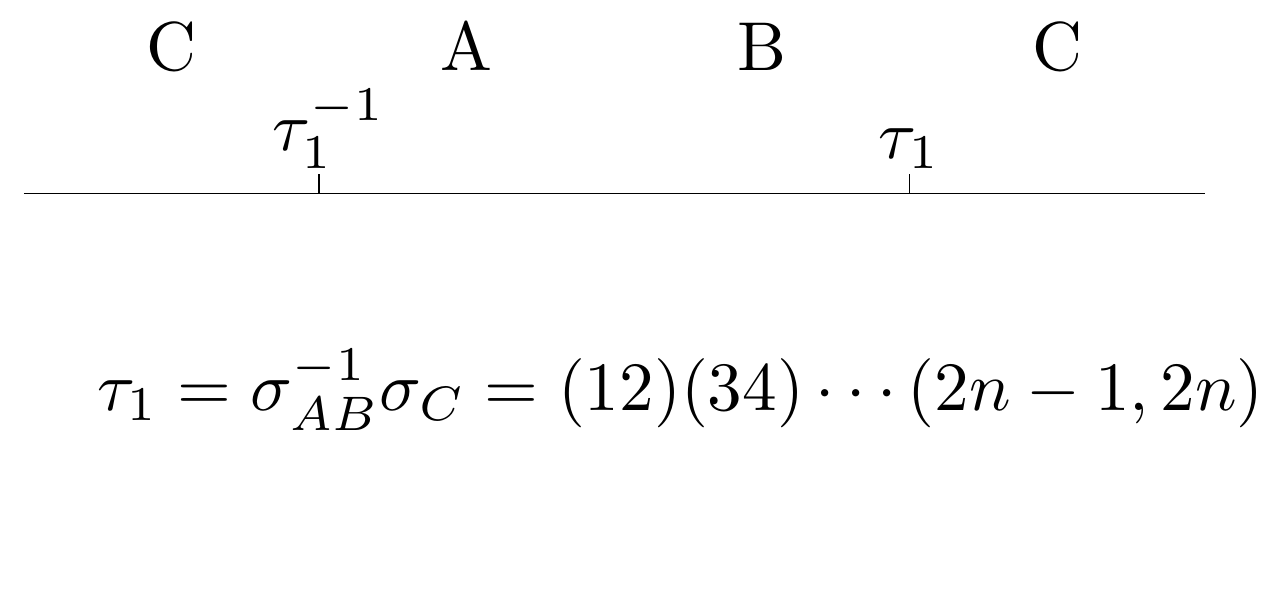}
\label{fig:dw_2} 
}

\caption{Domain walls in state and operator R\'enyi entanglement. The spin boundary conditions at the top boundary of the tenor network are shown for (a): operator R'enyi entanglement and (b): state entanglement. (c)/(d): The corresponding domain walls at the top boundary for operator/state entanglement. Notice the resemblance of domain walls to the twist fields in Eq.~\eqref{eq:twist_field}. }
\label{fig:perm_struc}
\end{figure}

\begin{figure}[h]
\centering

\subfigure[]{
\includegraphics[width=0.16\columnwidth]{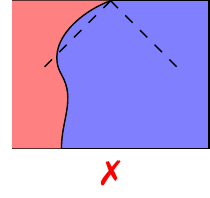}
\label{fig:dw_rule_1} 
}
\subfigure[]{
\includegraphics[width=0.16\columnwidth]{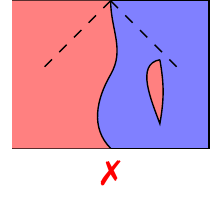}
\label{fig:dw_rule_2} 
}
\subfigure[]{
\includegraphics[width=0.16\columnwidth]{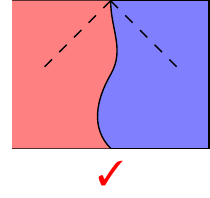}
\label{fig:dw_rule_3} 
}
\subfigure[]{
\includegraphics[width=0.16\columnwidth]{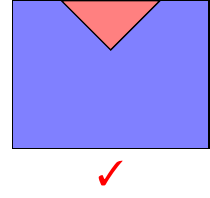}
\label{fig:dw_rule_4} 
}
\subfigure[]{
\includegraphics[width=0.16\columnwidth]{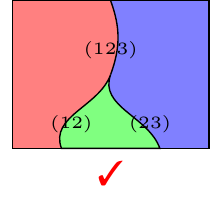}
\label{fig:dw_rule_5} 
}


\caption{Domain wall rules: (a) domain wall must not go outside the light cone. (b) ``bubble'' is forbidden. (c) legitimate configuration: an elementary domain wall travel inside the light cone. (d) legitimate configuration: two domain walls meet and annihilate. (3) domain wall split according to multiplication rule.}
\label{fig:dw_rules}
\end{figure}

This picture can be extended to more generic entropic quantities such as the operator R\'enyi entropy we are interested, which corresponds to imposing more complicated boundary conditions, see Fig.~\ref{fig:perm_struc}. In this case, the top boundary can host more generic domain walls. They can decomposed into elementary domain walls, which are defined to consist only of single transpositions. The elementary domain walls are the building blocks for the underlying dynamics. For example, the domain wall $(123)$ consists of two elementary domain walls. The locality of the interaction and unitarity of the evolution restricts the allowed behaviors of elementary domain walls (see Fig.~\ref{fig:dw_rules}). When propagating downward, an elementary domain wall must remain inside the light cone and can not go back. When two elementary domain walls meet, they combine according to the group multiplication rule; and the same applies when a composite domain wall split. 

As mentioned, the domain walls need to minimize the free energy while satisfying the boundary conditions imposed at the top and bottom of the circuit. An elementary domain wall will cost $s_{\rm eq} = \ln q $ amount of free energy for each discrete time step it takes. Therefore at the leading order in large $q$ expansion, the minimizing configuration is simply given by free elementary domain walls taking the shortest paths\footnote{Entropic corrections are of order $\frac{1}{\log q}$. }. At higher orders in large $q$, elementary domain walls can have two types of interactions. One is the statistical interaction between two non-commutative domain walls, such as those with transpositions $(12)$ and $(13)$. It corrects the entanglement velocity in order $\mathcal{O}(1 / \log q )$. For the special case of R\'enyi index $n = 2$, this effect is absent. Another weaker interaction at $\mathcal{O}(1/q^4 \ln q )$ order appears between commutative domain walls like $(12)$ and $(34)$. The suppression of domain wall interaction at $q = \infty$ effectively reduce the quench disorder to annealed disorder. We can thus equating $\overline{e^{ - (n-1) S_n}}$ and $e^{- (n-1) \overline{S_n}}$. 

\subsection{Domain walls in operator entanglement}
\begin{figure}[h]
\centering

\subfigure[]{
\includegraphics[width=1\columnwidth]{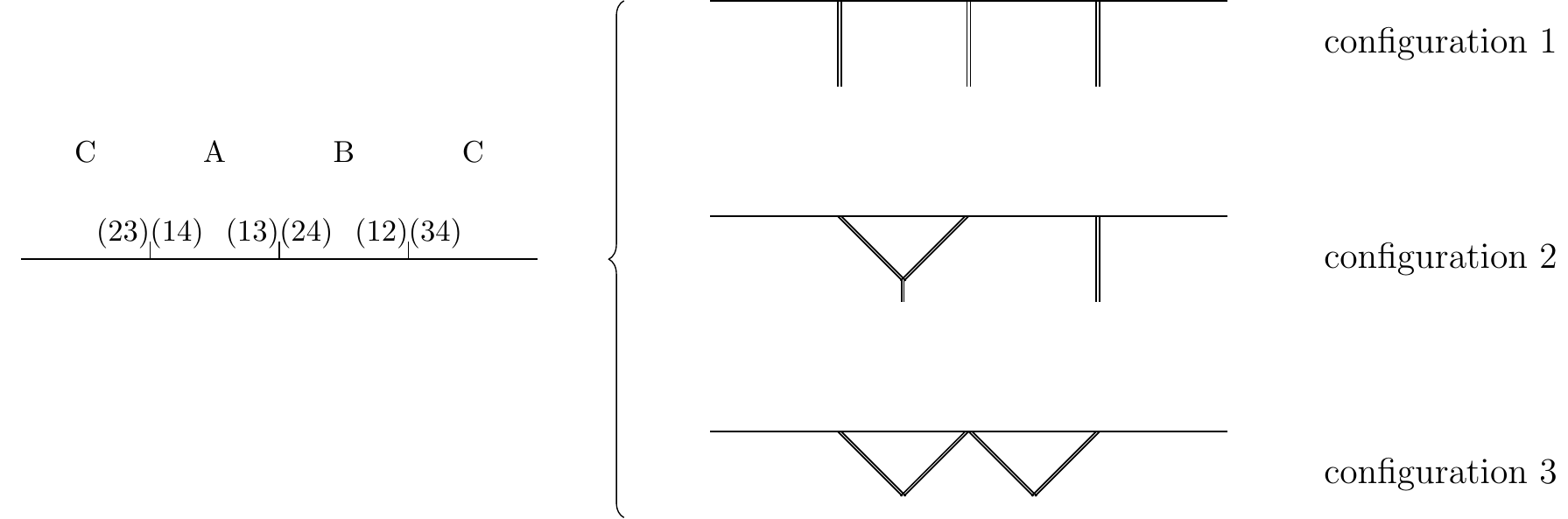}
\label{fig:dw_rho_op} 
}\\
\subfigure[]{
\includegraphics[width=1\columnwidth]{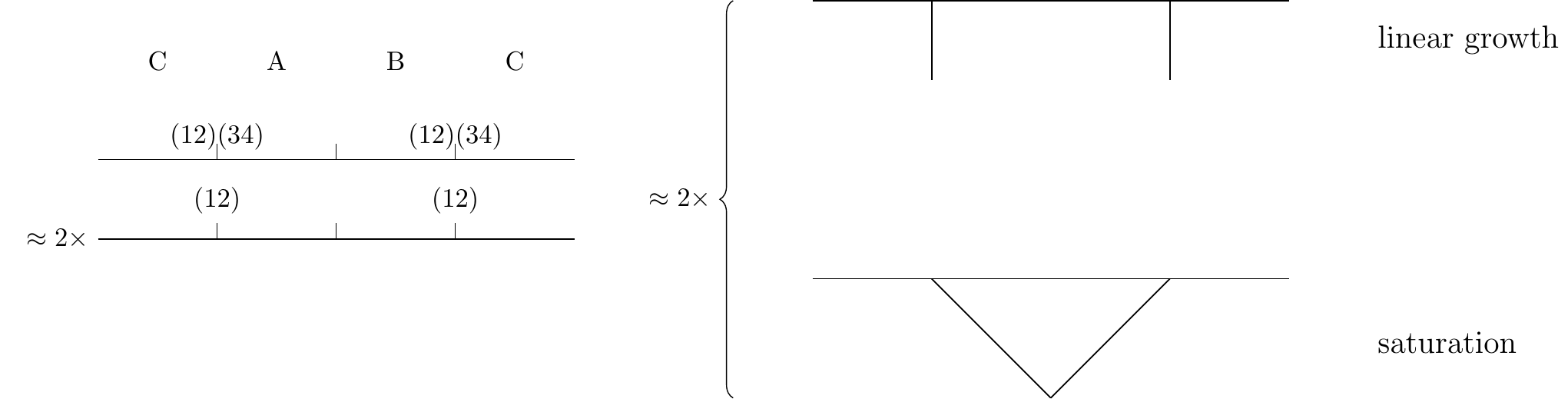}
\label{fig:dw_rho_state} 
}


\caption{Space time domain wall structures for the two terms in Eq.~\eqref{eq:s2}. According to our convention, the domain wall starts from the top boundary and go downward. (a) Domain wall configurations for the operator R\'enyi operator entanglement of unnormalized state (1st term in Eq.~\eqref{eq:s2}). There are six incoming (elementary) domain walls on the top boundary (see Fig.~\ref{fig:dw_1}). Three configurations that possibly occur at different stages of evolution are shown on the right of the brace.
(b) Domain wall configurations for the state entropy $S_2(AB)$ of region AB. There are four incoming (elementary) domain walls on the top boundary (see Fig.~\ref{fig:dw_2}). Their free energy is twice of a single pair of domain walls, due to their independence in the large $q$ limit. The two configurations for the two phases of state quench entanglement are shown on the right of the brace: 
In the linear growth phase, the two domain walls microscopically going down vertically; in the saturation phase, the two domain walls meet and annihilate.  }
\label{fig:rho_op_EE_dw}
\end{figure}

In this section, we study time evolution for the operator entanglement of the random unitary circuit. We proceed by listing the domain wall configurations that minimize the free energy at different stages of the evolution. We take $n = 2$ to simplify the domain wall structure, and work at leading order in $q\to \infty$ so that the elementary domain walls can be treated independently. This brings in the reduction
\begin{equation}
\label{eq:s2}
\overline{S_{2}^{ \rm op}(A, \rho_{AB})} =  - \log \overline{\tr_A\tr^2_B( | \rho_{AB}\rangle \langle  \rho_{AB} |   )} - 2\overline{S_2 (AB )}
\end{equation}
where the two terms corresponds have the initial domain wall configurations shown in Fig.~\ref{fig:rho_op_EE_dw}.

To warm up, let us first consider the evolution of the 2nd R\'enyi entropy of the state, which is subtracted in Eq.~\eqref{eq:s2}. When $t$ is small, the two domain walls at the entanglement cut cost the same amount of energy when going down in any direction within the light cone. Due to entropy consideration \cite{jonay_coarse-grained_2018,zhou_emergent_2018}, macroscopically they go down vertically. The free energy increases linearly with rate $v_E \ln q \approx \ln q$. At time $t > \frac{\ell}{2}$, the two domain walls can meet and annihilate at the intersection points of their light cones. The free energy of this configuration is $\ell \ln q $, which will be the minimal when $t \gtrsim \frac{\ell}{2}$. This switch from the configuration 1 to configuration 2 in Fig.~\ref{fig:dw_rho_state} gives rise to the linear growth and saturation behaviors of entanglement in a quenched state.

On the other hand, the first term in Eq.~\eqref{eq:s2} has six elementary domain walls to begin with. Three types of (macroscopic) configurations will occur in their life time. Configuration 1 is the beginning stage, where all domain walls going vertically down\footnote{Free random wall will predominantly go vertically, with fluctuation of size $\sqrt{t}$.} when $t$ is small. The growth rate is $6 \ln q $. In configuration 2, the four domain walls from the boundaries of region A first meet and fuse into two domain wall. These two go down vertically together with the other two domain walls, giving rise to a growth rate $4 \ln q$. In configuration 3 the middle domain walls split into two domain walls on the left and right, which then annihilate the remaining domain walls, giving the saturation value $2 \ell \ln q = 2S_2(AB)$. By equating their free energies, we can determine the respective transition time: 
\begin{equation}
\label{eq:trans_time}
\begin{aligned}
  &6 t_1 \ln q  = [2 t_1 + 2 \ell_A  + 2( t_1 -  \frac{1}{2} \ell_A )] \ln q  \implies t_1  = \frac{\ell_A}{2} \\
  &[2 t_2 + 2 \ell_A  + 2( t_2 -  \frac{1}{2} \ell_A )] \ln q = 2 \ell \ln q \implies t_2 = \frac{\ell}{2} - \frac{\ell_A }{4}
\end{aligned}
\end{equation}

\begin{figure}[h]
\centering
\subfigure[]{
\includegraphics[width=0.5\columnwidth]{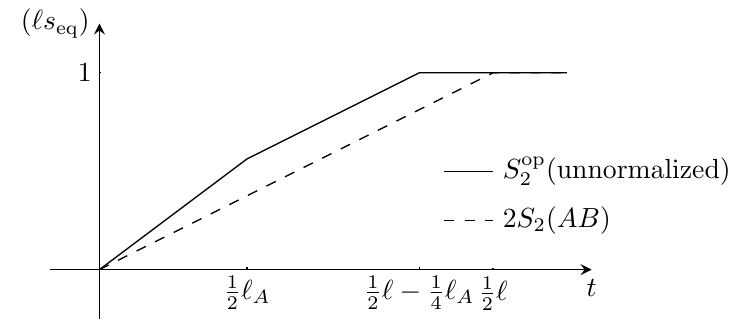}
\label{fig:op_EE_as_diff} 
}
\subfigure[]{
\includegraphics[width=0.4\columnwidth]{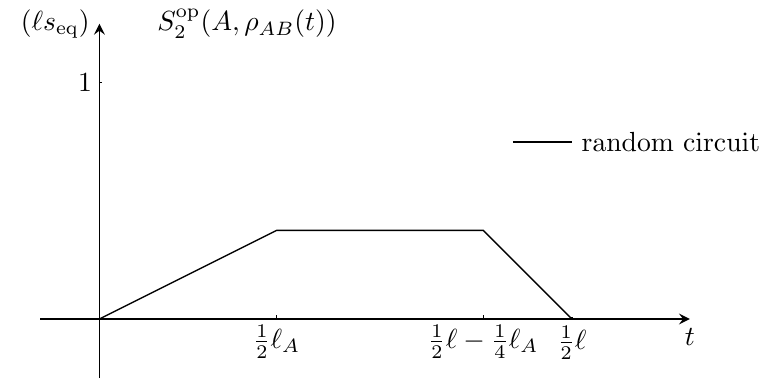}
\label{fig:op_EE_itself}
}
\caption{$S^{\rm op}_2(A, \rho_{AB})$ when $\ell_A = \ell_B = \frac{\ell}{2}$. (a). Operator entanglement as the difference of two terms in Eq.~\eqref{eq:s2}. As analyzed in the text, the solid curve has transitions between 3 configurations while the dashed curve representing the state entropy has only linear growth and saturation phases. (b). $S^{\rm op}_2(A, \rho_{AB} )$ as a function of time. }
\label{fig:rho_op_EE_curve}
\end{figure}

In Fig.~\ref{fig:op_EE_as_diff} we draw the time evolution of both terms in Eq.~\eqref{eq:s2}. The 1st term grows faster than the 2nd term in configuration 1 and then at the same rate in configuration 2. It eventually come to a tie with the state entanglement when reaching configuration 3. The operator entanglement is the difference between these two curves, which we draw in Fig.~\ref{fig:op_EE_itself}. It contains three stages of development: linear growth, plateau and linear decrease. 

We can generalize these results to integer $n$ and $\alpha$ assuming the $q \rightarrow \infty$ limit. The domain wall at the top boundary becomes identical to the permutation data in the twist fields in Eq.~\eqref{eq:twist_field}. There are again two terms
\begin{equation}
S^{\rm op}_n(A, \rho^\alpha_{AB} (t)) =  \frac{1}{1 - n} \log \tr_{\mathcal{H}_A\otimes \mathcal{H}_A}\tr^n_{\mathcal{H}_B\otimes \mathcal{H}_B}( |\rho^{\alpha}_{AB}(t) \rangle \langle \rho^{\alpha}_{AB} (t)| ) - \frac{1}{1 - n} \log \rho^{2\alpha n }_{AB}(t)
\end{equation}
The first term is a generalization of Fig.~\ref{fig:dw_rho_op}: on the top boundary, both the cuts between A,C and B,C host $(2\alpha - 1)n$ elementary domain walls, the cut between A, B host $2(n-1)$ elementary domain walls. Its dynamics also contain the 3 configurations in Fig.~\ref{fig:dw_rho_op}. The second term is a generalization of Fig.~\ref{fig:dw_rho_state}, the difference is only that the number of elementary domain walls is $(2 \alpha - 1)n$ rather than $2$ at the entanglement cut. From the geometry, one can see immediately that the transition time scales do not dependent on $n$ and $\alpha$, as long as the order of the configurations to appear is the same as $n = 2, \alpha = 1$. This can be confirmed by an explicit computation
\begin{equation}
\begin{aligned}
[ 2(2\alpha - 1)n + 2(n -1)  ]& t_1 \ln q   = [(2\alpha - 1)n t_1 + (2\alpha - 1)n \frac{\ell_A}{2} \\
&+ 2(n-1) \frac{\ell_A}{2} + (2\alpha-1)n ( t_1 -  \frac{1}{2} \ell_A )] \ln q  \implies t_1  = \frac{\ell_A}{2} \\
[(2\alpha - 1) n ] \ell \ln q  &= [(2\alpha - 1)n t_2 + (2\alpha - 1)n \frac{\ell_A}{2} \\
&+ 2(n-1) \frac{\ell_A}{2} + (2\alpha-1)n ( t_2 -  \frac{1}{2} \ell_A )] \ln q  \implies t_2 = \frac{\ell}{2} - \frac{\ell_A }{4}
\end{aligned}
\end{equation}
Further computation shows that the plateau value of $S_n^{\rm op}(A, \rho_{AB}^\alpha(t) )$ is also $n$ and $\alpha$ independent. 

We then claim that for a given integer $n \ge 2$ and $\alpha \ge 1$, the $q \rightarrow \infty$ collapses all the operator entanglement $S_n^{\rm op}(A, \rho_{AB}^\alpha(t) )$ to the curve in Fig.~\ref{fig:op_EE_itself}. 

The results above are derived for the large $q$ random unitary circuit. In a realistic system or even random circuit with finite $q$, we argue that such a generalized domain wall picture still exist, and the we can incorporate the non-universal feature into a coarse grained line tension function -- the free energy of the domain wall per unit length. 

We can not justify this assumption to the full extent, even when restricted to chaotic systems. Instead we motivate this assumption from the following aspects. 

Microscopically, the domain walls are the twist of the permutations. These permutations, as we draw in Fig.~\ref{fig:block_aver}, represents different ways to contract among copies of $U$ and $U^*$. These pairwise contraction gives a real deterministic value given in this case by formula shown in Fig.~\ref{subfig:block_aver}. There will contributions other than these contractions if we take a unitary gate in Fig.~\ref{fig:circle_block}. But these will usually be complex numbers, can consequently dephase in time in chaotic systems. This is also the reason why the domain wall picture works so well in the random average of the random unitary circuit: the random average makes the dephasing exact. 

Macroscopically, we can generally assume that the entanglement growth of a given cut can be written as a integral of a local quantity in space time. In the examples where the minimal cut works, it would be the integral of the increment of entanglement along the cut. Because of the translational invariance in space and time (true in the random average sense in the random circuit), we imagine that such minimal cut should be a line (or membrane in higher dimension), and the growth rate should depend only on the slope of this line. This motivates the line tension function $\mathcal{E}_n(v)$ for R\'enyi index $n$ and slope $v$. 

Ref.~\cite{jonay_coarse-grained_2018} motivates the existence of such line tension function by first assuming the entanglement growth rate $\partial_t S$ to be dependent only the a function of the slope $\Gamma(\partial_x S)$ at the leading order in coarse graining. The assumption only uses the local information of the entanglement (the derivative) which is reasonable in a system with only local interactions. The authors then deduce that the line tension function is the Legendre transform of $\Gamma(x)$. By constraints of entanglement (e.g. subadditivity imposed on $\Gamma(x)$), they further derive that $\mathcal{E}_n(v)$ has to satisfy several properties
\begin{equation}
\mathcal{E}_n( 0 ) = v^{(n)}_E \quad \mathcal{E}(v)  \ge v \quad \mathcal{E}( v_B)  = v_B 
\end{equation}
where $v_B$ is the butterfly velocity that characterizes the speed of operator spreading and many-body chaos \cite{jonay_coarse-grained_2018}. When entanglement saturates in Fig.~\ref{fig:rho_op_EE_curve}, the free energy cost is $ 2 \times \ell \mathcal{E}_n(v) / v $, which is minimized at slope $v = v_B $ rather than $v = 1$, though the saturation value is still the maximal entanglement.

In the configurations we considered in Fig.~\ref{fig:rho_op_EE_dw}, the domain walls either go vertically or at slope $v_B$ (which maximizes the line tension), we can then only use $\mathcal{E}(0) = v_E^{(n)}$ and $\mathcal{E}(v_B) = v_B$ to estimate the transition points. 

The random unitary circuit at the $n = 2$ case is a known example that the line tension function works well in practice even for $q = 2$. This is because after random average, the finite $q$ correction only introduces weak interactions between the domain walls in Fig.~\ref{fig:rho_op_EE_dw}. In leading order, we have 
\cite{nahum_dynamics_2017,nahum_operator_2017,von_keyserlingk_operator_2017,zhou_emergent_2018}
\begin{equation}
  v_B = \frac{q^2 - 1}{q^2 + 1}  \quad v_E ^{ (2) } = \frac{1}{\ln q} \left(  \ln \frac{q^2 + 1}{2q} + \cdots \right) 
\end{equation}
where $v_E$ is a perturbative series containing higher $\mathcal{O}(1/q^8)$ term due to the interactions of the domain walls.

Using these data, we can determine the transition points. 
\begin{equation}
\begin{aligned}
s_{\rm eq} \Big[ 6 t_1 v_E^{(2)} \Big] = s_{\rm eq}\Big[ 4 \frac{\ell_A}{2} +2 (t_1 - t_0 )v_E^{(2)}   + 2 t_1 v_E^{(2)} \Big] \quad t_0 = \frac{\ell_A}{2 v_B} \implies \quad  t_1 = \frac{\ell_A}{ v_E^{(2)}} \left( 1 - \frac{v_E^{(2)}}{2v_B} \right) \\
s_{\rm eq} \Big[4 \frac{\ell_A}{2} +2 (t_2 - t_0 )v_E^{(2)}   + 2 t_2 v_E^{(2)} \Big] = s_{\rm eq}\Big[ 4 \frac{\ell_A}{2} + 4  \frac{\ell_B}{2} \Big] \quad \implies \quad  t_2 = \frac{\ell_A}{2 v_E^{(2)} }\left(  \frac{\ell_B}{\ell_A} + \frac{v_E^{(2)}}{2v_B} \right)\\
\end{aligned}
\end{equation}

\begin{figure}[h]
\centering
\includegraphics[width=0.6\columnwidth]{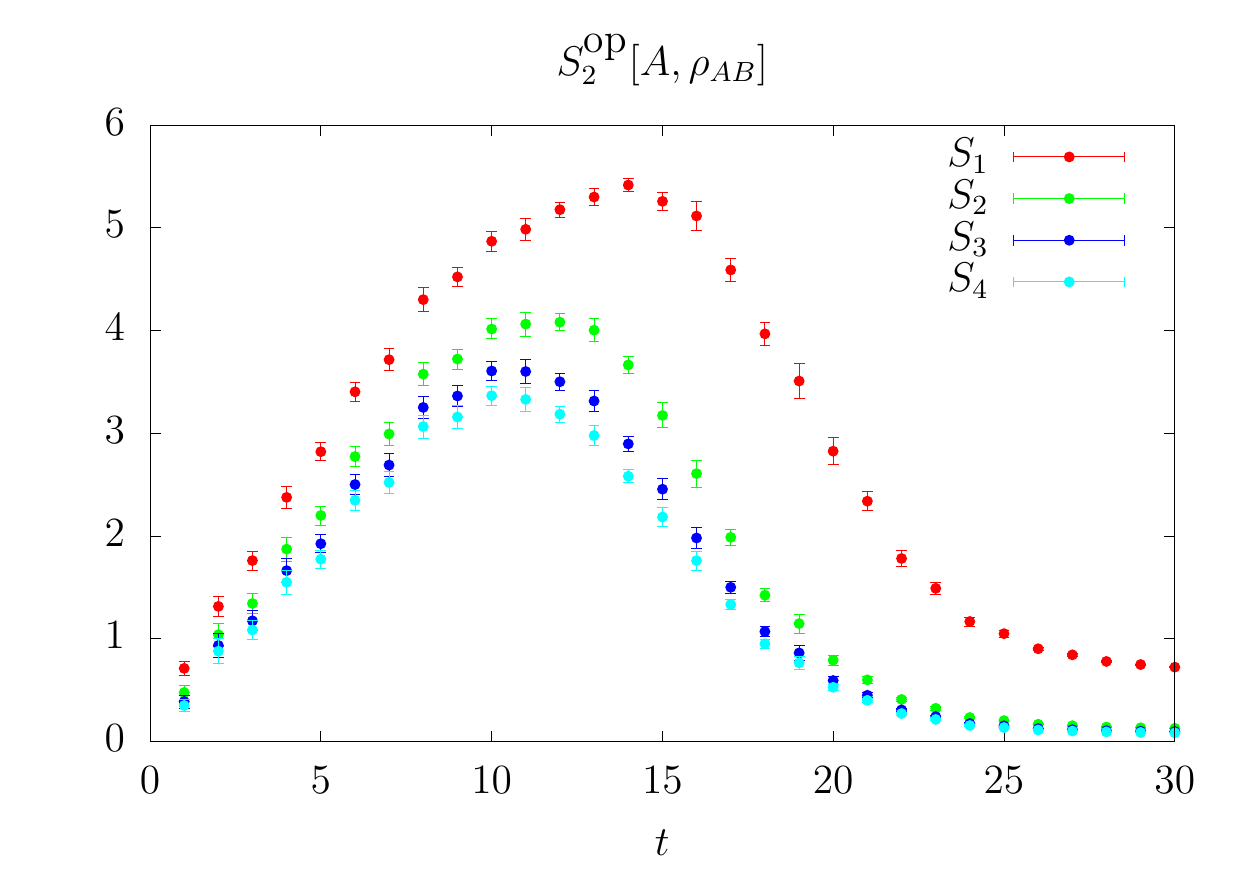}
\caption{Numerical results of random unitary circuit at $q = 2$ averaged over 10 samples. System parameters: $\ell = 28$, $\ell_A = \ell_B = 6$.  }
\label{fig:S}
\end{figure}

We look at the symmetric case $\ell_A = \ell_B$. When $q = 2$, we have $t_1 > t_2$, which means that configuration 2 will be skipped. There is only one transition time from configuration 1 to configuration 3
\begin{equation}
6 t_3 v_E^{(2)}  = 4 \left(\frac{\ell_A }{2} + \frac{\ell_B }{2}\right) \implies t_3 = \frac{\ell }{3 v_E^{(2)} }
\end{equation}
This is verified in the numerical results in Fig.~\ref{fig:S}, where $\ell = 28$, $\ell_A = \ell_B = 6$ and $q = 2$. As we can see the plateau phase found in large $q$ is absent, instead only a peak occurs at about $t_3 \approx 12.5$. 

It therefore suggests that the persistence of the plateau regime is a signature of the strongly chaotic system, and likely will only occur with large on-site degrees freedom.


\section{Holography}
\label{sec:holo}

In this section, we look at holographic CFTs which represent systems that are maximally chaotic in terms of saturation of the chaos bound \cite{maldacena_bound_2015}. In AdS/CFT, the entanglement entropy is conveniently captured geometrically by the area of the RT/HRT surfaces \cite{RT, HRT}. To study the time evolution for operator entanglement in holographic CFTs, our strategy is to start from a bulk description for operator states, whose asymptotic boundaries are two-copied subsystems, and then apply the HRT prescription. The results are obtained using related pictures derived in Ref.~\cite{tom:2019}. We first review the bulk picture for the stationary operator states, and then generalize to the dynamical case for generic $\alpha$. After that we study the operator entanglement entropy for $\alpha=1/2$, which is the case considered in \cite{tom:2019}, and turns out to be the natural case for constructing the bulk operator state. We will see that the operator entanglement of the reduced density matrix undergoes three phases in the bulk, which is partly driven by the background geometry representing the density matrix operator state.

\subsection{Stationary states}
We begin by introducing the prescription derived in Ref.~\cite{tom:2019} using the language of Euclidean path-integral, which works explicitly for stationary states, or more generally states that are time reflection symmetric under: $t\to -t$.  

The goal is to construct the bulk geometry dual to the operator state $|\rho_{AB}\rangle $, associated with the reduced density matrix:
\begin{equation} 
\rho_{AB}= \text{tr}_C |\psi\rangle \langle\psi|
\end{equation}
We assume that the global state $|\psi\rangle$ in the boundary CFT has a smooth bulk geometry. Furthermore, it is stationary with a reflection symmetry $t\to -t$ and thus admits a Euclidean path-integral definition. The reduced density matrix element between two field configurations $\phi_{1,2}$ on $AB$ is then given by:
\begin{equation} 
\left(\rho_{AB}\right)_{\phi_1,\phi_2}=\int_{\Phi_{AB}^+ =\phi_1,\Phi_{AB}^-=\phi_2} \mathcal{D}\Phi \;f(\Phi)^\dagger f(\Phi)\;e^{-S_E(\Phi)}
\end{equation}
where $\Phi$ denotes collectively the relevant quantum fields, and $f(\Phi)$ denotes the state-creating sources for $|\psi\rangle$. The path integral is over configurations that has a branch-cut along $AB$, and the field configurations along the branch-cut are fixed to be $\phi_1$ and $\phi_2$ from above and below respectively.

Alternatively, this defines the (un-normalized) wave-functional of the operator state $|\rho_{AB}\rangle$ on the doubled subsystem $\lbrace AB,A'B'\rbrace$:
\begin{equation}
\langle \phi_1,\phi_2|\rho_{AB}\rangle =\left(\rho_{AB}\right)_{\phi_1,\phi_2} 
\end{equation}

\begin{figure}[h!]
\centering
\includegraphics[scale=0.2]{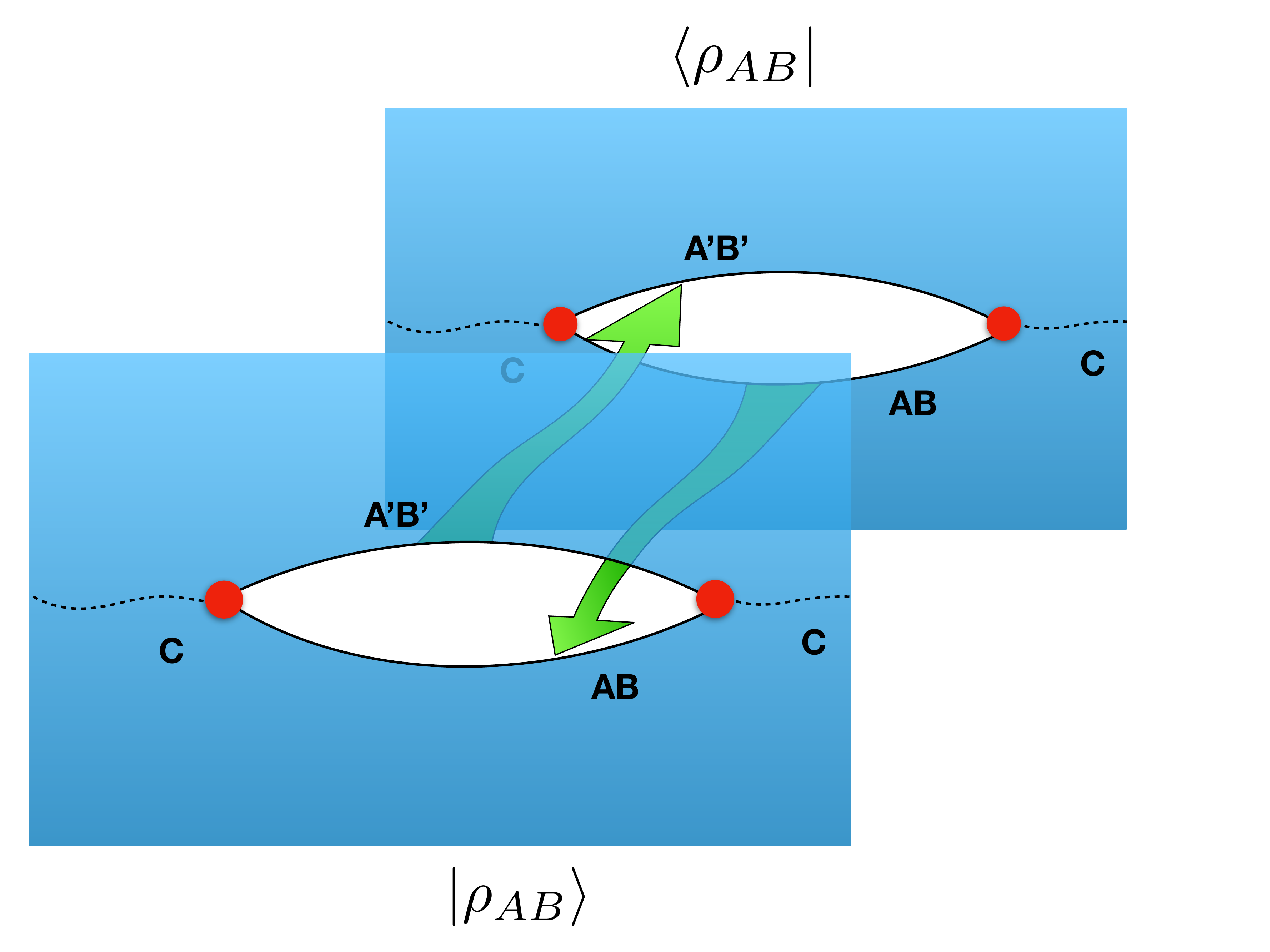}
\includegraphics[scale=0.2]{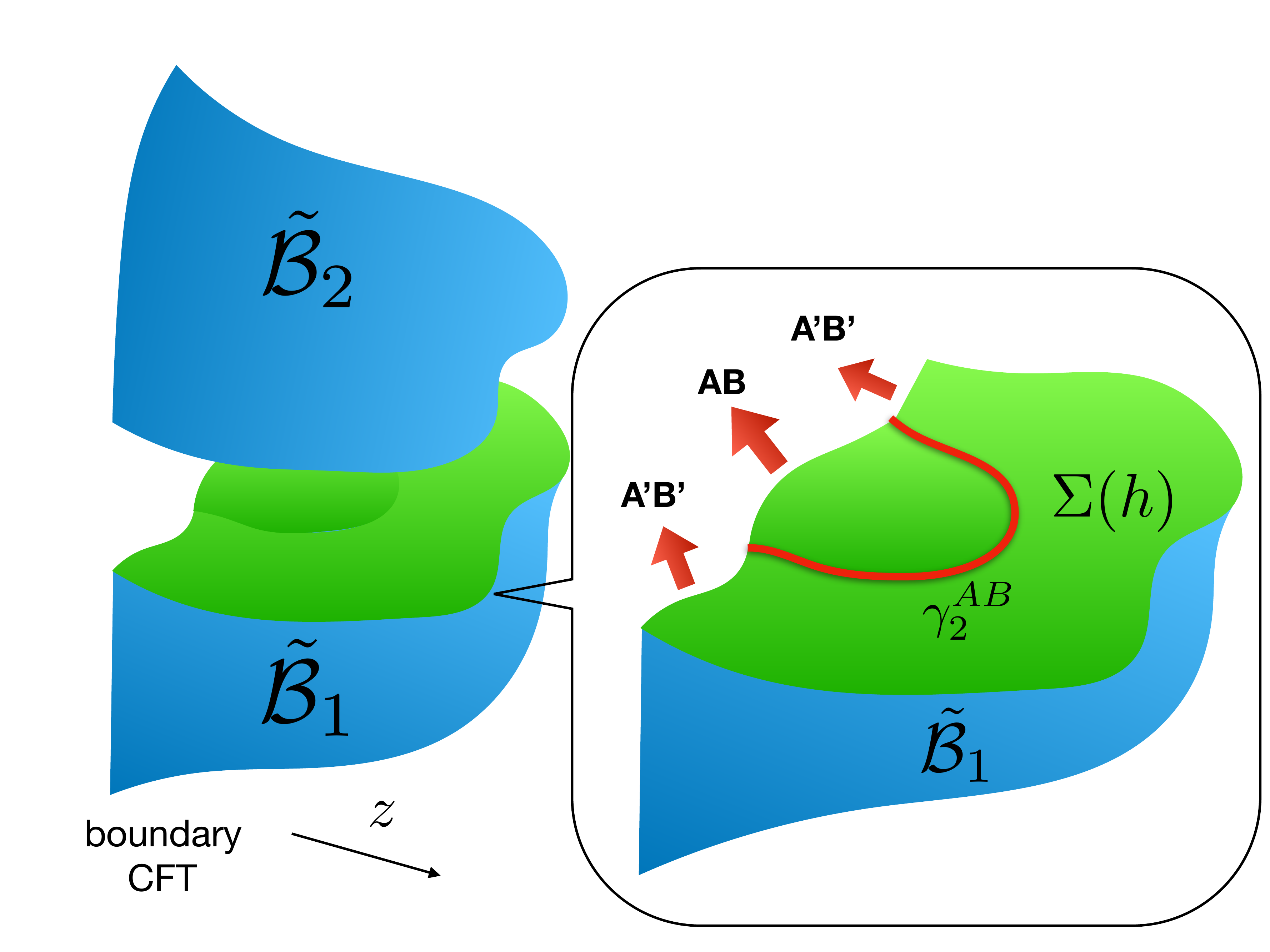}
\caption{Computation of the operator state norm $\mathcal{N}=\langle \rho_{AB}|\rho_{AB}\rangle$. Left: Euclidean path-integral on the doubly-branched manifold in the boundary CFT; Right: bulk saddle point contribution $\mathcal{B}=\mathcal{\tilde{B}}_1\cup \mathcal{\tilde{B}}_2$ to $\mathcal{N}$. The two wedges intersect at the fixed Cauchy surface $\Sigma(h)=\mathcal{\tilde{B}}_1\cap \mathcal{\tilde{B}}_2$ under reflection, which contains the cosmic-brane $\gamma^{AB}_2$. Due to the brane backreaction, the bulk portions matching to $AB$ and $A'B'$ towards the boundary only extend an angular range of $\pi$ around $\gamma^{AB}_2$ in the bulk.} 
\label{fig:norm_1}
\end{figure}

In order to find the bulk dual of $|\rho_{AB}\rangle$, let us consider computing its norm: 
\begin{equation} 
\mathcal{N} =\langle \rho_{AB}|\rho_{AB}\rangle = \text{tr} \rho_{AB}^2
\end{equation} 
For stationary states this is given by a boundary Euclidean path integral over a boundary manifold $\partial \mathcal{M}$ doubly branched across $AB$ (see Fig.~\ref{fig:norm_1}). Through the dictionary of AdS/CFT, this fixes the boundary conditions for the corresponding Euclidean path-integral over bulk fields and geometries, which matches the CFT result.  Let us further suppose that the bulk Euclidean path integral is dominated by a particular saddle point contribution, denoted by $\left(\mathcal{M}, g, \varphi\right)$ where $g, \varphi$ are respectively the metric and the collection of bulk fields: 
\begin{equation}
\text{tr} \rho_{AB}^2 = \int_{\partial \mathcal{M}} \mathcal{D}\Phi\;\Pi_{i=1,2} f^\dagger_i (\Phi)f_i(\Phi)\; e^{-S_E\left(\Phi,\partial \mathcal{M}\right)} \approx \exp{\left[-I^{bulk}_E\left(\mathcal{M},g,\varphi\right)\right]}
\end{equation} 
For simplicity we omit writing $\varphi$ in specifying bulk solutions from now on. On the other hand, path-integrals over such branched manifolds are related to the refined 2nd Renyi-entropy \cite{xi:2016}. For holographic theories and assume replica symmetry, the dominant saddle $\mathcal{B}$ can be constructed as in Ref.~\cite{xi:2016} and takes the form of two identical ``wedges" (see Fig.~\ref{fig:norm_1}):
\begin{equation}
\mathcal{B}=\mathcal{\tilde{B}}_1\cup \mathcal{\tilde{B}}_2
\end{equation}
where each wedge $\mathcal{\tilde{B}}_{1,2}$ can be obtained by solving under the original single-copied boundary condition, the back-reacted geometry upon inserting a ``cosmic brane" $\gamma^{AB}_2$ that anchors at the boundary along the entangling surface $\partial{(AB)}$, and has the tension 
\begin{equation}
T_n = \frac{n-1}{4nG_N},\;\;n=2 
\end{equation}
This creates a conical deficit of angle: $\Delta \phi = \pi$, so each wedge extends an angular range of $\pi$ around $\gamma^{AB}_2$. The two wedges are glued at a bulk Cauchy surface $\Sigma(h)$ with induced metric $h$: 
\begin{equation}
\Sigma(h) = \mathcal{\tilde{B}}_1 \cap \mathcal{\tilde{B}}_2
\end{equation}
$\Sigma(h)$ also contains the cosmic brane $\gamma^{AB}_2$. The reflection symmetry characterizing the stationary state acts by reflecting across $\Sigma$ and interchanging the two wedges: $\mathcal{\tilde{B}}_1\leftrightarrow \mathcal{\tilde{B}}_2$. Semi-classically, we can interpret $\Sigma(h)$ as the bulk spatial geometry at $t=0$ dual to the operator state $|\rho_{AB}\rangle$. 

\begin{figure}[h!]
\centering
\includegraphics[scale=0.137]{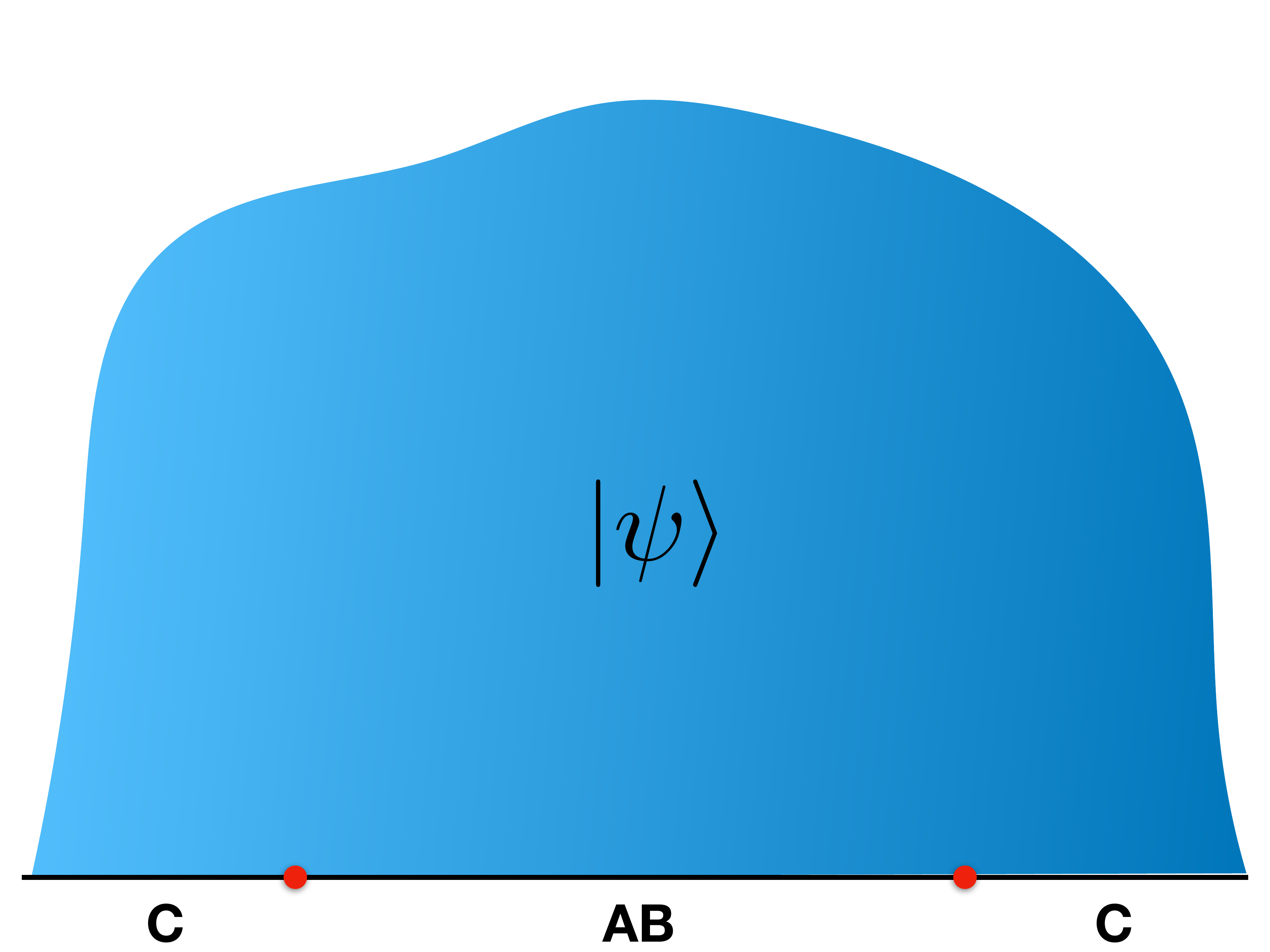}
\includegraphics[scale=0.137]{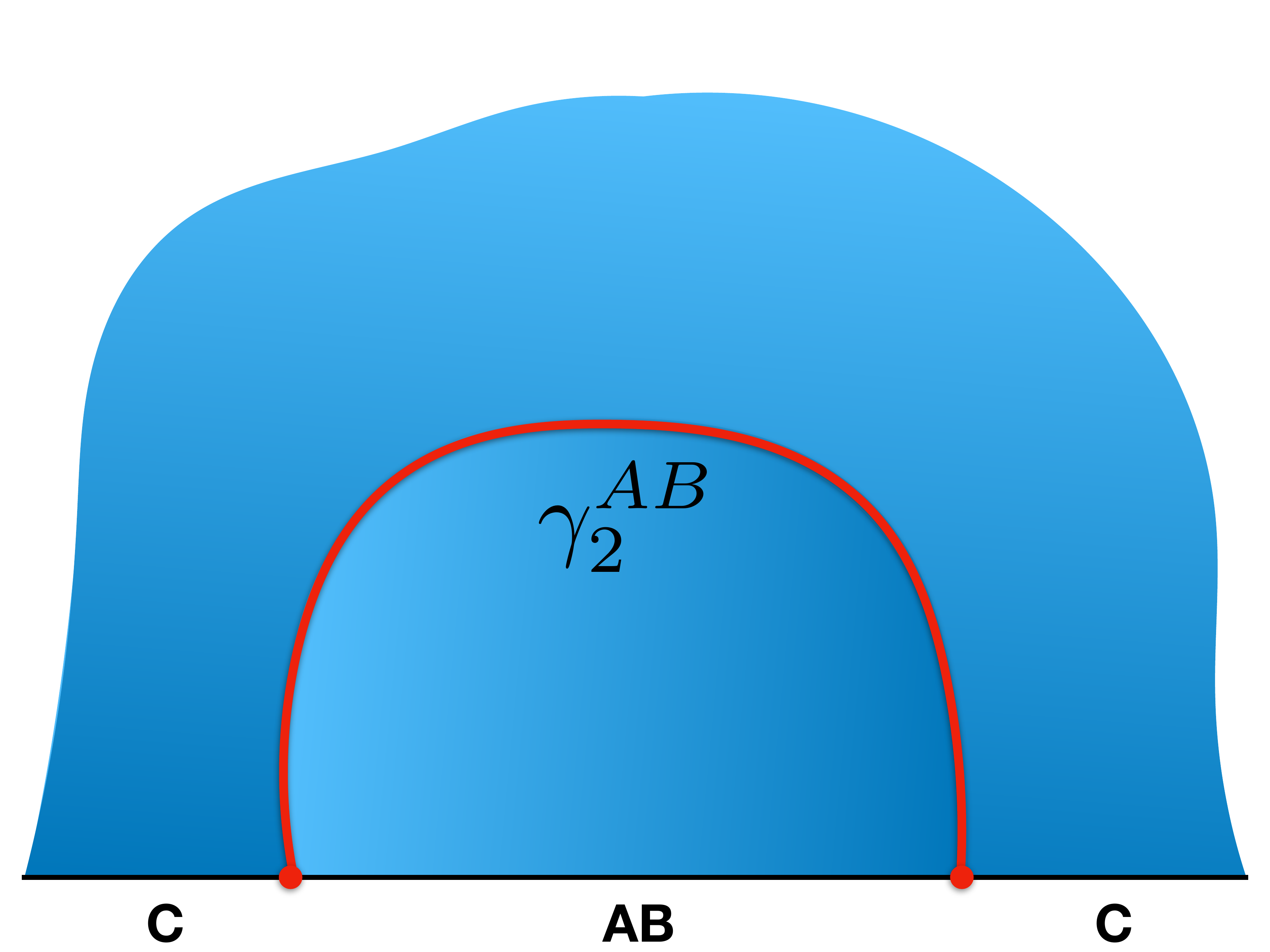}
\includegraphics[scale=0.137]{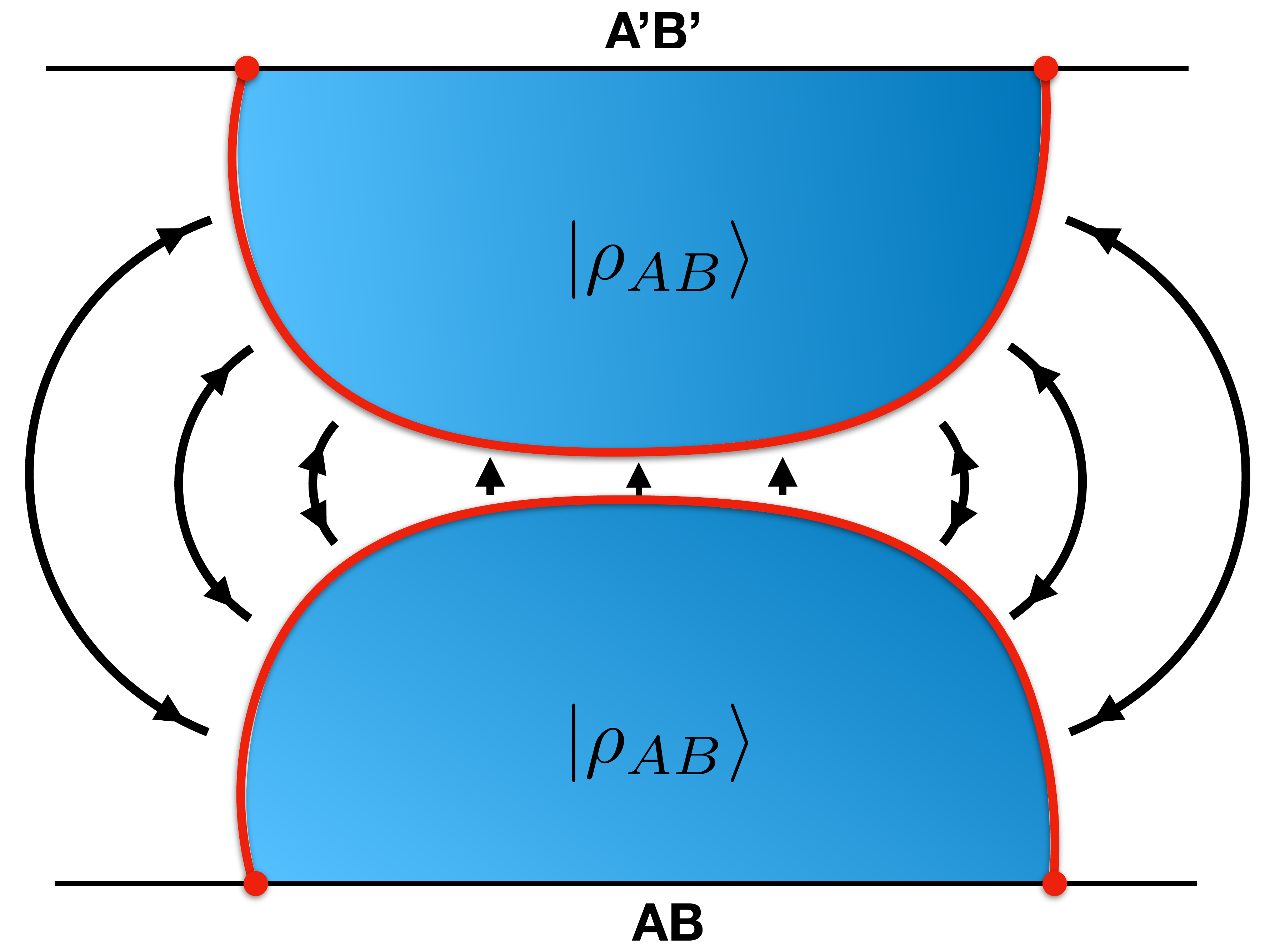}
\caption{Prescription for obtaining the bulk geometry dual of operator state $|\rho_{AB}\rangle$ from that of $|\psi\rangle$. Left: start from the bulk dual of $|\psi\rangle$; Center: insert a cosmic brane $\gamma^{AB}_2$, let it back-react on the dual geometry and settle down; Right: spatial geometry dual to $|\rho_{AB}\rangle$ is obtained by identifying two copies of the relevant bulk regions across $\gamma^{AB}_2$.}
\label{fig:steps}
\end{figure}

Let us now describe the prescription operationally (see Fig.~\ref{fig:steps}). To find the bulk geometry dual to the stationary operator state $|\rho_{AB}\rangle$, we start from the bulk dual of the original global state $|\psi\rangle$, insert a cosmic brane $\gamma^{AB}_2$ of tension $T=\frac{1}{4G_N}$ anchoring towards the boundary at $\partial(AB)$, let it back-react and settle down. The spatial geometry dual to $|\rho_{AB}\rangle$ is then obtained by identifying two copies of the back-reacted spatial geometries across the settled-down position of the cosmic brane $\gamma^{AB}_2$. 

\subsection{Dynamical states}
\begin{figure}[h!]
\centering
\includegraphics[scale=0.2]{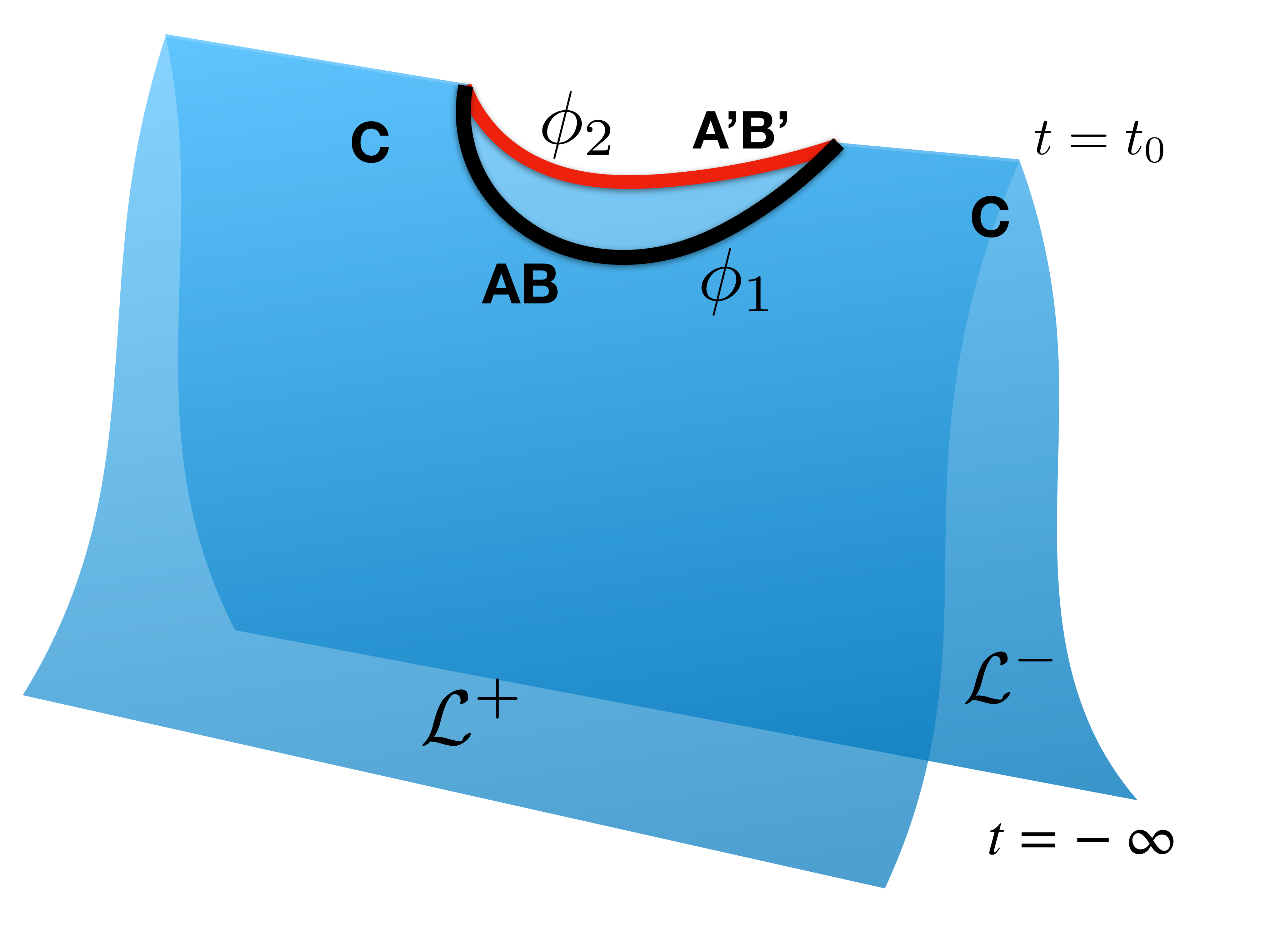}
\includegraphics[scale=0.2]{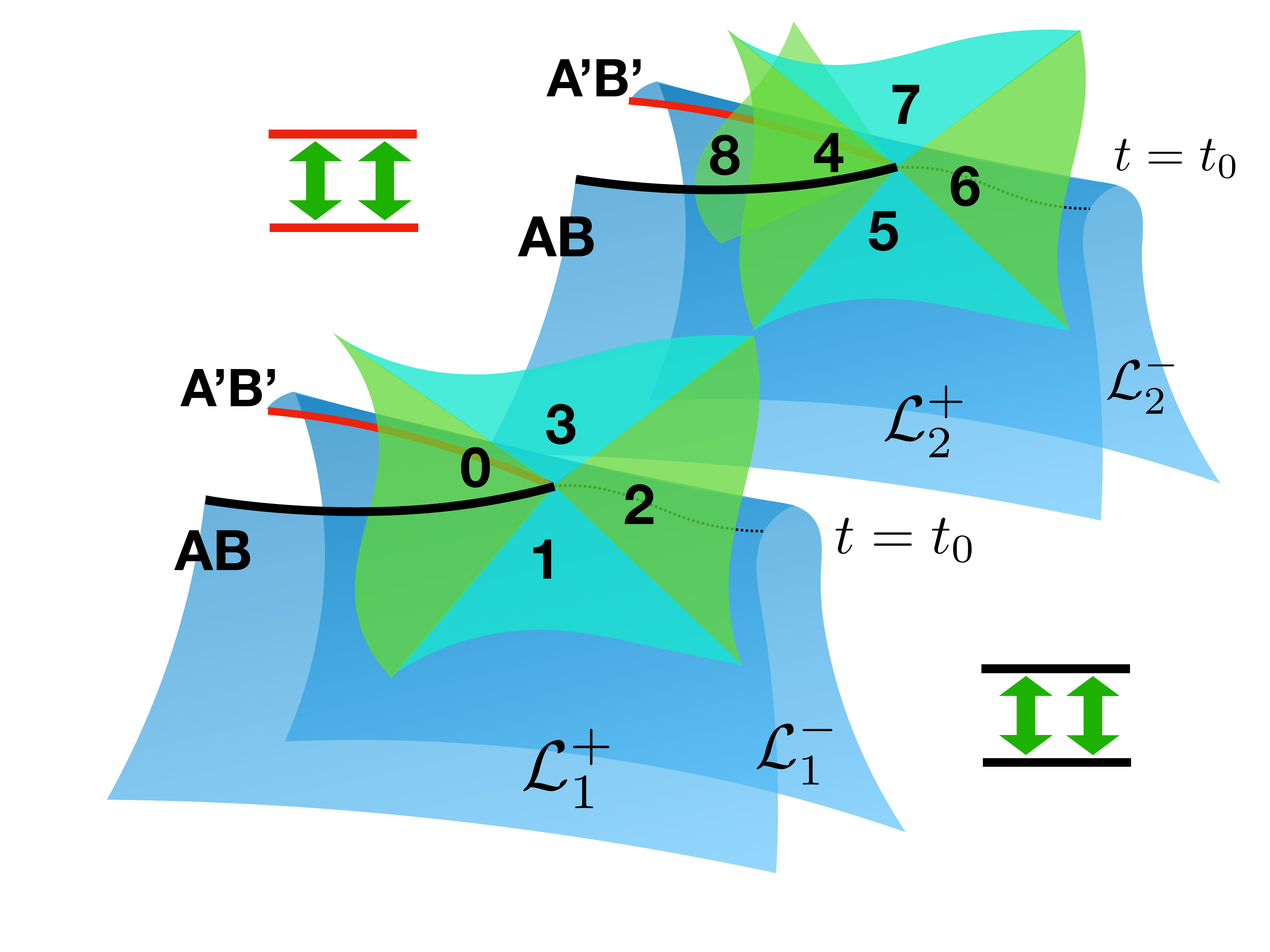}
\caption{Left: density matrices element $\left[\rho_{AB}(t_0)\right]^{\phi_1}_{\phi_2}$ defined by path-integral on Schwinger-Keldysh two-folds $\mathcal{L}^{\pm}$. Right: norm of the operator state $\mathcal{N}(t_0)$ as a path-integral over two Schwinger-Keldysh contours (i.e. four time-folds $\mathcal{L}^{\pm}_{1,2}$) glued as shown by the green arrows. The local singularity arising from the gluing is demonstrated by the structure of Rindler wedges, indexed by $m=0,...,8$. The index can be understood as the imaginary part of the local Rindler time $ds^2 = -r^2 d\tau^2+dr^2+d\omega^i d\omega^i$ near $\partial{(AB)}$: $\text{Im}\;\tau=\frac{m\pi}{2}$. Identifying $m=0$ and $m=8$ puts the local Rindler temperature at $T_{\text{Rindler}}=1/4\pi$. }
\label{fig:dynamic_boundary}
\end{figure}

We can extend the prescription to the time-dependent case, i.e. $|\psi\rangle$ is a dynamical state. In this case, the (un-normalized) density matrix elements, i.e. the wave-functional of the operator state, need to be constructed using the Schwinger-Keldysh formalism \cite{SK1, SK2}. At the instant $t=t_0$, it involves a path-integral over two Lorentzian time-folds $\mathcal{L}^+$ and $\mathcal{L}^-$, ranging from $t=(-\infty,t_0)$ and $t=(t_0,-\infty)$ respectively and glued at $t=t_0$ across $C$ (see Fig.~\ref{fig:dynamic_boundary}) : 
\begin{equation}
\label{eq:SK_def}
\rho_{AB}(t_0)\left(\phi_1\cup\phi_2\right) = \left(\rho_{AB}\right)^{\phi_1}_{\phi_2}= \int_{\Phi^+_{AB}(t_0)=\phi_1, \Phi^-_{AB}(t_0)=\phi_2} \mathcal{D}\Phi\; f^\dagger(\Phi)f(\Phi)e^{i S\left(\Phi,\mathcal{L}^+\cup \mathcal{L}^-\right)}
\end{equation}
where $f$ denotes the state-creating sources. Now let us follow the previous route and compute the norm: 
\begin{equation}
\mathcal{N}(t_0)=\langle \rho_{AB}(t_0)|\rho_{AB}(t_0)\rangle  = \int_{\partial \mathcal{M}} \mathcal{D}\Phi\;\Pi_{i=1,2} f^\dagger_i (\Phi)f_i(\Phi)\; e^{-i S\left(\Phi,\partial \mathcal{M}\right)} 
\end{equation}
where $\partial\mathcal{M}$ is a boundary (Lorentzian) manifold branched over two Schwinger-Keldysh contours \cite{Mukund:2016}, i.e. four time-folds $\mathcal{L}^{\pm}_{1,2}$, and glued at $t=t_0$ (see Fig.~\ref{fig:dynamic_boundary}). The gluing creates a local singularity on the branched manifold, in the form of a non-standard Rindler temperature: $T_{\text{Rindler}}=1/4\pi$.   

\begin{figure}[h!]
\centering
\includegraphics[scale=0.22]{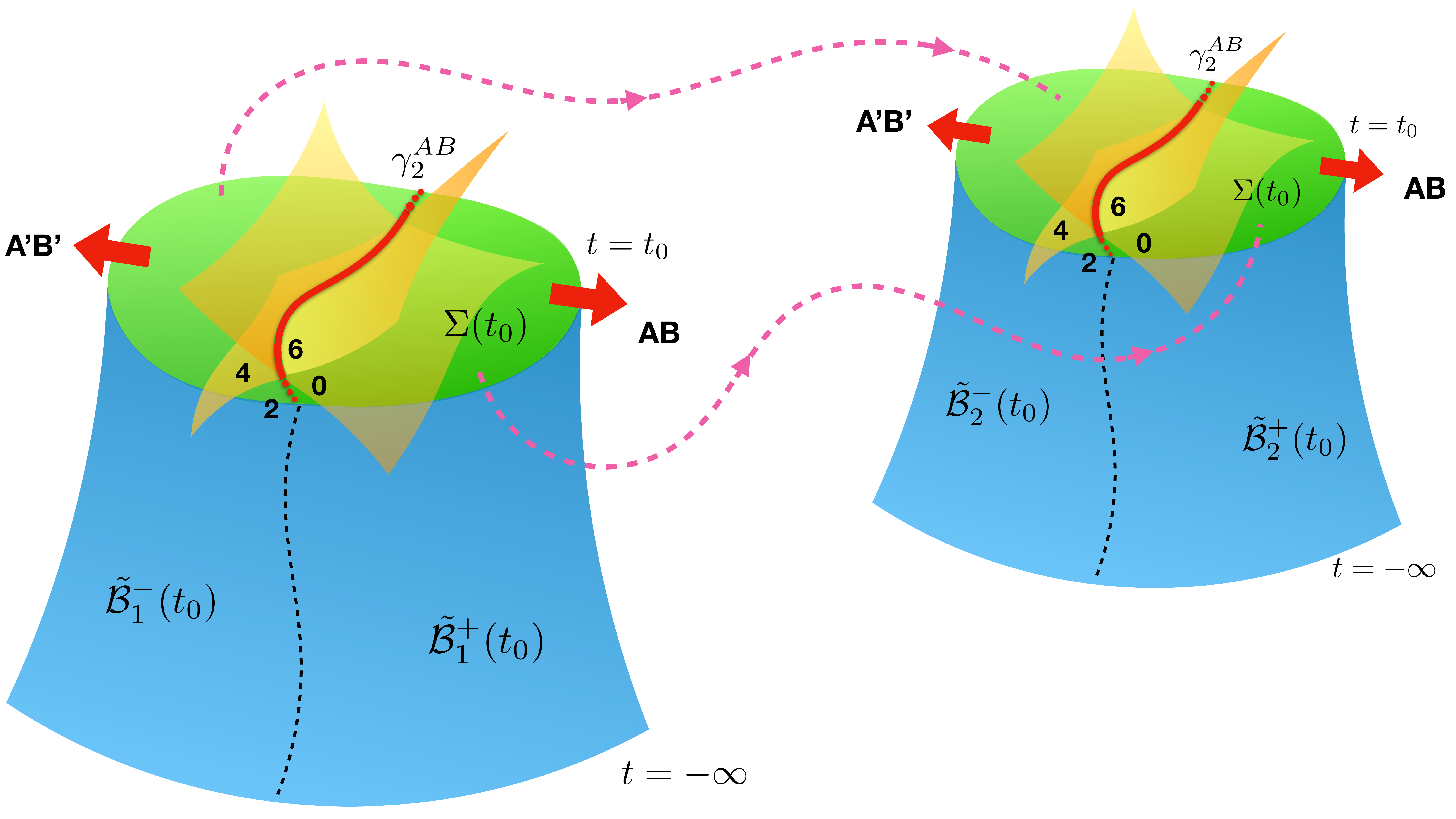}
\caption{Bulk Lorentzian saddle in the form of two Schwinger-Keldysh two-folds: $\mathcal{B}(t_0)=\cup^{a=\pm}_{i=1,2}\;\tilde{\mathcal{B}}^a_i(t_0)$, glued at $\Sigma(t_0)$. The structure of Rindler wedges near the cosmic brane $\gamma^{AB}_2$ are shown, whose indices $m$ indicate the corresponding boundary Rindler wedges they asymptote to in Fig.~\ref{fig:dynamic_boundary}. However, the bulk is smooth in the interior, the imaginary part of the local Rindler time near $\gamma^{AB}_2$ in the wedges are $\text{Im}\;\tau = \frac{m\pi}{4}$, corresponding to a standard Rindler temperature $T_{\text{Rindler}}=1/2\pi$. As a result, a single two-folds (e.g. $\tilde{\mathcal{B}}_1$) has a Rindler temperature of $T_{\text{Rindler}}=1/\pi$ near $\gamma^{AB}_2$ after identifying $m=0$ and $m=4$.}
\label{fig:dynamic_bulk}
\end{figure}

Similar to before, we can push this path-integral into the bulk via AdS/CFT, and suppose further that there exists a dominant contribution to this from the Lorentzian saddle-point geometry $\left(\mathcal{B}(t_0),g\right)$ \cite{Mukund:2016, Skenderis:2009}:
\begin{equation}
\mathcal{N}(t_0) \approx \exp{\left(i I^{\text{bulk}}(\mathcal{B}(t_0),g)\right)} 
\end{equation}
Strictly speaking, the existence of such a dominant Lorentzian saddle stands on shakier ground compared to the Euclidean counterpart. We shall not dwell on this too much in this paper other than acknowledging the potential subtlety, and proceed assuming such a saddle exists. The cosmic-brane construction has a natural generalization to the dynamical case.  In particular, let us still assume that the replica symmetry on the boundary, which in the dynamical case interchanges the two Schwinger-Keldysh two-folds, remains a symmetry of the bulk saddle. There exists a co-dimension 2 bulk hypersurface $\gamma^{AB}_2$ of fixed points under this symmetry, which ends on the boundary at $\partial(AB)$ at $t=t_0$. The bulk saddle $\mathcal{B}(t_0)$ takes the form of two identical bulk Schwinger-Keldysh two-folds, glued at a bulk Cauchy surface $\Sigma(t_0)$ that contains $\gamma^{AB}_2$ (see Fig.~\ref{fig:dynamic_bulk}):
\be 
\mathcal{B}(t_0)=\tilde{\mathcal{B}}_1(t_0)\cup \tilde{\mathcal{B}}_2(t_0),\;\ \tilde{\mathcal{B}}_1(t_0)\cap \tilde{\mathcal{B}}_2(t_0) = \Sigma(t_0)
\end{equation}
Furthermore, being a saddle point, the bulk geometry near $\gamma^{AB}_2$ should be locally smooth. For Lorentzian manifolds this implies that the local Rindler temperature takes the standard value $T_{\text{Rindler}}=1/2\pi$. This refers to the branched four-folds, and as a result, each of the two-folds $\tilde{\mathcal{B}}_{1,2}(t_0)=\tilde{\mathcal{B}}^+_{1,2}(t_0)\cup \tilde{\mathcal{B}}^-_{1,2}(t_0)$ has a local Rindler temperature near $\gamma^{AB}_2$ of $T_{\text{Rindler}}=1/\pi$. This is the back-reacted bulk solution obtained by inserting a cosmic brane of tension $T_2=\frac{1}{8G_N}$ ending on $\partial(AB)$ at $t=t_0$, with the boundary being the Lorentzian two-fold $\mathcal{L}^+\cup\mathcal{L}^-$ that defines the density matrix in Eq.~\eqref{eq:SK_def}. The settled-down location of the cosmic brane is $\gamma^{AB}_2$. The bulk solution $\tilde{\mathcal{B}}_1(t_0)$ obtained this way then defines the leading order description of the bulk state dual to the operator state $|\rho_{AB}(t_0)\rangle$. We can understand the instantaneous spatial geometry $\Sigma(t_0)$ as being obtained by identifying two copies of the back-reacted spatial geometries ending on $AB$ at $t=t_0$ across the cosmic brane $\gamma^{AB}_2$.

It is worthwhile to clarify the notions of dynamics. The operator state $|\rho_{AB}(t_0)\rangle$ ``evolves" as we change $t_0$. However, this evolution is not unitary, as manifested by the $t_0$-dependent norm $\mathcal{N}(t_0)$.  As a result, such $t_0$-dependence is not encoded by a single Lorentzian bulk solution; instead, $t_0$ serves as a ``label" for the one-parameter family of bulk saddles $\left\lbrace \tilde{\mathcal{B}}_1(t_0),\;t_0\in \mathbb{R}\right\rbrace$. On the other hand, for a fixed $t_0$ one can evolve the state $|\rho_{AB}(t_0, s)\rangle=U(s)|\rho_{AB}(t_0)\rangle $ using some unitary operator $U(s)$ and study the dynamics in $s$. A canonical candidate is  $U(s)=e^{is\left( K^\psi_{AB}+K^\psi_{A'B'}\right)}$, where  $K^\psi_{AB}$ is the half-side operator modular Hamiltonian, and is related to the original state Hamiltonian $K^\Psi_{AB}=-\ln{\rho_{AB}}$ by: 
\be 
 K^\psi_{AB}=2 K^\Psi_{AB}
 \ee
$K^{\psi}_{A'B'}$ is an identical copy of $K^{\psi}_{AB}$. Notice that $U(s)$ is not the full modular-flow, which acts as a symmetry of the operator state. In the vicinity of $\gamma^{AB}_2(t_0)$, the action of $U(s)$ can be approximated geometrically by two identical rindler boosts about $\gamma^{AB}_2(t_0)$, with $s$ being the rapidity \cite{tom:2018, xi:2018}. This effectively generates a time-like evolution in the near-$\gamma^{AB}_2(t_0)$ region of the bulk operator state.

\subsection{Generalization to $|\rho^\alpha_{AB}(t)\rangle $}
One can generalize the construction discussed before, and consider operator states of the form considered in section \ref{sec:CFT}:
\begin{equation}
\langle \phi_1,\phi_2|\rho^\alpha_{AB} (t)\rangle = \left(\rho^\alpha_{AB}(t) \right)_{\phi_1,\phi_2}
\end{equation}
where we take $\alpha$ as an integer to begin with. By making the same assumptions (e.g. replica symmetry, existence of a dominant saddle, etc) it is easy to see that the construction for the bulk dual proceeds the same, with the only change being the tension of cosmic-brane:
\begin{equation}
T_\alpha = \frac{2\alpha -1}{8\alpha G_N} 
\end{equation}
Next, we can analytically continue $\alpha$ away from integral values. A special case occurs at the value $\alpha = \frac{1}{2}$. In this case $T_{\alpha}=0$, i.e. the cosmic brane is tension-less, and the location of the brane is given by the HRT surface in the original geometry \cite{HRT}. More importantly, since the cosmic brane does not generate time-dependent back-reaction, the ``parent" bulk geometry for constructing the dual of the operator state $|\rho^{\frac{1}{2}}_{AB}(t)\rangle$ remain the same as we evolve in $t$. The only $t$-dependent ingredient is the HRT surface across which we make the identification. 

\subsection{Operator Entanglement of $e^{-\beta/2H}|B\rangle$} 
Let us now consider the holographic picture for the operator entanglement of the generalized operator state $|\rho^{\frac{1}{2}}_{AB}(t)\rangle$, where we take the global state $|\psi\rangle$ to be pure B-state $e^{-H\beta/2 }| B\rangle$. In a holographic $\text{CFT}_d$, this is dual to half side of a two-sided external black hole with the metric:
\begin{equation}
ds^2 = -\frac{4 r_0^2}{d^2}\left(\cosh{\frac{d\rho}{2}}\right)^{4/d}\tanh^2{\left(\frac{d\rho}{2}\right)} dt^2+ \frac{4 r_0^2}{d^2}\left(\cosh{\frac{d\rho}{2}}\right)^{4/d} dy_{d-1}^2+d\rho^2,\;r_0 = \frac{2\pi}{\beta}\nonumber
\end{equation}
Furthermore it contains an end-of-the-world (EoW) brane cutting through the interior \cite{maldacena_eternal_2003}. The EoW brane can be understood as arising from extending the boundary condition on the CFT into the bulk \cite{BCFT}. A particular type of boundary condition in the bulk is that the HRT suface can end perpendicular onto the EoW brane. 

We are interested in the operator entanglement entropy: 
\begin{equation}
S^{\rm op}_A(t) = S^{\text{op}}_1(A, \rho^{\frac{1}{2}}_{AB}(t) )=-\text{Tr}\tilde{\rho}_A(t)\ln{\tilde{\rho}_A(t)},\;\tilde{\rho}_A(t)=\text{Tr}_{\mathcal{H}_B\otimes \mathcal{H}_B} |\rho^{\frac{1}{2}}_{AB}(t)\rangle \langle\rho^{\frac{1}{2}}_{AB}(t)|
\end{equation}
This coincides with the reflective entropy considered in Ref.~\cite{tom:2019}, where for static cases it was shown to be equal to twice of the entanglement wedge cross section, the latter has been conjectured as a holographic realization of the entanglement of purification \cite{EoP}. Via the HRT prescription this is obtained by finding the extremal surface $\gamma^A(t)$ in bulk geometry dual to the operator state. For $\alpha=\frac{1}{2}$, the bulk dual is given by the original dual (i.e. half-sided eternal black brane with the EoW brane) identified across the the HRT surface $\gamma^{AB}(t)$. This allows us to extract $S^{\rm op}_A(t)$ directly in the original bulk geometry by modifying the rule of finding the HRT surface $\gamma^A(t)$ as follows:
\begin{itemize}
\item{A HRT surface $\gamma^{A}(t)$ in the original geometry remains a candidate HRT surface}
\item{A candidate HRT surface can also end perpendicularly on $\gamma^{AB}(t)$.}
\item{$S^{\rm op}_A(t)  $ is given by twice the smallest area among all candidate HRT surfaces.}
\end{itemize}
In other words, $S^{\rm op}_A(t)$ in this case equals twice of the dynamical version of the entanglement wedge cross section for A inside AB. Let us focus on the case where both $AB$ and $A$ are flat stripes, ranging between $(-\ell_A,\ell_B)$ and $(-\ell_A,0)$ respectively along the spatial $y$ direction, where $0<\ell_A\leq \ell/2,\;\ell=\ell_A+\ell_B$. 

\begin{figure}[h!]
\centering
\includegraphics[scale=0.137]{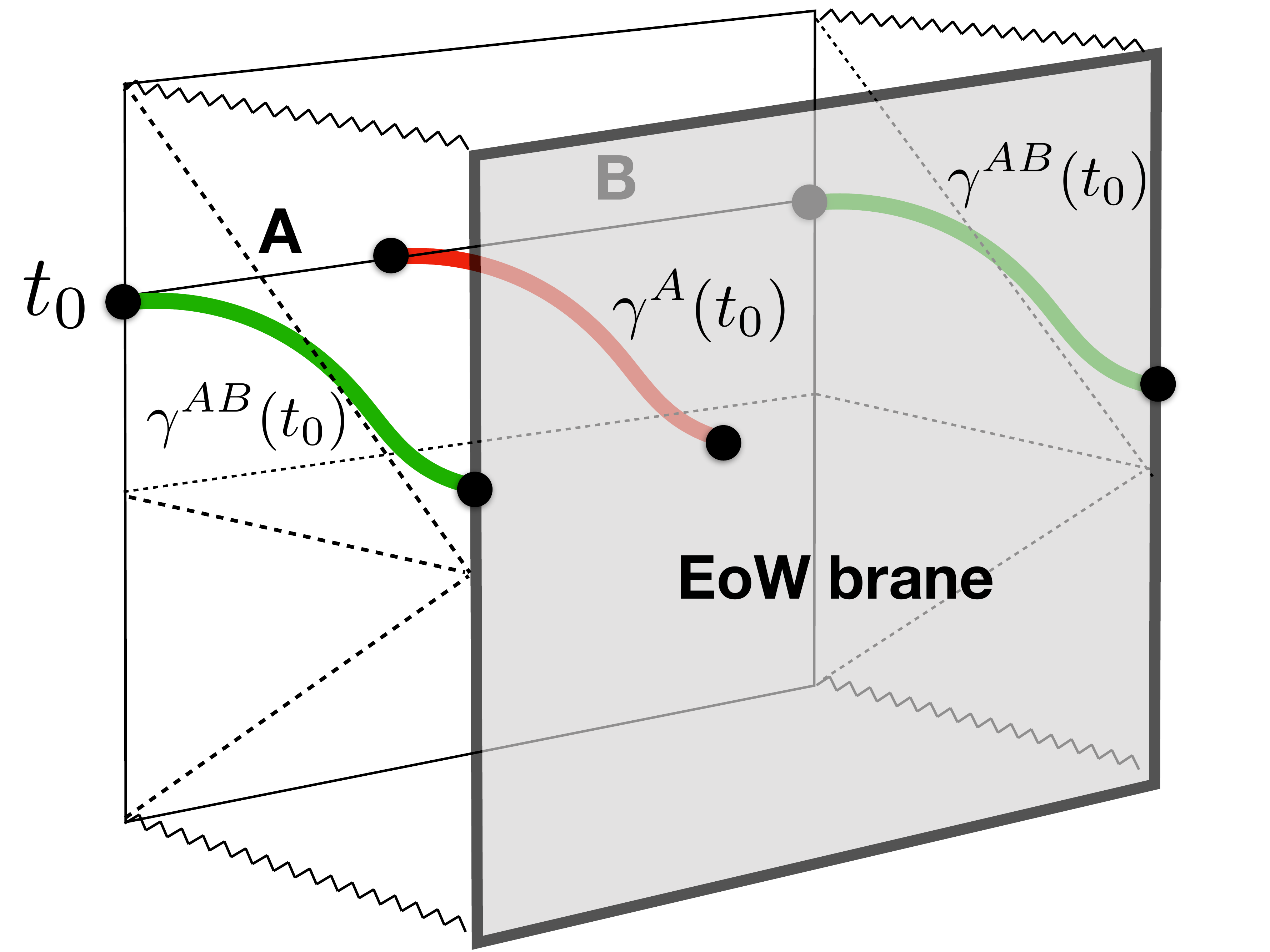}  
\includegraphics[scale=0.137]{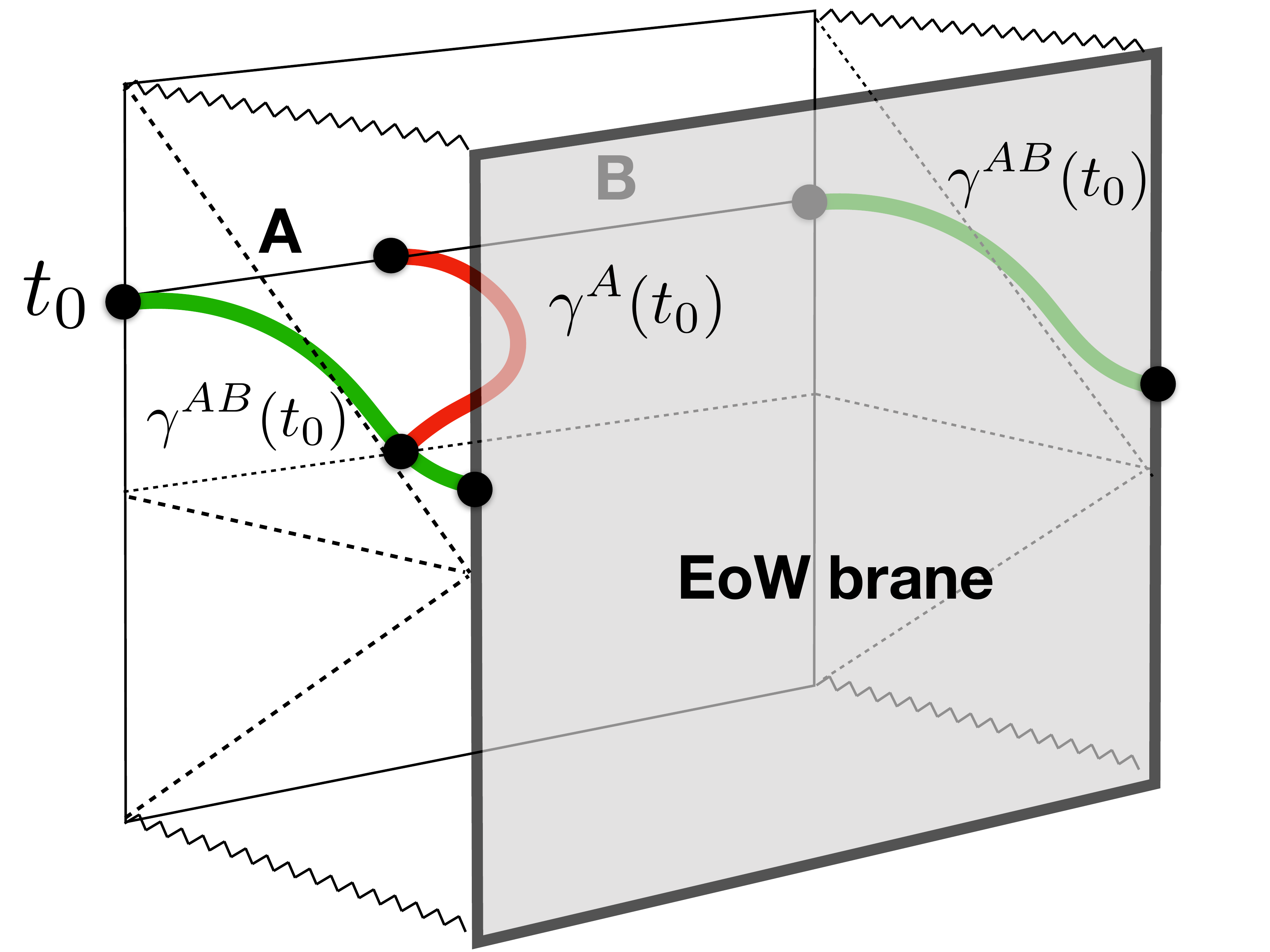}  
\includegraphics[scale=0.137]{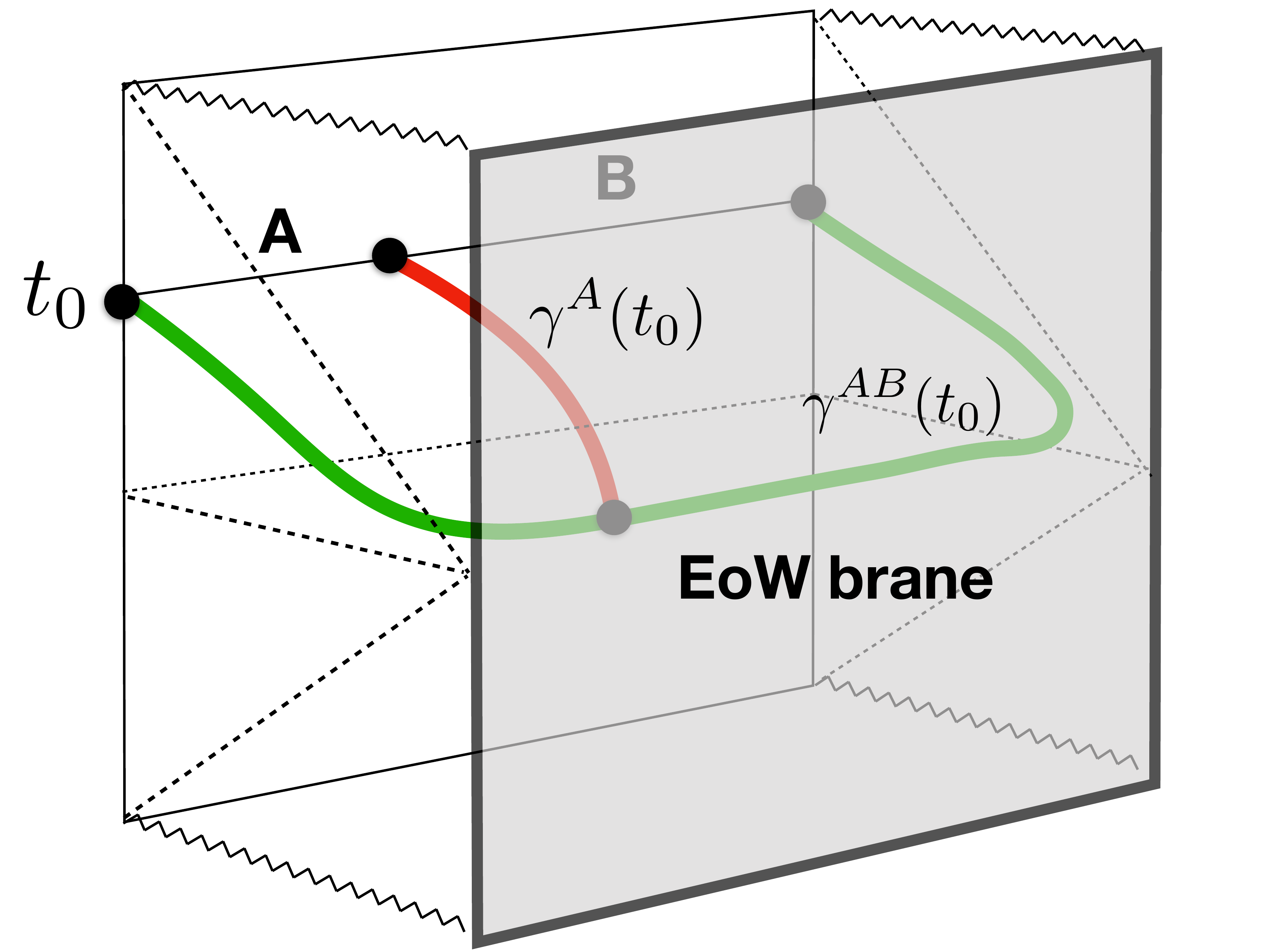}  
\caption{Phases for the HRT surfaces $\gamma^{AB}(t)$ (green) and $\gamma^A(t)$ (red) in the holographic pure B-state. Left: for $v_d t<f\ell_A$, the corresponding entanglement entropy $S^{\rm op}_A(t)$ grows linearly in $t$; Center: for $fL_A<v_d t<\ell/2$, after a phase transition for $\gamma^A(t)$ itself, $S^{\rm op}_A(t)$ settles down to be volume-law; Right: for $v_d t>\ell/2$, due to the phase transition of the underlying bulk state, i.e. the identification surface $\gamma^{AB}(t)$, $\gamma^A(t)$ is forced to jump discontinuously to be short-ranged.
}
\label{fig:holographic_phases}
\end{figure}

The first step is to determine the HRT surface $\gamma^{AB}(t)$ in the parent one-sided black brane geometry, on which  the HRT surfaces $\gamma^A(t)$ of the operator state can end. The time dependence of $\gamma^{AB}(t)$ has been studied in \cite{hartman:2013}, and exhibits two phases in the high temperature limit $\beta \ll 1$: 
\begin{itemize}
\item{For $v_d t<\ell/2$, $\gamma^{AB}(t)$ takes the form of two planes at fixed values of $y=-\ell_A$ and $\ell_B$ respectively, extending into the bulk and ending perpendicularly on the EoW brane, with the total proper area growing with time $\text{Area} \left(\gamma^{AB}(t)\right)\propto \frac{8\pi}{\beta}v_d t$;}
\item{For $v_d t>\ell/2$, $\gamma^{AB}(t)$ is a connected extremal surface that extends in the $y$ direction, most of which lie very close to the bifurcating surface $\rho=0$ and has a volume-law proper area $\text{Area} \left(\gamma^{AB}(t)\right)\propto \frac{4\pi}{\beta}\ell$.}
\end{itemize}
where the velocity $v_d=\frac{\sqrt{d}(d-2)^{1/2-1/d}}{\left[2(d-1)\right]^{1-1/d}}$ depends on the boundary CFT dimension $d$ \cite{hartman:2013}. On top of this evolution, we expect three phases for the behavior of the HRT surface $\gamma^A(t)$, summarized as in Fig.~\ref{fig:holographic_phases}:
\begin{itemize}
\item{For $v_d t< f \ell_A$ with a proportionality constant $f$ to be determined, $\gamma^A(t)$ is similar to each component of $\gamma^{AB}(t)$ at the same time, and takes the form of a plane at fixed $y=0$, extending into the bulk and ending perpendicularly on the EoW brane. During this phase, the entanglement entropy grows linearly in time: $S^{\rm op}_A(t)  \propto \frac{4\pi}{\beta}v_d t $.}
\item{For $f \ell_A<v_d t<\ell/2$, $\gamma^A(t)$ latches onto one of the components of $\gamma^{AB}(t)$. During this phase the entanglement entropy is approximately time-independent and exhibits volume-law scaling:  $S^{\rm op}_A(t)\propto \frac{4\pi}{\beta} f \ell_A$.}
\item{For $v_d t>\ell/2$, due to the transition of $\gamma^{AB}(t)$, $\gamma^A(t)$ undergoes a discontinuous transition, changing from volume-scaling to being short-ranged.}
\end{itemize}

\begin{figure}[h]
\centering
  \includegraphics[width=0.55\columnwidth]{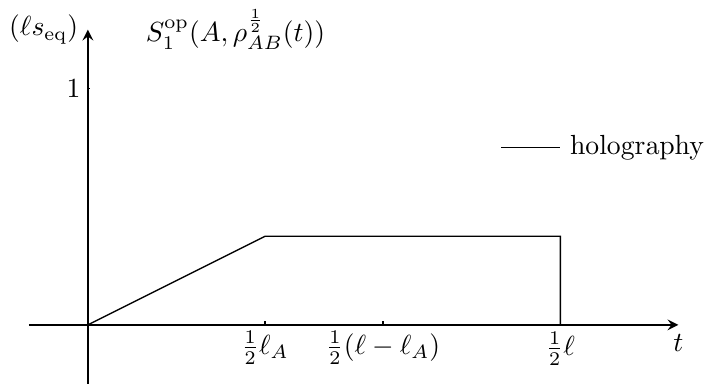}
  \label{fig:holo_res}
\caption{Results of holographic theory in AdS${}_3$/CFT${}_2$.}
\label{fig:holographic_plot}
\end{figure}
As we can see, the decline phase is absent for holographic theories, instead the plateau barrier persists until the very end of thermalization. In appendix (\ref{app:AdS3}), we calculate the operator entanglement $S^{\rm op}_A(t)$ explicitly in AdS${}_3$/CFT${}_2$, where $v_2=1$. After subtracting the usual UV divergent contribution, the results are summarized below, and plotted in Fig.~\ref{fig:holographic_plot}:
\begin{equation}
  \label{eq:AdS_3}
  S^{\rm op}_A (t) 
  = \left\lbrace 
    \begin{aligned}
      & \frac{c}{3}\ln{\left(\frac{2\pi}{\beta}\right)} +\frac{2\pi c}{3\beta} t  & \quad  t < \frac{\ell_A}{2}  \\
      &\frac{c}{3}\ln{\left(\frac{2\pi}{\beta}\right)} + \frac{\pi c}{3\beta} \ell_A & \quad \frac{\ell_A }{2} < t < \frac{\ell}{2} \\
      & \frac{c}{3}\ln{\left(\frac{2\pi}{\beta}\right)} +\mathcal{O}\left(e^{-\frac{2\pi}{\beta}\ell_{A,B} }\right) & \quad  t > \frac{\ell}{2}  
    \end{aligned} \right.
\end{equation}

This can be compared with the rational CFT results by sending $n\to 1, \alpha \to 1/2$ in Eq.~\eqref{eq:cft_qp_res}. In particular, we can identify explicitly that $f=1/2$, agreeing with the CFT results. We make a comment about this. The constant $f$ is determined by the HRT surface $\gamma^{A}(t)$ in the second phase, which has a proper area that scales with the volume $\sim \ell_A$, see Fig.~\ref{fig:holographic_phases}. Na\"ively an HRT surface that scales with volume-law would have its most part sticking to the bifurcating surface, like (half of) $\gamma^{AB}(t)$ in the third phase, see Fig.~\ref{fig:holographic_phases}. Let us call surfaces like this the thermal type. It is easy to see that the thermal type would give the incorrect answer $f=1$.

So $\gamma^{A}(t)$ has smaller proper area than the thermal type by a factor of $f$, which is $1/2$ in AdS${}_3$/CFT${}_2$. Geometrically this can be understood as follows. The thermal type for $\gamma^{A}(t)$ would emerge if we extremize all surfaces that end perpendicularly on the full time-like hyperplane $y=-\ell_A$; while the actual HRT surface comes from extremizing those that end perpendicularly on $\gamma^{AB}(t)$, which is a particular Cauchy slice of the $y=-\ell_A$ hyperplane. The extremizing procedure yielding the thermal type can be represented by a maxmin procedure \cite{Wall_2014}: foliate the relevant spacetime region into Cauchy surfaces; minimize on each Cauchy surface; then maximize the minimal values over all Cauchy surfaces. In this picture, the actual HRT surface  corresponds to a particular minimal configuration on the Cauchy surface whose boundary is $\gamma^{AB}(t)$. In the maxmin construction, it has smaller proper area than the thermal type. 

\begin{figure}[h]
\centering
\includegraphics[width=0.77\columnwidth]{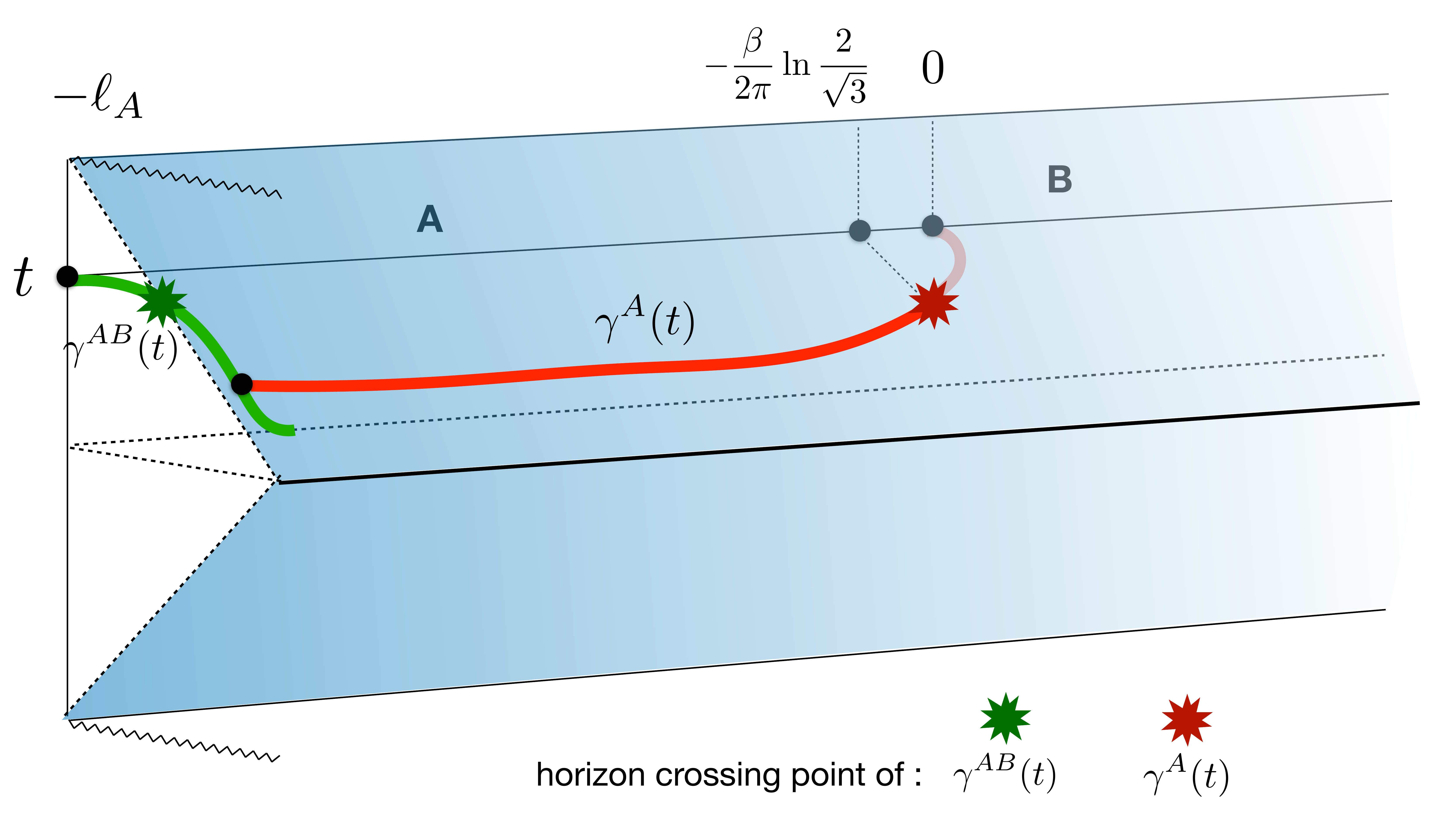}
\caption{Decomposing $\gamma^{A}(t)$ into two parts: a part outside the horizon (transparent red); a part behind the horizon (solid red).}
\label{fig:behind_horizon}
\end{figure}

As a result, $\gamma^A(t)$ does not stick to the bifurcating surface, and in fact passes through the horizon. One can then investigate what fraction of $\gamma^A(t)$ is in the interior of the black brane. In appendix~\ref{subsec:behind_horizon} we check this explicitly, and found that the entire volume-scaling part of $\gamma^A(t)$ is from behind the horizon, see Fig.~\ref{fig:behind_horizon}. More explicitly, the portion of of $\gamma^A(t)$ that is behind the horizon covers the interval in $y$ coordinate:
\begin{equation}
y\in \left(-\ell_A,-\frac{\beta}{2\pi}\ln{\frac{2}{\sqrt{3}}}\right)
\end{equation}
Such an extensive amount of entanglement from the black hole interior is only accessible from the operator entanglement. 

\section{Discussion}
\label{sec:discuss}

In this paper, we study the operator entanglement of the reduced density matrix of a quenched state in three representative systems: random unitary circuit with local interactions, rational CFT and holography. We find that in all those cases, the operator entanglement grows linearly from short-range value to an extensive value, stays constant for some time thus forming a plateau, and then drops to short-range value. These three phases has been summarized in Fig.~\ref{fig:growth-plateau-drop}, which we reproduce below in Fig.~\ref{fig:growth_plateau_drop_2} for the readers' convenience. 
\begin{figure}[h]
\centering
\includegraphics[scale =1.3]{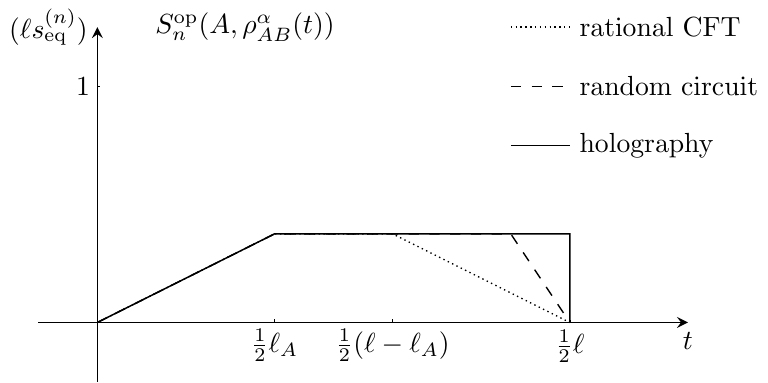}
\caption{The growth-plateau-drop pattern in three classes of models.}
\label{fig:growth_plateau_drop_2}
\end{figure}

To answer the question raised in the introduction as motivation, our results suggest that for generic non-integrable systems such as the random unitary circuits or the holographic CFTs, there exists an operator entanglement barrier of volume-law value during the thermalization of a quenched state. Hence algorithms that use a low entanglement representation can not faithfully reproduce the reduced density matrix throughout the process. This is a natural expectation for chaotic systems. On the other hand, the rational CFTs also share the operator entanglement barrier despite the evolution Hamiltonian being integrable. The reason lies in our choice of the initial global quenched state $|\psi\rangle = e^{-\beta H/4}|B\rangle$, which is a conformal invariant boundary state perturbed by the energy density operator $T_{tt}$. Such perturbation always exists in CFTs, and it has been shown in \cite{cardy_quantum_2015,wen_entanglement_2018} that the reduced density matrix $\rho_{A}(t)$ under the global quench converges to the thermal density matrix $\frac{1}{Z}e^{- \beta H} $ as $t\rightarrow \infty$. \footnote{More precisely, \cite{cardy_quantum_2015} shows that their overlap converges to 1; \cite{wen_entanglement_2018} derives the convergence in operators for geometries in which the modular Hamiltonian $K_A(t)=-\log(\rho_{A}(t))$ can be explicitly calculated.} The fate of thermalization for the initial states we considered puts the rational CFTs in equal footing as the other two chaotic models for comparison. However, it is also known that other irrelevant perturbations of the boundary state (for example by the descendants of the stress tensor) could lead to a Generalized Gibbs ensemble (GGE) \cite{rigol_relaxation_2007} at late times \cite{cardy_quantum_2015}. The breakdown of thermalization could evade the decay phase, unless the chemical potentials are all tuned to be small. It would be interesting to study how the operator entanglement behave by quenching from the general initial states.

Strictly speaking, the explicit computations we performed in different models are for different operator entanglement quantities: in rational CFT and random unitary circuit we obtained the results for general $n$-th operator R\'enyi entropy, and for operator state with general $\alpha$; in holographic CFTs what we computed is the reflective entropy, i.e. operator entanglement entropy $n\to 1$ with $\alpha=\frac{1}{2}$. That these results plotted on the same figure are showing the same plateau values therefore requires some technical explanation. First of all, the time scales of the systems are calibrated so that the effective speeds of light controlling the causality of local interactions are all set to $1$. Furthermore, we have re-scaled the entanglement quantities by their corresponding $n$-dependent equilibrium R\'enyi entropy density. More specifically, what we plot is the ratio $S^{\rm op}_n ( A, \rho^{\alpha}_{AB}(t) ) / (\ell s^{(n)}_{\text{eq}}  ) $, where $s^{(n)}_{\text{eq}}$ is the equilibrium R\'enyi entropy density of the quenched state at $t\rightarrow \infty$. More importantly, one might worry that transition time could depend on the R\'enyi index $n$, and therefore the comparison between different models in Fig.~\ref{fig:growth_plateau_drop_2} may be meaningless. We acknowledge this potential subtlety. However we point out that for integer $n$ and $\alpha$, both the CFT and random unitary circuit results are explicitly given: they are independent of $\alpha$ and $n$ only enters as a pre-factor $s_{\rm eq}^{(n)}$ (in fact, it is also $n$ independent in random circuit when $q\rightarrow \infty$.\footnote{Strictly speaking, when computing operator entanglement entropy, this is true if we take $q\rightarrow \infty$ before the replica limit. }). Therefore one can analytically continue them to $n = 1$, $\alpha = half$ and compare with holographic CFTs, and this should agree with Fig.~\ref{fig:growth_plateau_drop_2}. However, an honest comparison for generic $n$ would require considering the cosmic-brane back-reactions in holography, which is beyond our current computational control. It is possible that Fig.~\ref{fig:growth_plateau_drop_2} fails to accurately capture the case for generic $n$ (for example, the transition to the decay phase in holography maybe continuous due to back-reactions. ). We leave such investigations for the future. 


Now let us discuss the physics behind the three segments of the "growth-plateau-decay" pattern. 

During the linear growth phase, we observe that the growth rates (in units of the equilibrium entropy density) are the same among the three models. This is very similar the the linear growth behaviors of quenched state entanglement, where the rates are also equal among the models despite different entanglement dynamics. However, we should caution that the linear growth of the operator entanglement is not solely a consequence of the corresponding state entanglement increase for the subsystem $A$. During this phase, the operator state $| \rho_{AB}^{\alpha} (t) \rangle \in \mathcal{H} \otimes \mathcal{H} $ is building up entanglement between the subsystems $A \cup B$ and $A' \cup B'$, thus is drifting away from the form of a product state in $\mathcal{H}\otimes \mathcal{H}$. As a result, the operator entanglement growth rate is not twice of, but is equal to, the state rate. The linear growth stops at $t = \ell_A / 2$. This coincides with the thermalization time for A, after which the operator entanglement barrier emerges. This suggests that the transition between the two phases is possibly related to the thermalization of region A.

The plateau phase represents an intriguing "equilibrium" in the course towards thermalization. Despite the operator entanglement taking constant value, the operator state is certainly not static, for example the state R\'enyi entropy of AB is still linearly increasing. This excludes the possibility of describing the reduced density matrix by any kind of static (equilibrium) ensemble. We can ask a few questions regarding this phase. Firstly, is the calibrated plateau value universal? For example, could it represent the saturation to the maximal operator entanglement subject to the given constraints on the density matrix? Secondly, how universal is the plateau phase? For example, does it survive in chaotic systems away from the large $q$ or large $N$ limits we are considering? And finally, what are the implications for the observables? For example, does it implies any properties in the local correlation functions within region A? 

The existence of the decay phase for operator entanglement is dictated by the quenched thermalization of these models. In the high temperature limit, the thermalized reduced density matrix is approximately proportional to identity. Viewed as an operator state, it is therefore short-range entangled with correlation length of order $\beta$. However, the specific behaviors are highly model dependent; in fact, the models we study have different mechanisms for triggering the decay phase. 

The linear growth phase and the early plateau phase are the same among all three classes of models. It is the point of exit from the plateau and the subsequent decay phase that serve to differentiate among them. A similar story for the two-interval entanglement/mutual information in the quenched state is discussed in \cite{asplund_entanglement_2015}, where it is also found to encode the distinction between the chaotic and rational CFTs. However, if we consider the same quantity for the random unitary circuits, at large $q$ in large separation it exhibits the same behavior as the holographic CFTs \footnote{On the other hand, they were shown to differ for the case of $n=2$ R\'enyi mutual inormation.}. In contrast, by probing the later time behavior of the operator entanglement, the three models are fully distinguished. 

Phenomenologically we observe that how late the system exits the plateau phase is correlated with how chaotic it is: the rational CFTs exit at $t=\frac{\ell_B}{2}$; the random unitary circuits exit at $t=\frac{\ell}{2}-\frac{\ell_A}{4}=\frac{\ell_B}{2}+\frac{\ell_A}{4}$; the holographic CFTs exit at $t=\frac{\ell}{2}=\frac{\ell_B}{2}+\frac{\ell_A}{2}$. By the underlying quenched dynamics, the reduced density matrix is fully thermalized at $t=\frac{\ell}{2}$, after which the operator state becomes short-range entangled and the operator entanglement has to drop to short-range value. Therefore the exit has to occur before $t=\frac{\ell}{2}$. This is saturated by a class of maximally chaotic systems: the holographic CFTs. It is therefore tempting to suggest that the persistence of the operator entanglement barrier, calibrated using the standard quenched state $e^{-\beta H/4}|B\rangle$, can serve as an alternative measure for the quantum chaoticity of the system. 

To better check and understand the possible relation between the duration of the operator entanglement plateau and quantum chaoticity, it could be helpful in the future to study more examples of intermediate models, such as chaotic CFTs that are not holographic. Here based on the two chaotic models we study, it is still interesting to compare their mechanisms of exiting the plateau phase. In both models they are driven by geometrical transitions of the domain walls and the HRT surfaces respectively. The difference is as follows. In the random unitary circuits, the exit corresponds to a first order transition of the effective statistical mechanics, the operator entanglement is the free energy itself and thus is continuous across the exit. In particular, all the domain walls have the same tension in contributing to the free energy. In the holographic CFTs, the exit is driven by a phase transition of the bulk dual of the operator state, i.e. $\gamma^{AB}(t)$. The operator entanglement (or more specifically, reflective entropy) is given by the area of the HRT surface in the bulk, which changes by responding to the bulk dual transition, and thus is not continuous across the exit. In particular, contrary to the case in the random unitary circuits, $\gamma^{AB}(t)$ and $\gamma^A(t)$ play highly distinct roles in determining the operator entanglement. This distinction between the two models is not present when computing two-interval mutual information. This explains why the mutual information cannot distinguish between these two models as pointed out before. In the future, it would be interesting to connect this distinction with the different levels of quantum chaoticity between the two models.


\acknowledgments

We acknowledge the accommodation and interactive environment of the KITP program ``The Dynamics of Quantum Information", where the question of ``entanglement barrier'' was raised up and stimulated this work. We acknowledge Xiao Chen, Thomas Faulkner, Yuya Kusuki, Andreas W.W. Ludwig, M\'ark Mezei, Adam Nahum, Frank Pollmann, Shinsei Ryu, and Kotaro Tamaoka for insightful discussion and previous collaborations on related projects. HW is supported by the Gordon and Betty Moore Foundation
through Grant GBMF7392; 
TZ is supported by the Gordon and Betty Moore Foundation, under the EPiQS initiative, Grant GBMF4304, at the Kavli Institute for Theoretical Physics. This research was supported in part by the National Science Foundation under Grant No. NSF PHY-1748958. We acknowledge support from the Center for Scientific Computing from the CNSI, MRL: an NSF MRSEC (DMR-1720256) and NSF CNS-1725797.

\appendix
\section{AdS${}_3$/CFT${}_2$ Calculations}
\label{app:AdS3}

In this appendix we carry out the holographic calculations explicitly in AdS${}_3$/CFT${}_2$, where one can use the fact that BTZ black holes are locally isometric to AdS${}_3$ and can be obtained as a quotient of it. This would allow us to explicitly solve for the HRT surfaces/geodesics. To simplify the discussion, we will study the TFD state, dual to the (unwrapped) two-sided BTZ black hole. The answer for the pure B-state is simply given by half of the TFD results. 

The right exterior of BTZ black hole geometry is: 
\begin{equation}
\label{eq:BTZ_metric}
ds^2 = -r_0^2 \sinh^2{(\rho)} dt^2+ r_0^2 \cosh^2{(\rho)} dy^2+d\rho^2
\end{equation}
The AdS scale is set to 1, and the temperature is given by $\beta = 2\pi/r_0$. The left exterior has the same geometry and can be obtained by analytically continue $r_0t\to r_0 t+i\pi$. In addition, the future interior of the BTZ black hole can be obtained by $r_0t\to r_0t+i\pi/2,\;\rho\to i\alpha$ and has the geometry:
\begin{equation}
 ds^2 =r_0^2 \sin^2{(\alpha)} dt^2+r_0^2 \cos^2{(\alpha)} dy^2-d\alpha^2 
\end{equation}
Had we wrapped the $y$ direction to be compact, there will be a compact null circle at $\alpha=\pi/2$ and this corresponds to the black hole singularity. The two exterior and interior regions are all isometric to parts of the poincare patch for AdS${}_3$: 
\begin{equation}
ds^2=\frac{1}{z^2}\left[-dx_0^2+dx_1^2+dz^2\right] 
\end{equation}
via the identification of coordinates: 
\begin{eqnarray} 
&&\sinh{\left(r_0 t\right)}\sinh{\rho}=\frac{x_0}{z}\nonumber\\
&&\cosh{\left(r_0 t\right)}\sinh{\rho}=\frac{x_1}{z}\nonumber\\
&&\tanh{\left(r_0 y\right)}=\frac{z^2-x_0^2+x_1^2-1}{z^2-x_0^2+x_1^2+1}
\end{eqnarray}
and analytic continuations thereafter. Towards the asymptotic boundary of the right exterior $z\to 0$ ($\rho\to\infty$), the boundary coordinates can be directly identified: 
\begin{equation}
x_1+x_0\to e^{r_0(y+t)},\;\;x_1-x_0\to e^{r_0(y-t)} 
\end{equation}
Similarly, towards the left exterior we have (by sending $t\to -t+i\pi$): 
\begin{equation}
x_1+x_0\to -e^{r_0(y-t)},\;\;x_1-x_0\to -e^{r_0(y+t)} 
\end{equation}
Denote the end points of the interval $AB$ on the right asymptotic boundary by $(P_1,P_3)$, and the other end point of the sub-interval $A$ by $P_2$; the corresponding end points on the left asymptotic boundary are denoted by $P_{i=1,2,3}'$. Taking the set up as in Sec.~\ref{sec:CFT}, they are mapped into the poincare-patch coordinates $(x_0,x_1)$ as: 
\begin{eqnarray}
 P_1 &=&e^{-r_0 \ell_A}\left(\sinh{r_0 t}, \cosh{r_0 t}\right),\;P_2=\left(\sinh{r_0 t}, \cosh{r_0 t}\right),\;P_3=e^{r_0 \ell_B}\left(\sinh{r_0 t}, \cosh{r_0 t}\right)\nonumber\\
 P_1'&=&e^{-r_0 \ell_A}\left(\sinh{r_0 t}, -\cosh{r_0 t}\right),\;P_2'=\left(\sinh{r_0 t}, -\cosh{r_0 t}\right),\;P_3'=e^{r_0 \ell_B}\left(\sinh{r_0 t}, -\cosh{r_0 t}\right)\nonumber
\end{eqnarray}

Via this map we can from now on carry out the analysis in the poincare patch, in which generic space-like geodesics $\left\lbrace x_0(z),x_1(z)\right\rbrace$ take the following form of ``circles":
\begin{equation}\label{eq:space-like}
x_0(z)=\sqrt{R^2- z^2}\sinh{\eta}+C_0,\;\;x_1(z)=\sqrt{R^2- z^2}\cosh{\eta}+C_1
\end{equation}
if the trajectories projected onto the $(x_0,x_1)$ plane are space-like; otherwise the geodesics take the form: 
\begin{equation}\label{eq:time-like}
x_0(z)=\sqrt{R^2+ z^2}\cosh{\eta}+C_0,\;\;x_1(z)=\sqrt{R^2+ z^2}\sinh{\eta}+C_1
\end{equation}
\subsection{Phase 1}\label{subsec:AdS3_phase_1}
In the first phase, the segments for the HRT surfaces $\gamma^{AB}(t)$ that construct the operator state $|\psi^{1/2}_{AB}(t)\rangle$, and the HRT surface $\gamma^A(t)$ that computes the operator entanglement, are half-circles $\gamma^{AB}(t)= \widehat{P_1 P_1'}\cup  \widehat{P_3 P_3'},\;\gamma^A(t)= \widehat{P_2 P_2'}$ that connect $(P_1,P_1'), (P_3,P_3')$ and $(P_2,P_2')$ respectively, as studied in Ref.~\cite{hartman:2013, shinsei:2018}. They all pass through the black hole interior, and are described by Eq.~\eqref{eq:space-like} with $\eta=0, C_1=0$ (see the left of Fig.~\ref{fig:AdS3_phases}): 
\begin{eqnarray} 
\widehat{P_1 P_1'}:\;\; x_0(z)&=& e^{-r_0 \ell_A}\sinh{(r_0 t)},\;\;x_1(z)=\sqrt{e^{-2r_0 \ell_A}\cosh^2{(r_0 t)}- z^2}\nonumber\\
\widehat{P_2 P_2'}:\;\; x_0(z)&=& \sinh{(r_0 t)},\;\;x_1(z)=\sqrt{\cosh^2{(r_0 t)}- z^2}\nonumber\\
\widehat{P_3 P_3'}:\;\; x_0(z)&=& e^{r_0 \ell_B}\sinh{(r_0 t)},\;\;x_1(z)=\sqrt{e^{2r_0 \ell_B}\cosh^2{(r_0 t)}- z^2}
\end{eqnarray}
The operator entanglement for the TFD state is given by twice the geodesic length of $\widehat{P_2 P_2'}$ multiplied by $\frac{1}{4G_N}=\frac{c}{6\ell}$:
\begin{eqnarray}
S^{\text{TFD}}_A(t)&=&\frac{2c}{3} \int^{\cosh^2(r_0 t)}_0 \frac{dz}{z}\frac{\cosh{(r_0 t)}}{\sqrt{\cosh^2{(r_0 t)}-z^2}}\nonumber\\
&=&\frac{2c}{3}\ln{\left[\cosh(r_0 t)\Lambda \right]}\approx \frac{4\pi c}{3\beta} t + S_{\text{div}}
\end{eqnarray}
where $\Lambda$ is the UV cut-off we placed towards the asymptotic boundary at $z=1/\Lambda$. The operator entanglement for the pure B-state is 
\begin{equation}
S_A(t)\approx \frac{2\pi c}{3\beta}t+S_{\text{div}} 
\end{equation}
The same result also holds for the geodesic length of $\widehat{P_1 P_1'}$ and $\widehat{P_3 P_3'}$, after taking into account of a $y$-dependent UV $z$ cut-off, which corresponds to a uniform UV $\rho$ cut-off in the original geometry via $1/z\sim e^{\rho-r_0 y}$. These are the linear growth identified in Ref.~\cite{hartman:2013}.

\subsection{Phase 2}\label{subsec:AdS3_phase_2}
In this phase, $\gamma^{AB}(t)$ remains the same phase as before, while $\gamma^A(t)$ undergoes a transition from a half-circle $\gamma^A(t)=\widehat{P_2 P_2'}$ to a union of two segments $\gamma^A(t)=\widehat{P_2 \zeta}\cup \widehat{P_2' \zeta'}$ (see the center of Fig.~\ref{fig:AdS3_phases}). The two segments end orthogonally on $\widehat{P_1 P_1'}$ at $\zeta, \zeta'\in \widehat{P_1 P_1'}$ respectively. Both segments are of the form Eq.~\eqref{eq:space-like} and one is the reflection of another across $x_1=0$, so we focus on the one of them $\widehat{P_2 \zeta}$ and study the orthogonality condition between $\alpha = \widehat{P_1 P_1'}, \beta = \widehat{P_2 \zeta}$: 
\begin{eqnarray}
\label{eq:geo_phase_2}
\widehat{P_1 P_1'}:\;\; x^\alpha eta_0(z)&=&\pm \sqrt{R^2- z^2}\sinh{\eta}+C_0,\;\;x^\beta_1(z)=\pm \sqrt{R^2- z^2}\cosh{\eta}+C_1
\end{eqnarray}
The $(\pm)$ branches in $\widehat{P_2\zeta}$ is determined by whether the segment has passed the turning point at $z=R$. In order to satisfy orthogonality as well as the boundary condition we must solve: 
\begin{eqnarray}
&&x^\beta_0 (0)= \sinh{(r_0 t)},\;x^\beta_1(0)=\cosh{(r_0 t)}\nonumber\\
&& x^\beta_0 (\zeta_z)= e^{-r_0 \ell_A}\sinh{(r_0 t)},\;x^\beta_1(\zeta_z)=x^\alpha_1(\zeta_z)\nonumber\\
&& \partial_z x^\beta_1(\zeta_z)\partial_z x^\alpha_1(\zeta_z)+1=0
\end{eqnarray}
The last equation is satisfiable only if $\zeta$ is on the $(-)$ branch of $\widehat{P_2\zeta}$. This is a set of 5 equations for 5 variables $\zeta_z, \eta, R, C_1, C_2$, so there are a discrete number of solutions. One can eliminate the irrelevant integration constants $C_{1,2}$ and obtain the following equations: 
\begin{eqnarray}
\cosh{(\eta)}\left(R+\sqrt{R^2-\zeta_z^2}\right)&=&\cosh{(r_0 t)}\sqrt{R^2-\zeta_z^2} \nonumber\\
\sinh{(\eta)}\left(R+\sqrt{R^2-\zeta_z^2}\right)&=&\sinh{(r_0 t)}\left(1-e^{-r_0 \ell_A}\right)\nonumber\\
R-\sqrt{R^2-\zeta_z^2}&=& e^{-r_0 \ell_A}\sqrt{1- e^{2r_0 \ell_A}\cosh^{-2}{(r_0 t)}\zeta_z^2}
\end{eqnarray} 
To obtain solutions, we define $\epsilon^2=e^{-r_0 \ell_A}$ and $ \kappa^2 = e^{r_0 \ell_A}\cosh^{-2}{(r_0 t)}$. In the high temperature limit, $\epsilon\ll 1$ is a small parameter, it is easy to solve to the first few orders in $\epsilon$ expansion of the solution:
\begin{eqnarray}
\zeta_z &=& \sqrt{2\left(\sqrt{\kappa^2+1}-\kappa\right)} \epsilon + \frac{3\kappa^2-\kappa-2+\sqrt{1+\kappa^2}}{2\sqrt{2}(\kappa+1)} \epsilon^3+ ...\nonumber\\
R &=&1 + \left(\frac{3\kappa^2-\kappa-2}{2}+\sqrt{1+\kappa^2}\right)\epsilon^2+...\nonumber\\
e^{\eta} &=&\frac{1}{\kappa \epsilon}+\left(\frac{1}{2}-\kappa-\frac{\sqrt{1+\kappa^2}}{2\kappa}\right)\epsilon+...
\end{eqnarray}
In the current regime, $t>\ell_A/2$ and therefore $\kappa \ll 1$, one can further expand in $\kappa$ the leading order solution in $\epsilon$: 
\begin{equation}\label{eq:sol_phase_2}
\zeta_z \approx \sqrt{2}e^{-r_0 \ell_A/2},\;R\approx 1,\;e^{\eta} \approx \cosh{(r_0 t)} 
\end{equation}
We can use this result to compute the geodesic length of $\widehat{P_2\zeta}$: 
\begin{eqnarray}\label{eq:length_phase_2}
\mathcal{L}(P_2,\zeta) &=& \int^R_0 \frac{dz}{z}\frac{R}{\sqrt{R^2-z^2}}+\int^R_{\zeta_z} \frac{dz}{z}\frac{R}{\sqrt{R^2-z^2}}\nonumber\\
&\approx &  2\ln{R}-\ln{\zeta_z}+\ln{\Lambda}
\end{eqnarray}
The operator entanglement for the TFD state is equal to 4 times $\frac{\mathcal{L}(P_2,\zeta)}{4G_N}$, from both the left exterior region and the other copy across $\gamma^{AB}(t)$:
\begin{equation}
S^\text{TFD}_A(t)\approx \frac{2\pi c}{3\beta} \ell_A  + S_\text{div}
\end{equation}
The operator entanglement for the pure B-state is thus: 
\begin{equation}
S_A(t)\approx \frac{\pi c}{3\beta} \ell_A  + S_\text{div}
\end{equation}
\subsection{Phase 3}\label{subsec:AdS3_phase_3}
\begin{figure}[h!]
\centering
\includegraphics[scale=0.137]{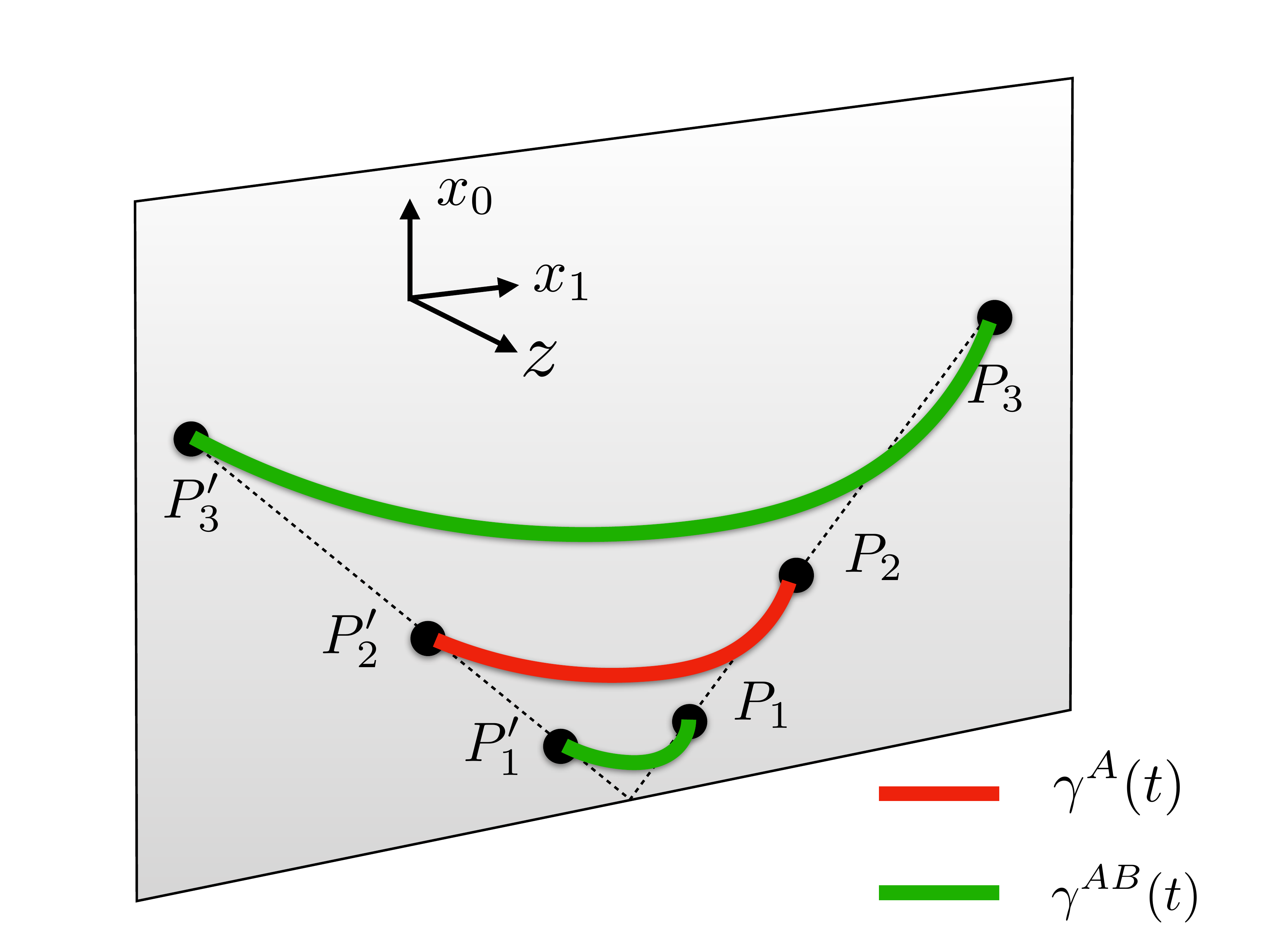}  
\includegraphics[scale=0.137]{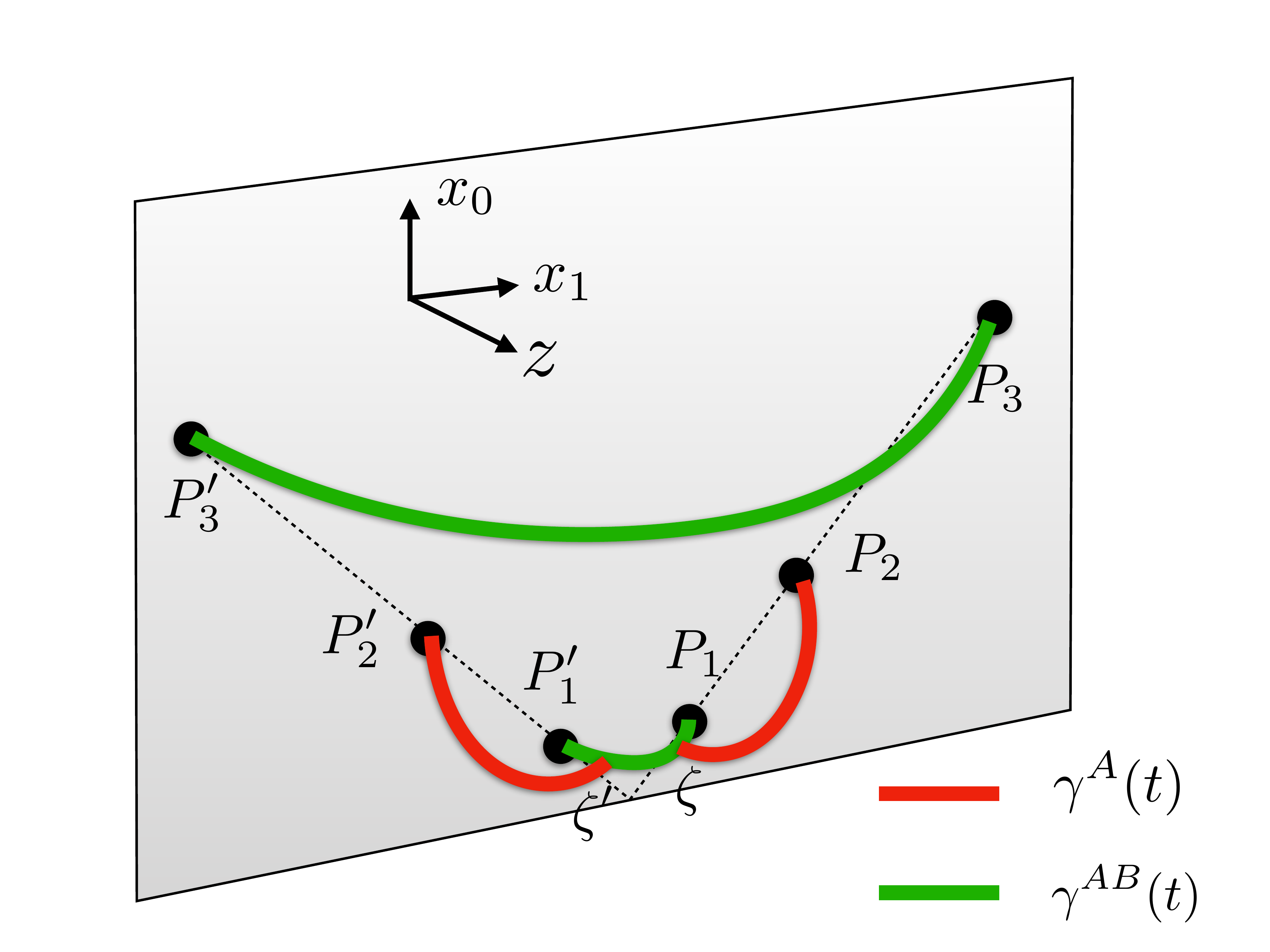}  
\includegraphics[scale=0.137]{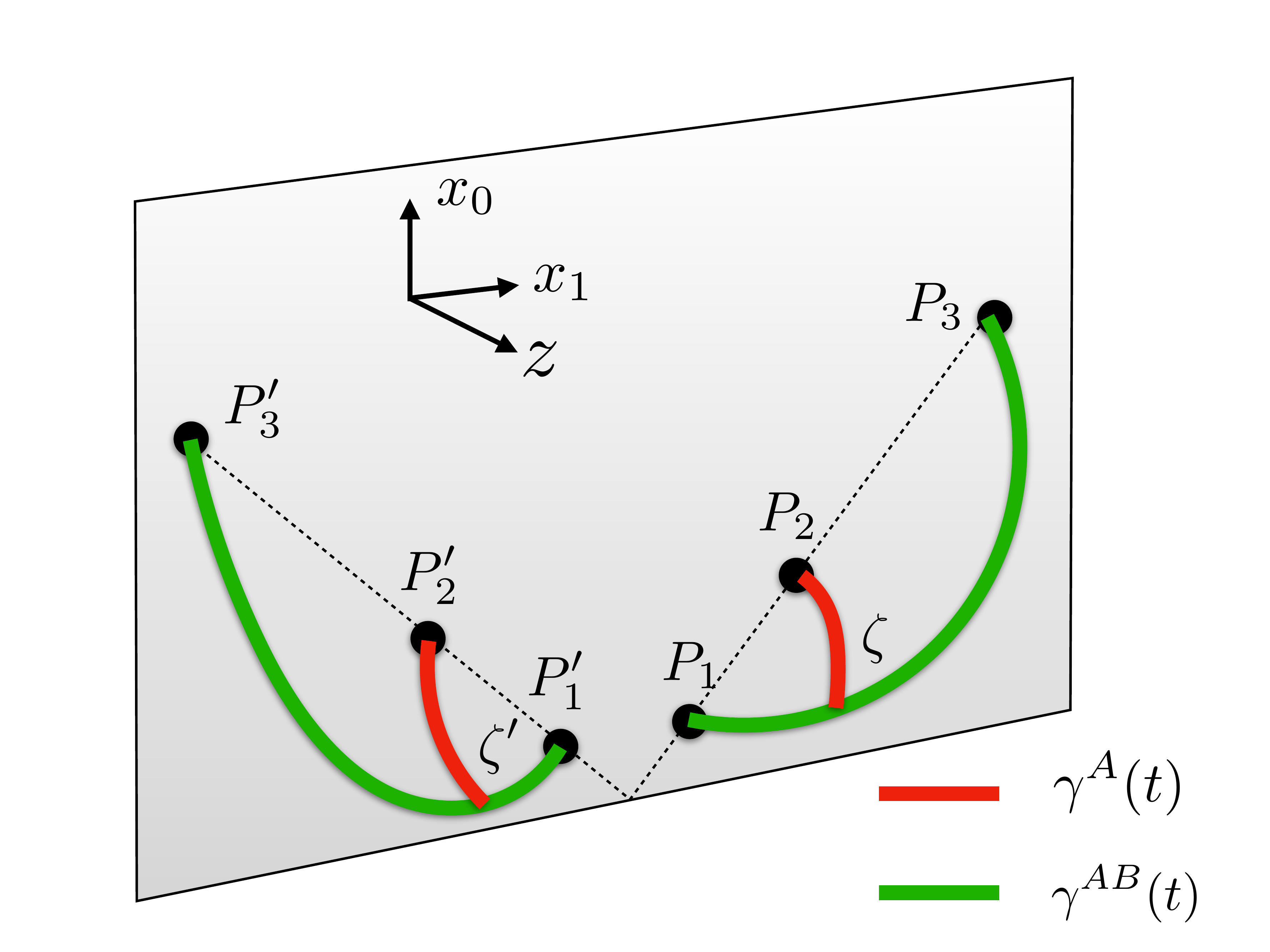}  
\caption{Phases for geodesics $\gamma^{AB}(t)$ (green) and $\gamma^A(t)$ (red) in the (unwrapped) two-sided BTZ black hole, mapped into circles of an AdS${}_3$ poincare patch.
}
\label{fig:AdS3_phases}
\end{figure}

In this phase, $\gamma^{AB}(t)$ undergoes a phase transition from $\gamma^{AB}(t)=\widehat{P_1 P_1'}\cup \widehat{P_3 P_3'}$ to $\gamma^{AB}(t)=\widehat{P_1 P_3}\cup \widehat{P_1' P_3'}$. Responding to this background transition, $\gamma^{A}(t)$ transits from $\gamma^{A}(t)=\widehat{P_2 \zeta}\cup \widehat{P_2' \zeta'}$ to $\gamma^{A}(t)=\widehat{P_2 \chi}\cup \widehat{P_2' \chi'}$ that end perpendicularly on $\widehat{P_1 P_3}$ and $\widehat{P_1' P_3'}$ at $\chi$ and $\chi'$ respectively (see the right of Fig.~\ref{fig:AdS3_phases}). 

We focus on the $\left(\widehat{P_1 P_3}, \widehat{P_2 \chi}\right)$ system. The geodesics are collinear projected onto the $(x_0, x_1)$ plane, so the geodesic solutions are: 
\begin{eqnarray}
\widehat{P_1 P_3}: x_1(z)&=&\cosh{(r_0 t)}\left(\pm\sqrt{R_1^2-z^2}+Q_1\right),\;x_0(z)=\sinh{(r_0 t)}\left(\pm\sqrt{R_1^2-z^2}+Q_1\right)\nonumber\\
\widehat{P_2 \chi}: x_1(z)&=&\cosh{(r_0 t)}\left(\pm\sqrt{R_2^2-z^2}+Q_2\right),\;x_0(z)=\sinh{(r_0 t)}\left(\pm\sqrt{R_2^2-z^2}+Q_2\right)\nonumber\\
R_1&=&\frac{e^{r_0 \ell_B}-e^{-r_0 \ell_A}}{2},\;Q_1=\frac{e^{r_0 \ell_B}+e^{-r_0 \ell_A}}{2},\;R_2+Q_2 = 1
\end{eqnarray}
The geodesic length of $\widehat{P_1 P_3}$ is: 
\begin{equation}
\mathcal{L}(P_1,P_3)= \ln{\left(R_1^2 \Lambda^2 e^{r_0 (\ell_A-\ell_B)}\right)}\approx \ell r_0 (\ell_A+\ell_B)+\mathcal{L}_{\text{div}}
\end{equation}
Following a similar procedure as before, we can solve for the geodesic solution for $\widehat{P_2 \chi}$ perturbatively in $e^{-r_0 \ell_B}$ and $e^{-r_0 \ell_A}$:
\begin{eqnarray} 
R_2 &=& 1-3 e^{-r_0 \ell_B}- e^{-r_0 \ell_A}+18 e^{-2r_0 \ell_B}+6 e^{-r_0(\ell_A+\ell_B)}+...\nonumber\\
\chi_z &=& 1-3 e^{-r_0 \ell_B}- e^{-r_0 \ell_A}+16 e^{-2r_0 \ell_B}+6 e^{-r_0(\ell_A+\ell_B)}+...
\end{eqnarray}
The intersection point $\chi$ is on the $(-)$ branch, so the geodesic length of $\widehat{P_2 \chi}$ is: 
\begin{eqnarray}
\mathcal{L}(P_2, \chi) &=& \int^{R_2}_0 \frac{dz}{z}\frac{R_2}{\sqrt{R_2^2-z^2}}+ \int^{R_2}_{\chi_z} \frac{dz}{z}\frac{R_2}{\sqrt{R_2^2-z^2}}\nonumber\\
&=& \ln{\left(\frac{R_2 (R+\sqrt{R_2^2-\chi_z^2})\Lambda}{\chi_z}\right)}\approx \mathcal{O}\left(e^{-r_0 \ell_{A,B}}\right)+\mathcal{L}_{\text{div}}
\end{eqnarray}
This is exponentially suppressed. To relate these results with the CFT result Eq.~\eqref{eq:cft_qp_res} more precisely, we need to define the UV cut-off scale with a factor of $r_0=2\pi/\beta$. This can be understood as coming from the factor $r_0^2$ in the $(dt,dy)$ part of the metric in Eq.~\eqref{eq:BTZ_metric}. 

\subsection{Extensive entanglement from behind the horizon}\label{subsec:behind_horizon}
An interesting question for the second phase \ref{subsec:AdS3_phase_2} is whether and how much of the HRT surface $\gamma^A(t)\sim \widehat{P_2 \zeta}$ remains in the black hole interior. To answer this, recall that the horizon for the BTZ black hole is mapped into the following null-sheets in the poincare patch of AdS${}_3$: 
\begin{equation}
\mathcal{H}^{\pm}= \left\lbrace (x_0,x_1,z): x_0 = \pm x_1, z\in \mathbb{R}^+\right\rbrace 
\end{equation} 
emanating from the bifurcating surface at $\lbrace x_0=x_1=0, z\in\mathbb{R}^+\rbrace$. The black hole interior corresponds to $\lbrace |x_0|>|x_1|, z\in\mathbb{R}^+\rbrace $. From Eq.~\eqref{eq:geo_phase_2} and Eq.~\eqref{eq:sol_phase_2} it is easy to see that the end point $\zeta$ is in the interior, although very close to the horizon from behind: 
\begin{equation}
\frac{x_1(\zeta)}{x_0(\zeta)}= \sqrt{\coth^2{(r_0 t)}-\frac{e^{2r_0 \ell_A}\zeta_z^2}{\sinh^2{(r_0 t)}}}\approx 1-8 e^{-r_0 (2t - \ell_A)}+...
\end{equation}
On the hand, it is also easy to work out where $\widehat{P_2 \zeta}$ crosses the horizon by solving $x^\beta_0(z^c)=x^\beta_1(z^c)$ of Eq.~\eqref{eq:geo_phase_2}:
\begin{equation}
z^c = \frac{\sqrt{3}}{2}+\mathcal{O}(\epsilon,\kappa),\;\; x^c_0 = x^c_1 = \frac{3}{8}e^{r_0 t}\left(1+\mathcal{O}(\epsilon,\kappa)\right)
\end{equation}
This is on the $(+)$ branch of $\widehat{P_2 \zeta}$, i.e. before the turning point. The contribution to the operator entanglement from the black hole interior can thus be computed from:
\bea 
\mathcal{L}_{\text{int}} &=& \int^R_{z^c} \frac{dz}{z}\frac{R}{\sqrt{R^2-z^2}}+\int^R_{\zeta_z} \frac{dz}{z}\frac{R}{\sqrt{R^2-z^2}}
\approx   r_0\ell_A/2+\mathcal{O}(1)
\eea 
This captures the entire volume-law contribution to $S_A(t)$. In this phase, we have extensive operator entanglement from behind the horizon. To have a better picture for what $\gamma^{A}(t)$ looks like in black brane geometry, let us express these in terms of the $y$ coordinate representing the spatial dimension:
\be 
r_0 y = \frac{1}{2}\ln{\left(z^2-x_0^2+x_1^2\right)}\nonumber
\end{equation}
Recall that the projection of $\gamma^A(t)$ onto the $y$-coordinate covers the interval $\lbrace y\in \mathbb{R}: -\ell_A \leq y \leq 0\rbrace$, the horizon-crossing occurs at: 
\begin{equation}
y_c \approx \frac{\beta}{2\pi}\ln{\left(\frac{\sqrt{3}}{2}\right)} 
\end{equation}
which stays fixed as we take $\ell_A\gg \beta$. This thus directly demonstrates that the interior segment $\lbrace y\in \mathbb{R}: -\ell_A \leq y \leq y_c \rbrace$ covers the entire volume contribution. Furthermore, the interior segment sticks very close to the horizon, though from behind. 
 
\bibliographystyle{JHEP}
\bibliography{tdvp_barrier_paper}

\end{document}